\newif\iflong
\longtrue

\iflong
\documentclass[10pt, conference]{IEEEtran}
\else
\documentclass[conference,compsoc]{IEEEtran}
\fi
\IEEEoverridecommandlockouts

\usepackage{tikz}
\usetikzlibrary{arrows.meta,decorations.pathmorphing,decorations.footprints,fadings,calc,trees,mindmap,shadows,decorations.text,patterns,positioning,shapes,matrix,fit, decorations.pathreplacing,calligraphy,backgrounds}
\usepackage{amsfonts}
\usepackage{makecell}
\usepackage{amsmath}
\usepackage{float}

\usepackage{amsmath}
\DeclareMathOperator*{\argmax}{arg\,max}

\usepackage[hyphens]{url}
\usepackage{hyperref} 
\usepackage{accents}
\usepackage{caption}
\usepackage{subcaption}
\usepackage{booktabs}
\usepackage{stackengine}
\usepackage{algorithm2e}
\RestyleAlgo{ruled}
\SetKwComment{Comment}{\#}{}
\usepackage{enumitem}
\setlist[enumerate]{align=parleft,left=0pt..1em}
\setlist[itemize]{align=parleft,left=0pt..1em}

\stackMath
\newcommand\tsup[2][2]{%
 \def\useanchorwidth{T}%
  \ifnum#1>1%
    \stackon[-.5pt]{\tsup[\numexpr#1-1\relax]{#2}}{\scriptscriptstyle\sim}%
  \else%
    \stackon[.5pt]{#2}{\scriptscriptstyle\sim}%
  \fi%
}

\newcommand{\new}[1]{\textcolor{black}{#1}}
\newcommand{\ournew}[1]{\textcolor{black}{#1}}

\newcommand{\ie}{{i.e.,~}}
\newcommand{\eg}{{e.g.,~}}
\newcommand{\etal}{{et~al.}}
\newcommand{\TT}[1]{``\textit{#1}''}

\newcommand\myeq{\mkern1.5mu{=}\mkern1.5mu}
\newcommand\mycdot{\mkern1.5mu{\cdot}\mkern1.5mu}

\newcommand\myin{\mkern1.5mu{\in}\mkern1.5mu}

\newcommand{\ks}{\mathcal{X}}
\newcommand{\ainf}{A_{\text{inf}}}

\newcommand{\sainf}{\tsup{A}_{\text{inf}}}

\newcommand{\lct}{\mathbf{L}_{\text{train}}}
\newcommand{\lcv}{\mathbf{L}_{\text{test}}}

\newcommand{\pwv}{v^w}
\newcommand{\cs}{\psi}
\newcommand{\dec}{f}
\newcommand{\enc}{\beta_{\theta}}
\newcommand{\senc}{\eta}

\newcommand{\hash}{\textit{H}}
\newcommand{\uncm}{\texttt{UNCM}}
\newcommand{\uncms}{\texttt{UNCMs}}

\newcommand{\targetedatt}{castelluccia2013privacy, targuess, li2017personal}

\newcommand{\uni}{\dec_{\dot{\Theta}}}
\newcommand{\seeded}{\dec_{\Theta | \cs}}

\newcommand{\citday}{\texttt{Cit0day}}

\begin{document}

\iflong
\title{Universal Neural-Cracking-Machines: \\ \huge{Self-Configurable Password Models from Auxiliary Data}$^*$}
\else
\title{Universal Neural-Cracking-Machines: \\ Self-Configurable Password Models from Auxiliary Data}
\fi

\author{
	\IEEEauthorblockN{Dario Pasquini}
	\IEEEauthorblockA{SPRING Lab, EPFL\\
		Lausanne, Switzerland\\ \url{dario.pasquini@epfl.ch}}
			\and
	\IEEEauthorblockN{Giuseppe Ateniese}
	\IEEEauthorblockA{George Mason University\\ Fairfax, Virginia, USA \\ 
		\url{ateniese@gmu.edu}}
			\and
	\IEEEauthorblockN{Carmela Troncoso}
	\IEEEauthorblockA{SPRING Lab, EPFL\\
		Lausanne, Switzerland\\  \url{carmela.troncoso@epfl.ch}}
}

\maketitle
\iflong
\renewcommand{\thefootnote}{\fnsymbol{footnote}}
\footnotetext{$^*$An abridged version of this paper appears in the proceedings of the 45th IEEE Symposium on Security and Privacy S\&P 2024.}
\renewcommand{\thefootnote}{\arabic{footnote}}
\fi

\begin{abstract}
We introduce the concept of \emph{\TT{universal} password model}---a password model that, once pre-trained, can automatically adapt its guessing strategy based on the target system. To achieve this, the model does not need to access any plaintext passwords from the target credentials. Instead, it exploits users' auxiliary information, such as email addresses, as a proxy signal to predict the underlying password distribution.
\par

 Specifically, the model uses deep learning to capture the correlation between the auxiliary data of a group of users (\eg users of a web application) and their passwords. It then exploits those patterns to create a tailored password model for the target system at inference time. No further training steps, targeted data collection, or prior knowledge of the community's password distribution is required.
\par

Besides improving over current password strength estimation techniques\iflong\ and attacks\fi, the model enables any end-user (\eg system administrators) to autonomously generate tailored password models for their systems without the often unworkable requirements of collecting suitable training data and fitting the underlying machine learning model. Ultimately, our framework enables the democratization of well-calibrated password models to the community, addressing a major challenge in the deployment of password security solutions at scale.

\end{abstract}
\IEEEpeerreviewmaketitle
\section{Introduction}
\label{sec:intro}
Passwords are still an integral part of our modern lives. Their true purpose is to provide a layer of protection for our personal data, which is always more susceptible to the risk of being \textit{hacked}~\cite{catt}.
For decades there has been a heated debate over whether or not we should ditch passwords and rely on more advanced forms of authentication, such as biometrics and hardware-based authentication~\cite{6234436, newauth}. However, that argument underestimates the potential for vulnerabilities in these techniques too. 
\iflong
Furthermore, relying solely on these methods may shoulder a heavy burden for better outfitted and affluent populations but ignore the segments of society who are unable to equip themselves with these newer technologies due to financial or institutional constraints. 
\fi

Password managers enable users to have unique, strong passwords for different accounts without having to remember them on their own. However, they have also been judged negligent in recent security breaches~\cite{lastpass, lastpass2}. If the company goes down or someone manages to hack into it, all the data encrypted with their systems may become exposed~\cite{recentattack} or inaccessible.
Users should still use password managers and ensure they have plenty of different, \textit{hard-to-guess} passwords. Yet, it is important for all parties involved to understand that this system requires more trust from the user than traditional passwords~\cite{lastpass2}.

Passwords are not dead, they are just evolving. We trust that they still provide an effective fallback method for security verification, even if biometrics or hardware authentication do not work.
In fact, passwords are still the most popular form of authentication in use today. However, it is widely recognized that human-generated passwords have inherent flaws. To ensure that authentication systems provide a meaningful level of resilience, the security community has been working to create a range of tools and methodologies centered around passwords. Those include but are not limited to: reactive security analysis~\cite{6234434, 190978}, proactive approaches such as Password Strength Meters (PSMs)~\cite{howdoes, doesmypassword}, breach alert systems~\cite{migp}, honey-encryption-based password vaults~\cite{honeyenc, 9833598}, and tailored hash functions~\cite{dash}. While explicating very different purposes, almost all these techniques rely on the same fundamental and critical component---a so-called \textbf{password model}. This is a data-driven adversary model that allows to estimate password strength by simulating password-guessing attacks.
\par

Unfortunately, password models are not universal tools. The composition habits of passwords vary from one community of users to another~\cite{psswd_lan}. As a result, in order to be effective,  \textbf{password models must be specifically designed to match the password distribution of the community where they will be used}. However, this is not an easy task.

Given a target credential database (\eg the one associated to a web application), the expected workflow requires the end-user (\eg the system administrator) to collect a suitable training set for the password model before deploying it. The system administrator must gather one or more public password leaks that have a password distribution  \textit{similar} to the target one, and use the collected data to train the password model; {typically a machine learning model}.
 
 Previous research has often assumed an \textit{intra-site} scenario, where the password model is trained on passwords from the same database where it will be used. However, this approach does not reflect any real-world scenario. When deploying a password model, system administrators can only access the stored credential hashes and do not have a legitimate means to view the original plaintext passwords.\footnote{The standard security principle is that plaintext passwords should not be stored on the server.} This prevents them from using the stored passwords as training data. More critically, it also precludes the use of any similarity metric to determine external password leaks with congruent distribution to the target one. 

In the real world, the only viable option is to approximate the concept of \TT{similar distribution} by relying on known macro-characteristics of the target community, such as the expected primary language(s) of users and/or the kind of service provided by the web-application, e.g., banking, e-shopping, etc. Pragmatically, if we want to train an accurate password model for a Swiss e-banking service, we would need to fetch a set of leaked passwords originating from a previously breached Swiss e-banking website. 

Unfortunately, in practice, the {system administrators} need to learn where to find such suitable password leaks. Those are often available only in underground forums or hidden behind a paywall or credit system. Ultimately, data with the requested properties may not be available in enough quantity to fit a password model. However, even if the data is simply there and easy to access, state-of-the-art password models are still built on top of machine learning models that must be trained before use~\cite{melichera}. Unfortunately, training and calibrating machine learning models is never a straightforward process.
This is especially true for system administrators of small web applications who might not have the expertise, experience and/or ability to afford the time (and cost) needed to train models that perform this job (which may take days of GPU time~\cite{melichera, pasquini2021improving}).
\iflong
Another thing to consider is that different training sets require different settings; while learning-based password models are more robust than simpler approaches such as dictionary attacks~\cite{reasoning}, they still rely on some degree of hyperparameter tuning that, whether successful or not, may end up with suboptimal password models regardless of the quality of the training set.
\fi
\ournew{\textbf{Ultimately, while password security depends on the existence of well-trained and calibrated password models, this requirement is inherently hard to meet in reality - especially by the average end-user.}}

In this work, we move towards solving this fundamental issue. We introduce the first \textbf{universal password model}; {a password model that, once trained, \new{can automatically optimize its guessing strategy based on the context, without requiring} users to run additional training steps or undertake targeted data collection.} At inference time, the model does not need access to any plaintext passwords from the target set. Instead, it relies on auxiliary data, such as email addresses and usernames, to build a predictive model for the underlying target passwords. We call this family of models \textit{Universal Neural-Cracking-Machines} or \uncms\ for short. 

\ournew{
The main intuition is that human-chosen passwords and personally identifiable information that a person typically shares at registration time (such as their name, address, and phone number) are naturally correlated with each other. A \uncm\ is a password model that exploits this natural correlation to adapt to the target password distribution. Going beyond just individual pairs of users and passwords~\cite{\targetedatt}, a \uncm\ extends this correlation to  whole communities. In particular, \textbf{a \uncm\ uses deep learning, specifically an attention mechanism~\cite{vaswani2017attention}, to combine and interpolate together information about users in the same group to define a robust and accurate prior over their password distribution.} This prior is then utilized to dynamically optimize the model's configuration at inference time, enhancing its capability to guess the user passwords within the target system.}
\par

\ournew{
 In the paper, we show that, although requiring no human intervention and having no explicit prior knowledge about the target, a \uncm\ can outperform equivalent password models that have been manually calibrated for the target distribution based on prior information  (\eg target users' language).}
 We also show that user information used for the configuration process can be efficiently anonymized via Differential Privacy, enabling the publication of the model with formal privacy guarantees. The automatic configuration process is lightweight and has sub-second latency, making it possible for every end-user to generate accurate password models. Furthermore, the resulting password models can be efficiently compressed~\cite{melichera} and used as client-side password strength meters, enabling a wide range of applications.
 \par

Our contributions can be summarized as follows:
\begin{itemize}
\item We introduce the concept of universal password model and demonstrate that auxiliary information associated with users can be combined and used to learn an accurate prior on the password distribution of the underlying user's community.
\item We develop a fully-automatic approach to exploit auxiliary information and instantiate context-aware password meters and guessing attacks without requiring any plaintext samples from the target password distribution.
\item We introduce a differentially-private version of our framework, which is well-suited for building client-side password strength meters. This provides the \textbf{formal} guarantee that no meaningful amount of information on the individuals who provided the auxiliary data for the model configuration is leaked upon model publication. The proposed framework 
 enables us to reach $\epsilon < 1$ (\ie strong privacy guarantees~\cite{Dwork_Smith_2010}) with only a limited loss in utility.
\iflong
\item Incidentally, our work also defines the new state-of-the-art for offline password guessing attacks as well as password strength metering.
\fi
\end{itemize}

We provide access to our code and pre-trained models.\footnote{\url{https://github.com/TheAdamProject/UniversalNeuralCrackingMachines}}\iflong\else\ A full version of the paper is available.\footnote{\url{https://arxiv.org/abs/2301.07628}}\fi

\section{Preliminaries and Background}
\label{sec:preliminaries}
In this section we go over some necessary background. We start by covering  password models\iflong\ and surveying relevant related works in password security. \else.\fi\ Section~\ref{sec:att} introduces the concept of \textit{attention} in machine learning. Finally, Section~\ref{sec:dpbck} gives a brief introduction to differential privacy. Section~\ref{sec:notation} concludes by formalizing the notation used within the paper.

\subsection{Passwords Models}
\label{sec:passmodel}
Human-chosen passwords do not distribute uniformly in the key space $\ks$ (the set of all possible passwords). Instead, users tend to choose easy-to-remember strings that aggregate in a relatively few dense clusters~\cite{bonneau, habitat}. Real-world passwords result in predictable distributions that can be easily modeled by an adversary, making authentication systems susceptible to \textbf{guessing attacks}~\cite{unix, MM}. In a guessing attack, an adversary aims to recover plaintext credentials by attempting several candidate passwords (\textbf{guesses}) till success or budget exhaustion~\cite{economics}; this happens by either searching for the pre-image of password hashes (offline attack) or attempting remote logins (online attack). In this process, the attacker relies on a so-called \textbf{password model} that defines which, and in which order, guesses should be tried to maximize the effectiveness of the attack \ie the expected number of recovered credentials per guess~\cite{bonneau}.
\par

In general, a password model can be understood as an estimate of the password distribution that allows for an informed search of the key space. However, the focus of this search can vary depending on the objective. It can be heavily conditioned on auxiliary information, as in the case of targeted online attacks against individual accounts~\cite{stuffing, targuess, targeted0}, or based on a less informative prior, as in the more common trawling offline setting~\cite{190978}. In this work, we focus on the latter case, and human-chosen passwords.
\iflong 
\paragraph{\textbf{Implementations}}
\label{app:addbkpsswwdml}
Existing password models construct over a heterogeneous set of assumptions and implementations. 

{Probabilistic password models} assume that real-world password distributions are sufficiently smooth to be accurately described by a suitable parametric probabilistic model~$\dec_\Theta$. Here, a password mass function is explicitly \cite{backoffmc, melichera, MM} or implicitly \cite{pasquini2021improving} derived from a set of observable data (\ie previously leaked passwords), typically, via maximum likelihood estimation~\cite{melichera, MM}. This is then used to assign a probability to each element of the key space.  During the guessing attack, guesses are produced by traversing the key space following the decreasing probability order imposed by~$\dec_\Theta$.

 Other hybrid approaches such as Probabilistic Context-Free Grammars (PCFG) and dictionary attacks~\cite{unix, 190978, reasoning, pasquini2021usenix}, instead, rely on simpler and more intuitive constructions, which tend to be closer to human logic. Generally, those assume passwords as realizations of templates and generate novel guesses by abstracting and applying such patterns on token collections. Those tokens are either directly given as part of the model configuration (\eg the dictionary and rules-set for dictionary attacks) or extracted from observed passwords in a training phase (\eg terminals/grammar for PCFG). In contrast with parametric models, these can produce only a limited number of guesses, which is a function of the chosen configuration.
\fi 

\paragraph{\textbf{Applications}} 
Password models are essential for password security. In addition to their use in simulated guessing attacks for reactive security analysis~\cite{6234434, 190978} (\eg penetration testing) and proactive approaches such as Password Strength Meters (PSMs), password models are widely used in a variety of password security applications, including breach alert systems~\cite{migp}, honey encryption/words~\cite{honeyenc, 9833598}, and distribution-aware hash functions~\cite{dash}. \iflong \else Additional background on password models is provided in Appendix~\ref{app:addbkpsswwdml}.\fi

In this work, we primarily focus on the application of our methodology as PSMs (see Section~\ref{sec:psm}). This is the most common application of password models in the real world~\cite{de2014very} and therefore has the most impact. However, we emphasize that the results achieved by our methodology apply equally to all other applications of password models, affecting the password security community as a whole.

\subsubsection{Password Strength Meters (PSMs)} 
\label{sec:psm}
Within the PSM context, every password model can be understood as a function that maps each password $x\myin \ks$ to a \textbf{guess number}~$g$---the rank of the password $x$ in the guesses generation process of the attack. Under the assumption that the employed password model captures an accurate adversary model, the guess number is universally recognized as the best estimator of password security~\cite{6234434, 190978}. Pragmatically, passwords with low guess numbers are attested to be weak, whereas passwords with high guess numbers are considered secure. 
\par

Once a password model has been fitted to model a suitable password distribution, it can be deployed as PSM; typically in web-applications~\cite{hfla, melichera, pasquini2020esorics, zxcvbn}. Here, the password model is embedded into the sign-up page of the target web app and used to meter users' passwords at composition time. Supported by tailored interfaces and feedback mechanisms~\cite{hfla, zxcvbn, pasquini2020esorics}, the meter dissuades users from picking passwords with low guess numbers. PSMs, when correctly developed~\cite{howdoes, doesmypassword}, have been shown to be one of the most effective tools to induce stronger passwords for the underlying user community.

\iflong
\subsubsection{Autoregressive password models}
\label{app:autoreg}
In the present work, we focus on a specific class of probabilistic password models: the autoregressive ones~\cite{melichera}. An autoregressive password model segments every password $x$ in the key space as a list of atomic components $x \myeq [c^1, c^2, \dots, c^q]$; typically, at character-level.\footnote{Other kinds of segmentation processes such as \textit{n-grams} or semantic tokens are possible~\cite{melichera}; however, in the paper, we focus on character-level models as this is the simplest and most natural implementation.} Then, the model assigns to each element $c^i$ a probability that is conditioned to all the elements that come before in the sequence $P(c^i) \myeq P(c^i| c^1, \dots,  c^{i-1})$. The product of all those local probabilities results in the joint probability of the string~$x$: $P(x) \myeq f_{\Theta}(x) \myeq f_{\Theta}(c^1)\cdot f_{\Theta}(c^2|c^1) \dots  f_{\Theta}(c^q|c^1, \dots c^{q-1}),$
where $f_{\Theta}$ is the password model---a parametric function (\eg a neural network) in this context. As for the other models, the set of parameters~$\Theta$ is derived/learned via maximum likelihood estimation~\cite{melichera, MM} from a set of observable data, which are typically set(s) of previously leaked (plaintext) passwords.
\par

Autoregressive models come with valuable practical properties. In contrast with other neural approaches such as GANs~\cite{passgan, pasquini2021improving} and similar non-autoregressive constructions~\cite{pasquini2021usenix}. Autoregressive models (1) offer enumeration algorithms that allow us to generate/traverse the key space in decreasing probability order and (2) can explicitly assign probabilities to the generated passwords. Such probabilities can then be efficiently mapped to the corresponding guess numbers via suitable estimation techniques~\cite{monte}. We stress that these properties are pivotal to implementing efficient password strength meters, cast password guessing attacks, and support security studies on plaintext passwords~\cite{howdoes, bonneau, psswd_lan}.
\fi
\subsection{Attention Mechanisms}
\label{sec:att}
\begin{figure}[t]
	\centering
	\resizebox{.25\textwidth}{!}{
		\begin{tikzpicture}		
			
			\tikzstyle{chosen} = [rectangle]
			\tikzstyle{other} = [rectangle]
			\tikzstyle{arrow} = [->,>=stealth]
			
			\node (t0) [chosen] {$q_i$};
			\node (t1) [other, right of=t0] {$v_1$};
			
			\node (fk0) [above of=t0] {$g_Q$};
			\node (fq0) [above of=t1] {$g_K$};
			
			\draw [arrow] (t0) -- (fk0);
			\draw [arrow] (t1) -- (fq0);
			
			\node (d0) [above of=fk0, xshift=.5cm] {\Large{$\odot$}};
			
			\draw [arrow] (fk0) -- (d0);
			\draw [arrow] (fq0) -- (d0);
			
			\node (t2) [chosen, right of=t0, xshift=1.5cm] {$q_i$};
			\node (t3) [other, right of=t2] {$v_2$};
			
			\node (fk1) [above of=t2] {$g_Q$};
			\node (fq1) [above of=t3] {$g_K$};
			
			\draw [arrow] (t2) -- (fk1);
			\draw [arrow] (t3) -- (fq1);
			
			\node (d1) [above of=fk1, xshift=.5cm] {\Large{$\odot$}};
			
			\draw [arrow] (fk1) -- (d1);
			\draw [arrow] (fq1) -- (d1);
			
			\node (t4) [chosen, right of=t2, xshift=1.5cm] {$q_i$};
			\node (t5) [right of=t4, other] {$v_3$};
			
			\node (fk2) [above of=t4] {$g_Q$};
			\node (fq2) [above of=t5] {$g_K$};
			
			\draw [arrow] (t4) -- (fk2);
			\draw [arrow] (t5) -- (fq2);
			
			\node (d2) [above of=fk2, xshift=.5cm] {\Large{$\odot$}};
			
			\draw [arrow] (fk2) -- (d2);
			\draw [arrow] (fq2) -- (d2);
			
			\node (sf) [above of=d1, yshift=-.1cm] {\textit{softmax}};
			
			\draw [arrow] (d0) -- (sf);
			\draw [arrow] (d1) -- (sf);
			\draw [arrow] (d2) -- (sf);
			
			\node (w0) [above of=sf, yshift=-.4cm] {$\cdot w_1$};
			\node (w1) [above of=w0, yshift=-.5cm] {$\cdot  w_2$};
			\node (w2) [above of=w1, yshift=-.5cm] {$\cdot  w_3$};
			\draw [arrow] (sf) -- (w0);
			
			\node (v0) [other, left of=w0, xshift=-2cm] {$v_1$};
			\node (v1) [other, above of=v0, yshift=-.5cm] {$v_2$};
			\node (v2) [other, above of=v1, yshift=-.5cm] {$v_3$};
		
			\node (fv0) [right of=v0, xshift=.5cm] {$g_V$};
			\node (fv1) [right of=v1, xshift=.5cm] {$g_V$};
			\node (fv2) [right of=v2, xshift=.5cm] {$g_V$};
			
			\draw [arrow] (v0) -- (fv0);
			\draw [arrow] (v1) -- (fv1);
			\draw [arrow] (v2) -- (fv2);
			
			\draw [arrow] (fv0) -- (w0);
			\draw [arrow] (fv1) -- (w1);
			\draw [arrow] (fv2) -- (w2);
			
			\node (agg) [right of=w1] {\large{$\Sigma$}};
			
			\draw [arrow] (w0) -- (agg);
			\draw [arrow] (w1) -- (agg);
			\draw [arrow] (w2) -- (agg);
			
			\node (res) [right of=agg, xshift=1cm] {$= \gamma(q_i,\ \{v_1, v_2, v_3\})$};
			
			\draw [arrow] (agg) -- (res);
			
		\end{tikzpicture}
	}
	\caption{Depiction of a partial execution of a simplified  attention-mechanism for a single query vector $q_i$  and the set of  values: $\{v_1, v_2, v_3\}$.}
	\label{fig:att}
\end{figure}
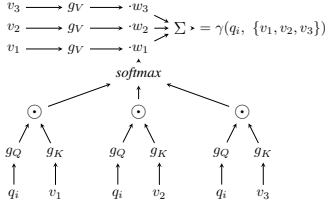
Attention mechanisms (AMs) are the driving force of modern deep learning. Since their seminal application on Natural Language Processing (NLP)~\cite{DBLP:journals/corr/BahdanauCB14}, AMs became critical components of state-of-the-art models, culminating in the current \TT{transformer revolution}~\cite{vaswani2017attention, gpt3, dosovitskiy2021an}. 
\par

In contrast to classic architectures (\eg RNN and CNN) that either imply spatial or temporal ordering on input data, AMs can be seen as a function defined over sets rather than sequences or images. Attention-based networks, unless intentionally amended for that propose (\eg adding positional encoding~\cite{vaswani2017attention}), operate on permutation-invariant inputs and naturally abstract operations on length-varying data. 
\par
  In the setup considered in this paper, an attention mechanism can be understood as a function $\gamma(\cdot, \cdot)$ defined over two sets of equal size vectors.\footnote{In the most general construction, an attention mechanism is defined over three input sets. In our paper, we consider \textit{keys} and \textit{values} to be the same.} Given a set $Q \myeq \{q_1, q_2, \dots, q_n\}$ (called \textbf{queries}) and a set $V\myeq\{v_1, v_2, \dots, v_d\}$ (called \textbf{values}), for each element $q_i \myin Q$, the AM computes a function of~$q_i$ and all the other vectors in $V$. Formally, for each pair ($q_i,\ v_j \myin V)$, $\gamma$ first maps $q_i$ and $v_j$ via two different dense layers ($g_K$ and $g_Q$, respectively) and it computes the dot product on their outputs. The resulting scalar values are then scaled via the \textit{softmax} function that returns a weight $w_j$ for each pair~$(q_i,\ v_j)$. The output of the attention mechanism is then the sum of all the input tokens~$V$ (mapped through a third dense layer $g_V$) weighted by the weights produced by the \textit{softmax}. Figure~\ref{fig:att} depicts the \textit{computational graph} for a {simplified} version of the Multi-Head Attention mechanism proposed in~\cite{vaswani2017attention} on a single query vector and set of three values. 

In the password security context, following the success of attention-based architectures in NLP, a growing body of work~\cite{ptransf0, trans2} is proposing to use them as a replacement for classic architectures such as RNN~\cite{melichera} to implement password models, obtaining mixed results. We highlight that, in our work, we do not seek incremental results harnessing improved architectures based on AMs; instead, we rely on their unique technical properties to implement modules of our framework that would be impossible to efficiently implement otherwise (see Section~\ref{sec:mixenc}).

\subsection{Differential Privacy}
\label{sec:dpbck}
Given a privacy budget $\epsilon \myin \mathcal{R}^+$, a randomized function $\mathcal{M}$ is  \mbox{$\epsilon$-differentially-private} if this verifies Eq.~\ref{eq:dp} for each possible input pair $(D,\ D')$ of neighbor sets \ie sets that differ for a single element.
\begin{equation}
	P\left[ \mathcal{M}(D) \in \mathbb{S} \right] \leq P\left[ \mathcal{M}(D') \in \mathbb{S} \right] \mycdot e^\epsilon,
	\label{eq:dp}
\end{equation} 
\new{for any subset of outputs $\mathbb{S}\subseteq \texttt{Range}(\mathcal{M})$.}
More pragmatically, being differential private implies that no adversary can infer the result of the predicate $d \myin D$ from~$\mathcal{M}(D)$   better than $\ln(\frac{TPR}{FPR}) \leq \epsilon$~\cite{dpbound}, where $TPR$ and $FPR$ are the \textit{true positive rate} and the \textit{false positive rate} of the attacker in inferring whether~$d \myin D$. \iflong This holds regardless of the amount of information and influences the attacker has on $D/ \{d\}$~and~$\mathcal{M}$. \fi Given a function $\mathcal{M}$, there are different techniques one can use to make it differentially private. In the paper, we use the \textbf{Gaussian mechanism}, which is the ubiquitous choice in~ML~\cite{dldp}. The Gaussian mechanism results in \mbox{$(\epsilon, \delta)$-differential-privacy}; a relaxed form of DP that accounts for a failure event in which Eq.~\ref{eq:dp} does not hold for~$\mathcal{M}$ with probability~$\delta$. Another necessary notion we use in the paper is the one of \textbf{sensitivity} of a function~$\mathcal{M}$. This is defined as: $\max_{D, D'} \| \mathcal{M}(D) - \mathcal{M}(D') \|$,
where $D$ and $D'$ are neighbors sets. The sensitivity quantifies the magnitude by which a single element can change the function~$\mathcal{M}$ in the worst case~\cite{dwork2014algorithmic}. 

\subsection{Notation \& Definitions}
\label{sec:notation}
In the paper, we model a \textbf{credential database} as a pair $S \myeq (\ainf,\ \hash(X))$ (\eg the credential database of a web application), where $X$ is the set of user-chosen passwords and $\ainf$ is the set of \textbf{auxiliary information} associated to the passwords in $X$: users' email addresses, usernames, etc. \iflong The letter $x$ is used to identify a single password in~$X$. \fi  With~$\hash$, we denote a secure cryptographic hash function, where~$\hash(X)\myeq \{\hash(x) | x \myin X\}$.
We abuse the term \textbf{\TT{community}} to refer to the group of users who originated the data in~$S$ and use~$n$ to express the cardinality of a set of passwords~$n \myeq |\ainf|\myeq |X|$. We use the letter $u$ to identify a single user in $S$ and $U$ for a set of users. In the paper, we must often refer to sets of credential databases $\{S_1, \dots, S_m\}$. For those, we use the term \textbf{\TT{leak collection}} and the notation~$\mathbf{L}$.
\par

Hereafter, without loss of generality, we use the notation~$\dec_\Theta$ to refer to a password model, where~$\Theta$ are the model's parameters. We focus on probabilistic password models; thus, $\dec_\Theta$ is intended as a mass function; a function that assigns a probability to each element of the key-space $\ks$ and $\sum_{x \in \ks} \dec_\Theta(x)\myeq 1$. 
Additionally,  we use the shorthand notation~$\dec_\Theta(X)$ to refer to~$\sum_{x \in X} \dec_\Theta(x)$. Given a password model, we call \textbf{\TT{guessing strategy}} the list of guesses $[x_0, x_1, \dots]$ generated by the model when running a guessing attack. For probabilistic models, the guessing strategy is the key-space sorted by decreasing probability according to $\dec_\Theta$.

\section{On Universal Password Models}
\label{sec:universal}
\begin{figure}[t!]

\centering
\includegraphics[trim = 0mm 116mm 0mm 0mm, clip, width=.6\columnwidth]{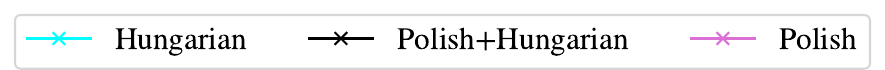}

\resizebox{.95\columnwidth}{!}{
\begin{subfigure}{.4\columnwidth}

\centering

\includegraphics[trim = 0mm 0mm 0mm 0mm, clip, width=.95\columnwidth]{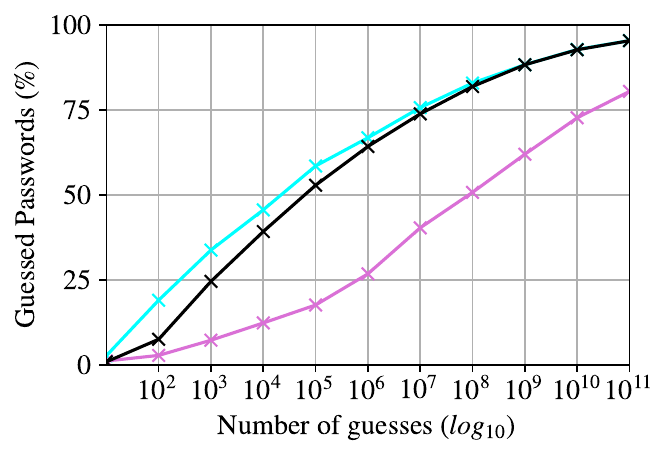}

\caption{Hungarian leak.}\label{fig:hu}


\end{subfigure}\begin{subfigure}{.4\columnwidth}

\centering

\includegraphics[trim = 0mm 0mm 0mm 0mm, clip, width=.95\columnwidth]{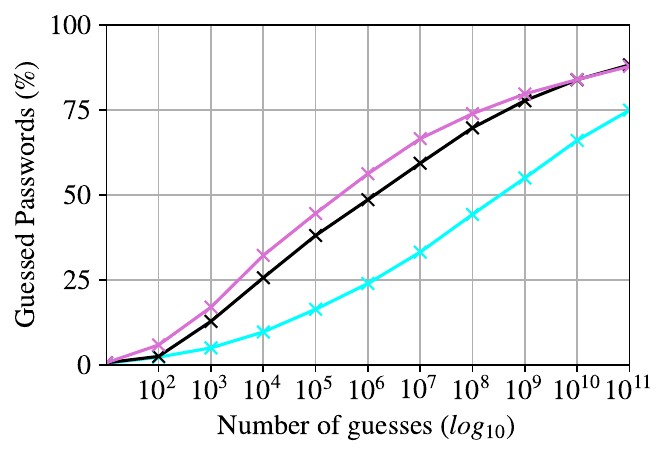}

\caption{Polish leak.}\label{fig:pl}


\end{subfigure}
}
\caption{Results of two guessing attacks for three password models~\cite{melichera} trained to model different language-specific password distributions.}

\label{fig:toy_example}

\end{figure}

\label{sec:uncm}

In this section, we introduce our main contribution---the concept of Universal Neural-Cracking-Machine (\uncm). To this end, we first review some notions and intuitions that are pivotal to motivating our work.




\subsection{Password Strength is not a Universal Property}

\label{sec:weakpassnonuni}


Password strength is not a universal concept. While passwords tend to have similar characteristics, different communities choose passwords arbitrarily, resulting in disparate password distributions~\cite{psswd_lan}. In other words, while there are guidelines for what makes a strong or weak password, these are not universal and will vary depending on the context in which the password is created and used. 

This has important implications for password models, which are fully data-driven and rely solely on the data used to train them. As a result, when a model is trained, it can only provide meaningful strength estimates for passwords that are distributed similarly to the ones used to train it~\cite{pdistribution}. More specifically, regardless of the size of the training set, if a password model is trained to be optimal for a given password distribution, it cannot also be optimal for a different distribution. 

The toy example in Figure~\ref{fig:toy_example} provides an illustration of this phenomenon. In this experiment, three instances of a password model~\cite{melichera} are trained on three different training sets: a collection of Polish password leaks (1,434,802 passwords in total), a collection of Hungarian password leaks (1,215,767 passwords in total), and the combination of the two. \ournew{When the Hungarian and Polish models are used to carry out a guessing attack on a Hungarian and a Polish credential database\footnote{{Which are disjointed from their respective training collection.}} (Figure~\ref{fig:hu} and Figure~\ref{fig:pl}, respectively), these models demonstrate optimal performance on their respective matching database, but their performance is suboptimal when applied to each other's database. This mutual exclusivity phenomenon is further demonstrated by the results obtained by third model, which is trained on the combined Hungarian and Polish password lists. While this model is trained on twice as much data than the previous two, it performs worse than the models trained specifically for the Hungarian and Polish distributions on both leaks. However, it does achieve better performance on average over the two password sets.}



Ultimately, the mutual exclusivity of password models entails that no single model can achieve optimal performance across all distributions. Although one can attain an optimal performance \textbf{on average} by approximating the universal password distribution (\ie training the model on the combination of all known password distributions), no model can meet the optimal thresholds for all password distributions simultaneously, at least with current implementation methods. In this work, however, {we show that universal password models can exist} and can be implemented through novel deep learning techniques. We call this family of models  \textbf{Universal Neural-Cracking-Machines}, or \uncms\ for short, and describe them in the following.

\subsection{A Universal Neural-Cracking-Machine (\uncm)}

\label{sec:uncm}

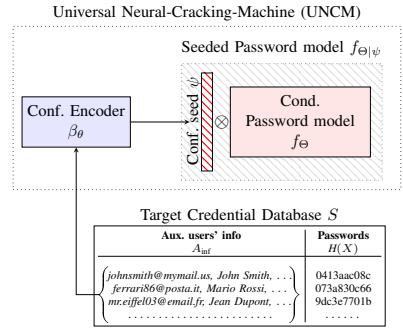
\begin{figure}[t]
	\centering
	\resizebox{.6\columnwidth}{!}{
		\begin{tikzpicture}		
			\tikzstyle{block} = [rectangle, minimum width=.5cm, minimum height=1cm, text centered, draw=black, fill=blue!10]
			\tikzstyle{vector} = [rectangle, minimum width=1cm, minimum height=.1cm,text centered, draw=black, fill=white]
			\tikzstyle{arrow} = [->,>=stealth]

		\node (boxe) [rectangle, draw=black, dotted,  minimum width=8.3cm, minimum height=3.5cm, yshift=.3cm, xshift=2.3cm, label=Universal Neural-Cracking-Machine (UNCM)]{};
			
		\node (box) [rectangle, draw=black, dotted,  minimum width=4.3cm, minimum height=2.5cm, yshift=0cm, xshift=3.9cm, opacity=.5, pattern=north west lines, pattern color=gray, dotted, label=Seeded Password model $\seeded$]{};

			\node (e) [block, xshift=-.5cm]{ \makecell{\text{Conf. Encoder} \\ $\enc$} };%

			\node (d) [block, xshift=4.3cm, fill=red!10, minimum width=3cm,]{ \makecell{\text{Cond.} \\ 
			\text{Password model}\\$\dec_\Theta$} };%

			\node (seed) [vector,  right of=e, xshift=1.8cm, minimum width=2.1cm, rotate=90, pattern=north west lines, pattern color=red, label={[yshift=1.1cm, xshift=-.15cm, rotate=90]{Conf. seed $\cs$}}]{};

			\node (fb) [yshift=-3.7cm, xshift=.14cm]{};%
			
			\draw [arrow, shorten <= 0cm, shorten >= 0cm] (fb.west) -| (e.south) node[] {};
			
			\draw [arrow, shorten <= 0cm, shorten >= .25cm] (e) -- (seed) node[] {};
			\node(pp)[right of=seed, xshift=-.68cm]{\large $\otimes$};

		\draw  [decorate, 
			decoration = {calligraphic brace, raise = 2pt, amplitude = .2cm, mirror}, thick] (0.25,-3.1) --  (0.25,-4.3);
			
		\draw [decorate, 
			decoration = {calligraphic brace, raise = 2pt, amplitude = .2cm}, thick] (4,-3.1) --  (4,-4.3);

\matrix[ampersand replacement=\&, below of=seed, yshift=-2.3cm, xshift=.7cm, label={[yshift=-.2cm, xshift=0cm]{Target Credential Database $S$}}] {
	\scriptsize
	\node (species1) [shape=rectangle,draw] {
		\begin{tabular}{c|c}
			\makecell{ \textbf{Aux. users' info} \\ $\ainf$} & \makecell{ \textbf{Passwords} \\ $\hash(X)$}\\
			 \hline
			 \\
			 \textit{johnsmith@mymail.us, John Smith, $\dots$} & 0413aac08c  \\
			 \textit{ferrari86@posta.it, Mario Rossi, $\dots$} &  073a830c66 \\
			 \textit{mr.eiffel03@email.fr, Jean Dupont, $\dots$} &  9dc3e7701b \\
			 $\dots\dots\dots\dots\dots\dots\dots\dots $ &  $\dots\dots$ \\
		\end{tabular}
	};
	\&  \\
};

\normalsize
			
			
		\end{tikzpicture}
	}
	\caption{Depiction of a \uncm~and its internal working at deployment time. 
	}
	\label{fig:uncm}
\end{figure}

It should be clear from the previous section that there is no single guessing strategy that is optimal for every guessing attack. Therefore, by definition, \textbf{a universal password model must be adaptive}; that is, it must conditionally change its guessing strategy based on the context. In a nutshell, a \uncm\ is a deep neural network built and trained to implement this functionality. Given a target credential database $S$, a \uncm\ can be understood as an \TT{adaptive password model} that is able to automatically optimize its guessing strategy based on~$S$, even when the passwords in $S$ are unknown at inference time. {Nonetheless, for this to work, the model must either have observed the optimal guessing strategy for $S$ as part of the training data or be able to derive it by interpolating similar known distributions.}
\par

 The general structure of a \uncm\ is reported in Figure~\ref{fig:uncm}; it is the combination of two main modules---two deep neural networks.

\paragraph{\textbf{The Conditional Password model}}
The first module is something that we can define as a \textbf{conditional password model}. Pragmatically speaking, this is just a probabilistic password model whose mass function can be altered by an additional input~$\cs$ that is disjointed from the set of parameters learned during the training phase. Within the paper, we call this additional input: \textbf{configuration} or \textbf{configuration~seed}. The main idea here is that, while the parameters~$\Theta$ of the password model~$\dec_\Theta$ are trained to capture multiple password distributions simultaneously (\ie all the password distributions), the input configuration seed  can instruct $f_\Theta$ to focus on specific modalities of the modeled distribution without acting on the parameters~$\Theta$, \ie without retraining the  underlying model. 

Given a target credential database~$S \myeq(\ainf, \hash(X))$, the role of a \uncm~is straightforward. The objective is to find a suitable configuration seed such that the conditional password model performs optimally on the (unknown) passwords in~$X$. Finding such a configuration is the role of the second module---the \textbf{configuration encoder}.

\iflong
\paragraph{\textbf{The Configuration Encoder - auxiliary data as a proxy for password distributions}}
\else
\paragraph{\textbf{The Configuration Encoder}}
\fi 

The configuration encoder~$\enc$ is another deep neural network (with parameters $\theta$) whose task is to generate a suitable configuration~$\cs$ for the conditional password model~$\dec_\Theta$, conditionally to $S$. In essence, the configuration seed generated by the encoder should be understood as a succinct representation of the target password distribution that the conditional password model can leverage to maximize the allocated mass on the target passwords~$X$.
\iflong
Ideally, the task of the configuration encoder is to find $\cs$ such that:
\begin{equation}
	\enc(X) \myeq \argmax_\cs \dec_\Theta(X \ | \ \cs).
\end{equation}
\fi
In other words, the objective of $\enc$ is to define a prior probability distribution over the passwords~$X$ for the given credential database~$S$. However, there is a fundamental problem: the passwords $X$ and the specific password distribution from which $X$ is sampled are entirely unknown \eg the passwords are hashed or completely inaccessible.
\par

\ournew{As mentioned earlier, the configuration encoder circumvents the issue of obtaining direct knowledge of the password distribution $X$ by using personal information of users as a proxy. Indeed, while obtaining any part of the plaintext passwords $X$ may not be feasible, it is always possible to derive insights into their distribution using other information available in the target credential database.} Password hashes are always accompanied by \textbf{auxiliary} data such as email addresses, usernames, and related tags. As demonstrated in previous studies on targeted guessing attacks~\cite{\targetedatt, stuffing, targeted0}, this auxiliary data and users' passwords are not independent. Users often incorporate personal information into their passwords, and an attacker with knowledge of a user's personal information can launch a targeted attack and guess the user's password more efficiently. 


The configuration encoder takes this correlation one step further by aiming to \textbf{capture general properties of the password distribution associated with the entire community of users in a target system}. Rather than trying to model the correlation between an individual user and their chosen password~\cite{\targetedatt, stuffing, targeted0}, the configuration encoder receives the auxiliary data of all users simultaneously and correlates these individual pieces of information to extract general patterns that can be used to model the password distribution for the whole community.

For example, returning to the toy example in Section~\ref{sec:weakpassnonuni}, if most of the email addresses in the target credential database have the domain  \TT{.pl} or/and usernames built on Polish words, the encoder should detect this general trend and instruct the password model to generate \TT{Polish passwords} as output. When given Hungarian-related auxiliary data, the same model should focus on \TT{Hungarian passwords} instead. In addition to simple and pre-defined characteristics like the language of the community, the configuration encoder is trained to identify and generalize these patterns (which may be latent) from the available data. At inference time, it applies them to the input auxiliary information $\ainf$ to define an accurate \textit{priori} over the unknown passwords in $X$, which the conditional password model then uses.\iflong\  Essentially, we can see a \uncm~as a \TT{password model factory} - an algorithm that takes the auxiliary information available for the target community as input and produces a password model tailored to the underlying password distribution in a fully autonomous way. \fi In Section~\ref{sec:imp}, we discuss how a \uncm~is implemented and trained for this purpose.
\iflong
\begin{table*}[ht!]
	\centering
	\caption{Number of leaks for the top-$25$ top-level domains appearing in \textit{Cit0day.in} before and after pre-processing.}
	\label{table:cit}
	\resizebox{1\textwidth}{!}{%
		\begin{tabular}{c|ccccccccccccccccccccccccc}

 \toprule
    
    \textbf{Raw} & \textit{.com} & \textit{.ru} & \textit{.net} & \textit{.org} & \textit{.de} & \textit{.kr} & \textit{.br} & \textit{.pl} & \textit{.uk} & \textit{.it} & \textit{.fr} & \textit{.cz} & \textit{.tw} & \textit{.in} & \textit{.hu} & \textit{.jp} & \textit{.info} & \textit{.ca} & \textit{.th} & \textit{.eu} & \textit{.nl} & \textit{.ua} & \textit{.au} & \textit{.es} & \textit{.ar}\\
    \# leaks & $9539$ & $1208$ & $1167$ & $982$ & $874$ & $798$ & $609$ & $551$ & $533$ & $467$ & $438$ & $400$ & $330$ & $322$ & $305$ & $226$ & $223$ & $193$ & $191$ & $175$ & $172$ & $163$ & $142$ & $139$ & $134$ \\
    \midrule
    \textbf{Clean} & \textit{.com} & \textit{.ru} & \textit{.net} & \textit{.org} & \textit{.de} & \textit{.kr} & \textit{.pl} & \textit{.br} & \textit{.uk} & \textit{.it} & \textit{.cz} & \textit{.fr} & \textit{.hu} & \textit{.tw} & \textit{.in} & \textit{.jp} & \textit{.fj} & \textit{.info} & \textit{.ca} & \textit{.eu} & \textit{.ua} & \textit{.ar} & \textit{.au} & \textit{.nl} & \textit{.ro}\\
    \# leaks & $4741$ & $567$ & $560$ & $464$ & $426$ & $403$ & $389$ & $299$ & $286$ & $238$ & $210$ & $200$ & $192$ & $175$ & $172$ & $112$ & $102$ & $102$ & $99$ & $91$ & $79$ & $79$ & $72$ & $72$ & $71$ \\
    \bottomrule

\end{tabular}
	}
\end{table*}
\fi
\subsubsection{\textbf{Deploying a \uncm~- seeded password models}}
\label{sec:deploy}

{Once the~$\uncm$~has been trained and made public\footnote{Following the \TT{foundation models} paradigm in modern ML~\cite{bommasani2021opportunities}, the~\uncm~is pre-trained and made public by a third party \eg our research group or an external company.}, end-users (\eg system administrators) can use it to generate accurate password models for the target distribution without any additional dataset collection or training step.}

Formally, let \textit{Bob} be the system administrator of a system~$\mathcal{S}$ (\eg a web application or a multi-user system) with a credential database~$S\myeq(\ainf, \hash(X))$. \textit{Bob} wants to deploy an accurate password model for $\mathcal{S}$. To this end, \textit{Bob} \textbf{(1)}~downloads the pre-trained~\uncm\ and \textbf{(2)}~uses the auxiliary information $\ainf$ associated with the users in $\mathcal{S}$ as input for the \uncm. 
This process results in a \textbf{seeded password model}~$\seeded$, which is the combination of the pre-trained conditional password model~$\dec_\Theta$ and the configuration seed~$\cs$ specifically generated for the system $\mathcal{S}$ by the configuration encoder.\iflong\ As demonstrated in Section~\ref{sec:result}, the seeded model is an accurate representation of $\mathcal{S}$'s password distribution and may even outperform equivalent password models that have been manually configured for the target. 
\fi Once the seeded password model for~$\mathcal{S}$ has been generated, it can be used like a standard password model, such as for deploying a client-side password strength meter or conducting a guessing attack on $\hash(X)$.
\par

In summary, starting from a pre-trained \uncm, \textit{Bob} can automatically generate a tailored, state-of-the-art password model for his system. This process is fast \iflong(see Section~\ref{sec:cotrain})\fi, requiring only users' auxiliary information (no external password leaks collection), and can be implemented with a single API call. \textit{Bob} does not need to share users' auxiliary data or password hashes with any external parties. Additionally, the configuration step only needs to be performed once, although it can be periodically updated to accommodate an evolving user community. \iflong In the differentially private setting (discussed in Section~\ref{sec:dp}), updates to the configuration seed must be carefully considered, as they will increase privacy loss.
\fi

\input{scheme_encoder}
\section{On the implementation of a \uncm}
\label{sec:imp}
In this section, we describe the relevant technical details about the implementation of our \uncm. We start by introducing the training set we use to develop and validate our approach. Then, we cover the architecture of the configuration encoder, and the one of the conditional password model we use to build the \uncm. Finally, we discuss how we train the \uncm~and the baseline models we compare~with.
\subsection{The training set (\textit{Cit0Day.in})}
\label{sec:trainingset}
To train and validate the proposed model, we rely on the credential leak collection originated by \textit{Cit0day.in}~\cite{cit0dayart0, cit0dayart1}. \textit{Cit0day.in} was 
a  service aimed at collecting leaked credential databases and provisioning them to malicious actors for a premium. In November 2020, \textit{Cit0day.in} suffered a security incident that resulted in the leakage of more than $22,500$ (previously breached) credential databases stored by the service. The originated data dump was shortly made publicly available after the incident. Hereafter, we refer to this leak collection as \citday. The complete list of the web applications appearing in \citday~is publicly available at~\cite{cit0dayartlist}.
\par

\iflong
Unlike similar leak collections harnessed in previous works (\eg the one identified by \textit{4iQ} in 2017~\cite{4iq} and used in~\cite{stuffing, pasquini2021usenix}), \texttt{Cit0day} stores the compromised credentials partitioned by source web-application rather than merging them in a single set.
\fi 

 The collection covers a heterogeneous set of leaks with broad geographic coverage. Table~\ref{table:cit} (raw) reports the most frequent top domains associated with websites appearing in the collection. Not all the password leaks reported in \texttt{Cit0day} are in plaintext.\iflong\footnote{Plaintext passwords in \citday~are likely to be the ones that have been recovered via offline guessing attacks or that were stored in plaintext on the source credential database.}\fi\ A non-negligible fraction of credentials appears in  hashed form or other noisy formats. After an extensive cleaning process\iflong\ (which is detailed in {Appendix~\ref{app:cleaning})\fi, the number of suitable leaks was reduced to a total of $11,922$. We report the updated distribution of top domains in Table~\ref{table:cit} (clean). The cleaned version of \citday~contains a total of $120,521,803$ compromised accounts with an average cardinality per leak (\ie number of users) of $10,088$.\iflong We report the exact leaks' cardinality distribution in Figure~\ref{fig:leakdis} in the appendix.\fi
	
\textbf{Available auxiliary data:} For every web application in \citday, credentials are reported in the format \textit{email address} and \textit{password}; no other personally identifiable information (PII) associated with users appears in the leaks. Ultimately, this fundamentally limits the amount of auxiliary information we can provide to the configuration encoder. Indeed, hereafter,  we consider the email addresses as the only modality of auxiliary data available for the configuration seed computation. We stress that, in web applications, email addresses are always associated with users' passwords; thus, this setting represents the minimal and most general setup for the application of our model. Nonetheless, we forecast that \uncm's performance should be improved by providing additional auxiliary data when available. 

\textbf{Pre-processing:}
 Given the cleaned leak collection, we partition it uniformly at random into two disjoint subsets ($90$-$10\%$): a training set composed of  $10054$ leaks (for a total of $102$ million accounts)  and a test set composed of  $1766$  leaks (for a total of $11$ million accounts). Additionally, we removed all the users who appeared in the training leaks from the test leaks. Hereafter, we refer to these two sub-collection of leaks as $\lct$ and $\lcv$, respectively. In the paper, the training set $\lct$ is used to train the proposed  \uncm~as well as the baselines we compare with. The test portion $\lcv$ is used to simulate the target credential databases during the evaluation phase. 
\subsection{The Configuration Encoder}
\label{sec:confencoder}
We now start discussing the implementation of the configuration encoder~$\enc$ used to build the \uncm.
The general structure of~$\enc$ is depicted in Figure~\ref{fig:model}. It comprises two main logical components: a \textbf{Sub-encoder} $\senc$ and a \textbf{Mixing-encoder}~$\gamma$.

\subsubsection{The Sub-encoder}
\label{sec:subencoder}
The role of the sub-encoder $\senc$ is to encode all the pieces of auxiliary information associated with a single user and map them to a standard format that can be interpreted by the mixing-encoder (\ie a dense vector of fixed size $d$). In the current setup, the sub-encoder is designed to process only users' email addresses. However, this can be extended to handle arbitrary auxiliary data modalities as described in Appendix~\ref{app:seaddm}.

Given an input email address (\eg \TT{johnsmith@mymail.us}), the sub-encoder  partitions it in three segments: \textbf{(1)} the username (\TT{johnsmith}), \textbf{(2)} the  provider (\TT{@mymail}), and \textbf{(3)} the domain (\TT{.us}). These segments are then processed separately, relying on three sub-models.
\par

A small RNN handles the username; a GRU in our implementation---the green rectangle at the bottom of Figure~\ref{fig:model}.\footnote{In the deployment phase, we noticed that utilizing a more complex architecture such as a LSTM did not yield any added advantages. Consequently, we decided to employ a simpler configuration, such as GRU for efficiency.} This sub-model iterates over the username string at the character level and returns the output obtained on the terminal character of the string. Instead, the provider and the domain strings are processed differently. Those are first discretized and then embedded via two separated embedding matrices (the two cyan squares in Figure~\ref{fig:model}). Each provider and domain string is associated with a row in the corresponding embedding matrix, which value is learned during the encoder's training process. To restrain the size of the matrices in our implementation, we exclude providers and domains that appear with low frequency. Specifically, we apply a frequency cutoff threshold of $300$, as this results in appropriately sized embedding matrices  ($2974$ and $268$ for providers and domains, respectively). Provider and domain strings that fall outside this range are mapped to a special \textit{out-of-the-vocabulary} embedding.
\par

Finally, to create the embedding for the complete email address, the vectors produced by each sub-model (username, provider, and domain) are concatenated together, forming a single vector of size~$d$ as depicted in Figure~\ref{fig:model}.
%
\subsubsection{The Mixing-encoder} 
\label{sec:mixenc}
The objective of the mixing-encoder is to \textit{mix} the outputs produced by the sub-encoder (one for each user in the target credential database) and condense them in a single dense vector---the configuration seed. More pragmatically, the mixing encoder is another deep neural network trained to interpolate over the provided (projected) auxiliary data and distill out the information that the password model can exploit to predict the password distribution of the underlying community. 
\par

  Given that the auxiliary input information does not follow any particular ordering and may drastically vary in size depending on the target credential database, a natural choice to implement the mixing encoder is to rely on an attention mechanism (Section~\ref{sec:att}). In our setup, we implement it with a single AM layer of the form described in Section~\ref{sec:att}. Here, every output produced by a (replica) of a sub-encoder is considered a value vector $v_i$, whereas there is a single query vector~$q$. The latter is an additional embedding vector learned during the training (the yellow rectangle in Figure~\ref{fig:model}). After the training, this vector is used as a constant input for the model.\footnote{Essentially, this is equivalent to the special \texttt{[CLS]} vector used in BERT-like models~\cite{bert}.}
\par

The main reason behind using a single attention mechanism is that this allows an efficient application of differential privacy, as we cover in Section~\ref{sec:dp}.
\begin{algorithm}[b]
\scriptsize
\caption{Seeded password model from \uncm.}\label{alg:mspm}
\KwData{\small  \uncm, Auxiliary information: $\ainf$}
\KwResult{\small Seeded password model: $\seeded$}
$(\senc,\ (\gamma,\ q)) ,\ \dec_{\Theta} = \uncm$\;
$\sainf \sim_{k} \ainf$ \Comment*[r]{Sample at most $k$ users}
$V = \{\}$\;
\For{$a_{\text{inf}} \in \sainf$}{ 
	$v_i = \senc(a_{\text{inf}})$ \Comment*[r]{Apply Sub. enc.}

	$V = V \cup \{v_i\}$\;
}
$\cs = \gamma(\{q\}, V)$ \Comment*[r]{Apply Mix. enc.}
$\dec_{\Theta|\cs} = \dec_{\Theta} \otimes \cs$ \Comment*[r]{Init. p. model with $\cs$}
\end{algorithm}
\subsubsection{Configuration seed computation}
\label{sec:inference}
Provided the set of auxiliary information $\ainf$ (\ie email addresses in the current setup), the sub-encoder $\senc$ and the mix. encoder $\gamma$ are used to compute the configuration seed $\cs$. This process is formalized in Algorithm~\ref{alg:mspm}.
\par

Credential databases can come with millions of users, and computing the seed using the whole population may result {computationally} impractical. Thus, rather than providing the entire set $\ainf$ to the configuration encoder, we sample a uniform set of~$k$~users' auxiliary information $\sainf$ from $\ainf$ which is then used as input for the encoders instead. We tested different values for~$k$; surprisingly, increasing it beyond $2048$ does not result in meaningful utility gains for the average case. Thus, eventually, we opt for $k\myeq 8192$  in the standard setup (as this value still provides better performance in some edge cases) and $k\myeq 2048$ for the private \uncm~so to better capitalize on the effect of \textit{privacy amplification by subsampling} (more on Section~\ref{sec:dp}).
 \par
  
As reported in Algorithm~\ref{alg:mspm}, once sampled the input subset of auxiliary information, the sub encoder $\senc$ is applied on every entry in $\sainf$, resulting in a set $V$ of at most $k$ vectors. This set is then provided to the mix. encoder $\gamma$ as the set of input values aside with the query vector~$q$ (see Section~\ref{sec:mixenc}). Regardless of the number of users $k$, the mix. encoder results in a single output vector of size~$d$; the configuration seed $\cs$, with $d\myeq 756$ in our setup. We tested configuration seeds with dimensionality higher than $756$ achieving no meaningful improvements. \iflong {Regardless of its apparent complexity, the configuration encoder can produce a configuration seed with a sub-second latency. When $k\myeq 8192$, our unoptimized implementation, on average, requires $0.65$ \textit{sec} on a GPU and $0.97$ \textit{sec} on CPU on a NVIDIA DGX-2.} It is worth noting that the configuration seed is generated only once before the seeded model is deployed, and a slower seed generation would not significantly affect the overall performance of the framework.\fi

Finally, the produced configuration seed can be applied to the conditional password model. Details on this final step are provided in the next section.

\subsection{The Conditional Password Model}
\label{sec:ppm}
\input{scheme_decoder}
In order to implement the conditional password model~$\dec_\Theta$ for the \uncm, {we extend the password model proposed by Melicher~\etal~\cite{melichera}. This is a standard, non-conditional, autoregressive password model implemented using a plain $3$-layers LSTM network.} 
The model has been shown to outperform most existing password models, in addition to providing the practical advantages described in \iflong Section\else Appendix\fi~\ref{app:autoreg}. More importantly, the model enables for heavy compression, and it can be used as a client-side password strength meter with limited loss in utility~\cite{melichera}.
\par

In order to convert this model into a conditional password model (see Section~\ref{sec:uncm}), we need to modify it to support the configuration seed $\cs$ as an additional input. Inspired by the classic NLP encoder-decoder paradigm for recurrent networks~\cite{seqseq}, we achieve this by providing  $\cs$ as the initial state of the LSTM cells. As depicted in Figure~\ref{fig:dec}, given the vector $\cs$, we apply two different transformations (one for the hidden state and one for the cell state) for each LSTM layer of the model. In our setup, those transformations are implemented using a pair of different dense layers $(g_h^i,\ g_C^i)$ for each layer~$i$. The resulting $3\cdot 2 \myeq 6$ vectors are then used as initial states for the 3-layer LSTM. We refer to this initialization process with the symbol $\dec_{\Theta} \otimes \cs$ (last line of Algorithm~\ref{alg:mspm}). The model resulting from this operation (\ie the password model with the custom initialized states as depicted in Figure~\ref{fig:dec}) is what we call a {seeded password model}~$\seeded$. 
\par

The initialization via the configuration seed is the only modification to the original model~\cite{melichera}, and both its functionality and internal workings remain unchanged. In the original model, the LSTM's states would be initialized to all zeros. Therefore, \textbf{the two models are virtually identical} and the seeded password models produced by the \uncm~can be compressed using the same techniques used for the original model~\cite{melichera}. Similarly, a seeded model has the same computational complexity as the original model~\cite{melichera}. Note that the dense layers $g_h^i$ and $g_C^i$ do not need to be included in the seeded password model at the deployment phase. Once the configuration seed is generated by~$\enc$, the outputs of those layers (\ie $g_h^0(\cs),\ g_C^0(\cs), \dots g_C^2(\cs)$) can be pre-computed and statically shipped with the model at the configuration phase: $\dec_{\Theta} \otimes \cs$. In total, without any form of compression, these layers account for only $24$KB, so they do not hinder the use of the model as a client-side PSM.


%
\subsection{Training a \uncm}
\label{sec:cotrain}
All existing password models~\cite{MM, melichera, passgan,  pasquini2021improving, pasquini2021usenix} are trained at the \textit{password granularity}. Given a set of plaintext passwords~$X$ (\ie the union of one or more password leaks), at each training round, a single or a batch of passwords is sampled from $X$ and used to tune the parameters of the model; typically, via maximum likelihood estimation. A \uncm, instead, is the first model trained at \textbf{\textit{password-leak granularity}}; that is, the atomic training instance for the model is a whole password leak rather than a single password. To be trained, therefore, a \uncm~requires a collection of credential databases. Hereafter, we use the training leak collection $\lct$ from \citday~presented in Section~\ref{sec:trainingset}. 
\par

The password-leak-based training process is defined as follows. At each training step, \textbf{(1)} we sample a credential database $S\myeq (\ainf, X)$ from the leak collection $\lct$. \textbf{(2)}~Following Algorithm~\ref{alg:mspm}, a subset of auxiliary information~$\sainf$ is given as input to the configuration encoder $\enc$ that produces a configuration seed~$\cs$ as output. \textbf{(3)}~The configuration seed~$\cs$ is then applied to the password model~$\dec_\Theta$ that, in turn, is  
 trained to assign a high probability to \textbf{all} the passwords in~$X$~\cite{melichera}, completing a single training iteration. In the process, \textbf{the configuration encoder (\ie the mixing-encoder and the sub-encoder) and the password model are jointly trained as a single network} to maximize the performance of the password model in the conditional password generation task. This objective is summarized by the following optimization problem:
  \begin{equation}
 	\argmax_{\theta,\ \Theta} \left\{ \mathbb{E}_{\tiny{(\ainf, X) \sim \lct}} \sum_{x\in X} \frac{\texttt{xen}(x,\ \dec_{\Theta | \enc(\ainf)}(x))}{|X|} \right\},
 	\label{eq:loss}
 \end{equation}
where \texttt{xen} is cross entropy defined at the character level, quantifying the discrepancy between the prediction of the seeded password model and the ground-truth password $x$.\footnote{This loss is computed via \textit{teacher forcing} for the RNN~\cite{tforcing}.} Intuitively, we are training the configuration encoder to retrieve suitable information from the auxiliary data and shift the mass function of~$\dec_\Theta$ towards the passwords in~$X$. At the same time, we train the password model $f_\Theta$ to generate the passwords~$X$ following the instructions provided by the configuration seed. At the next training round, the models are challenged  with another credential database, forcing them to generalize over to auxiliary data and password space, respectively. 
\iflong  
\par

\textbf{A meta-learning interpretation:} \new{Another pragmatic interpretation is that a \uncm~is just a meta-ML model~\cite{metalearning} trained to produce as output a password model $\dec_{\Theta | \enc(\ainf)}$ such that the probability $\dec_{\Theta | \enc(\ainf)}(X)$ is maximized for a set of passwords~$X$ defined at inference time. The core difference with standard meta-learning approaches (few-shot learning, mainly~\cite{pmlr-v162-zhmoginov22a}) is that, instead of directly providing a set of support examples to define the target task (this would be a subsample of passwords from $X$ in our setting), the meta-model gets only a noisy proxy signal as input (the user auxiliary information).}\fi

It is important to understand that, in Eq ~\ref{eq:loss}, {loss values computed on every $x\in X$ are accumulated in a single error signal that is then used to jointly train both the encoder and the password model within a single SGD iteration}. In our implementation, we rely on minibatch-SGD. Thus, the loss is computed over multiple leaks at each training iteration. \iflong To reduce memory consumption, we rely on gradient accumulation; we compute the gradient signal sequentially for each leak in the mini-batch. Then, we average and apply them to the models' parameters. As discussed in Section~\ref{sec:inference}, we sample only a subsample of at most $k$ users to compute the seed. This means that, at most, in a single iteration, we replicate the sub-encoder $8192$ times, run the attention mechanism on a value set of $8192$ vectors, and compute probabilities for $8192$ passwords with the password model. Nonetheless, despite the size of the models and the large sample size, we managed to train the proposed \uncm~using a single \textit{NVIDIA V100 GPU} with $32$GB of memory thanks to gradient-accumulation.\fi
\par

We train the model until the loss on the test set stops improving (\ie early stopping). We use \textit{Adam} as optimizer~\cite{adam} with the default setting, and we implement the code in \textit{TensorFlow2}. 

\subsection{Baseline: Approx. of the Universal Distribution}
\label{sec:untrain}
\label{sec:baseline}

Within the paper, we use a baseline model to empirically demonstrate the advantages provided by the proposed \uncm's models over previous methodologies. 
 This is a reimplementation of the state-of-the-art password model proposed by Melicher~\etal~\cite{melichera}. We use the same architecture used for the password model in the \uncm~(see Section~\ref{sec:passmodel}). However, we train it following the standard training process considered in~\cite{melichera} that can be summarized by the following optimization problem:
   \begin{equation}
 	\argmax_{\Theta} \left\{ \mathbb{E}_{x \sim \mathbf{X}} \left[ \texttt{xen}(\ x,\ \dec_\Theta(x)\ ) \right] \right\}.
 	\label{eq:lossun}
 \end{equation}
To train this baseline model, we use the same training and validation set used for the \uncm. In Eq.~\ref{eq:lossun}, the training set $\mathbf{X}$ is the union of all the leaks in $\lct$ \ie $\bigcup_{\{\ainf, X\} \in \lct}  X$. All the hyper-parameters are kept the same as for the~\uncm.

\par
 Eventually, given the heterogeneity of \citday, this model can be understood as a suitable approximation of the \textbf{universal password distribution} (see Section~\ref{sec:weakpassnonuni}). That is, an unconditional model optimized to perform well on the average case. Hereafter, we refer to this model as $\dec_{\dot{\Theta}}$. \iflong Pragmatically speaking, fixed the training data, $\dec_{\dot{\Theta}}$ represents the best \textit{off-the-shelf} model that system administrators can use for their systems when there is no possibility of producing tailored password models.\fi\ Details on the training process and architectures are reported in Appendix~\ref{app:params}.

\section{Evaluation}
\label{sec:result}
\begin{figure*}[ht!]

	\centering
	\captionsetup{justification=centering}

	\includegraphics[trim = 12mm 77mm 0mm 0mm, clip, width=.6\columnwidth]{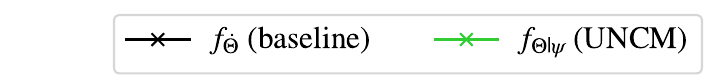}
	
\resizebox{1\textwidth}{!}{
	\begin{subfigure}{.4\columnwidth}
		\centering
		\includegraphics[trim = 0mm 0mm 0mm 0mm, clip, width=.95\columnwidth]{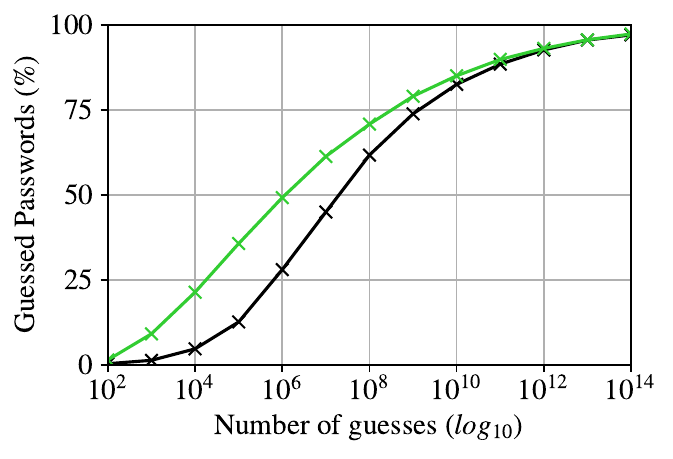}
		\caption{ \textit{edj.pl} \\ \tiny $|X| \myeq 3120$}\label{fig:gb}
	\end{subfigure}
	\begin{subfigure}{.41\columnwidth}
		\centering
		\includegraphics[trim = 0mm 0mm 0mm 0mm, clip, width=.95\columnwidth]{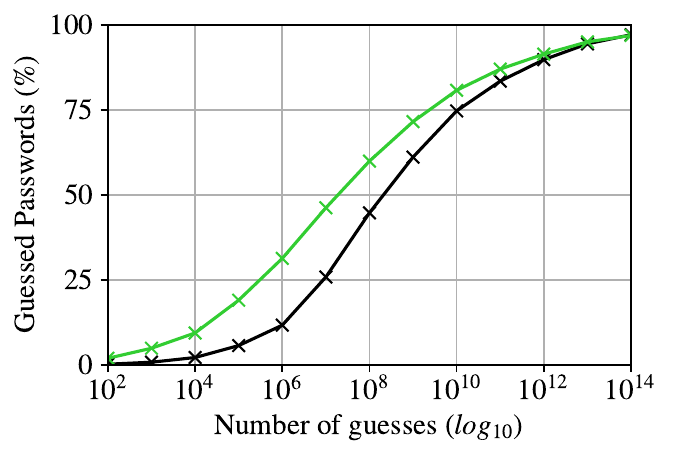}
		\caption{ \textit{atcenter.or.jp}  \\ \tiny $|X| \myeq 3135$}\label{fig:ga}
	\end{subfigure}\begin{subfigure}{.4\columnwidth}
		\centering
		\includegraphics[trim = 0mm 0mm 0mm 0mm, clip, width=.95\columnwidth]{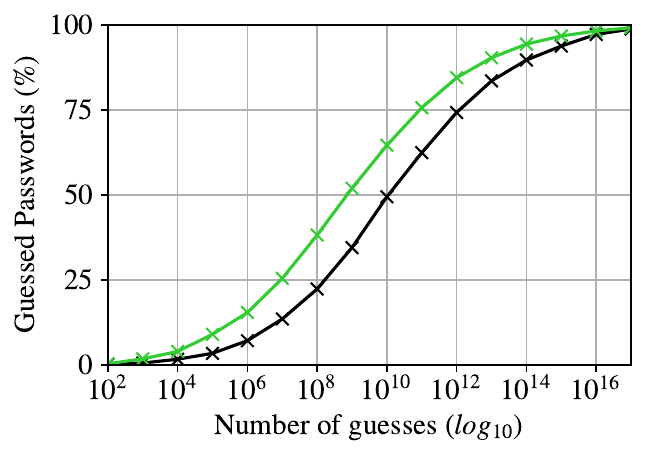}
		\caption{ \textit{fvchinese.com} \\ \tiny $|X| \myeq 21532$}\label{fig:gc}
	\end{subfigure}\begin{subfigure}{.4\columnwidth}
		\centering
		\includegraphics[trim = 0mm 0mm 0mm 0mm, clip, width=.95\columnwidth]{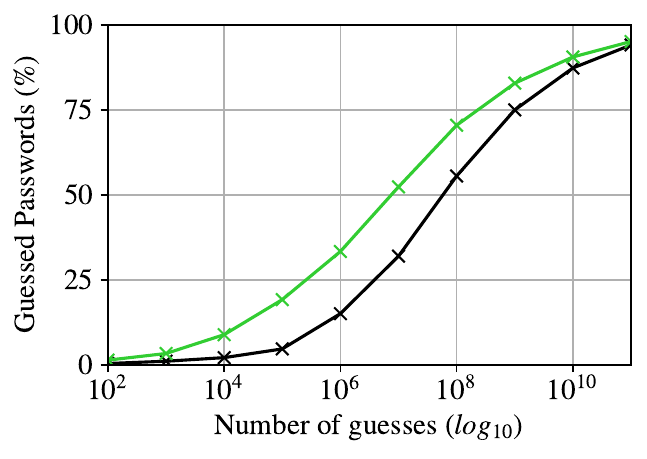}
		\caption{ \textit{chinaguide.co.kr}  \\ \tiny $|X| \myeq 4065$}\label{fig:gd}
	\end{subfigure}\begin{subfigure}{.4\columnwidth}
	\centering
	\includegraphics[trim = 0mm 0mm 0mm 0mm, clip, width=.95\columnwidth]{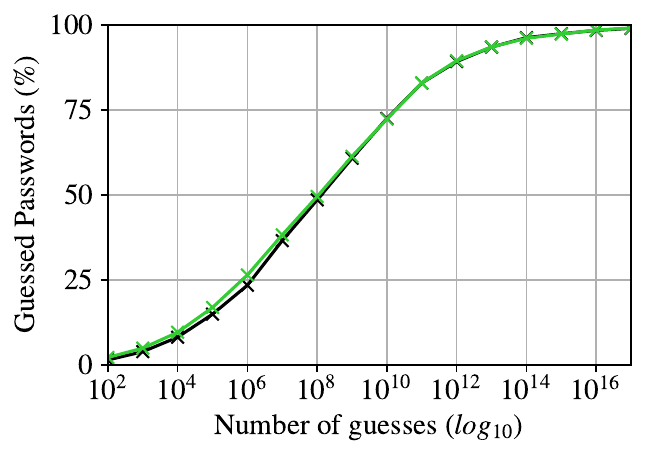}
	\caption{ \textit{everdrybas$\dots$.mobi} \\ \tiny $|X| \myeq 36020$}\label{fig:ge}
\end{subfigure}

}
\captionsetup{justification=raggedright}
	\caption{Comparison between the seeded password model and baseline on guessing attacks for $5$ example password leaks sampled from $\lcv$. 
	 In the caption of each plot, we report the size of the tested password leak aside from the leak~identifier.}
	\label{fig:ex_guess_global}
\end{figure*}
\begin{figure}[t!]
	\centering
	\resizebox{.95\columnwidth}{!}{
		\begin{subfigure}{.4\columnwidth}
		\centering
		\includegraphics[trim = 0mm 0mm 0mm 0mm, clip, width=.95\columnwidth]{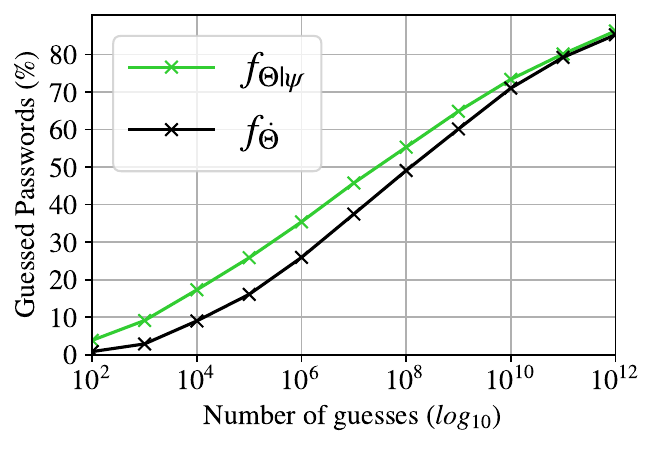}
		\caption{Average performance.}\label{fig:avgg}
	\end{subfigure}\begin{subfigure}{.41\columnwidth}
		\centering
		\includegraphics[trim = 0mm 0mm 0mm 0mm, clip, width=.95\columnwidth]{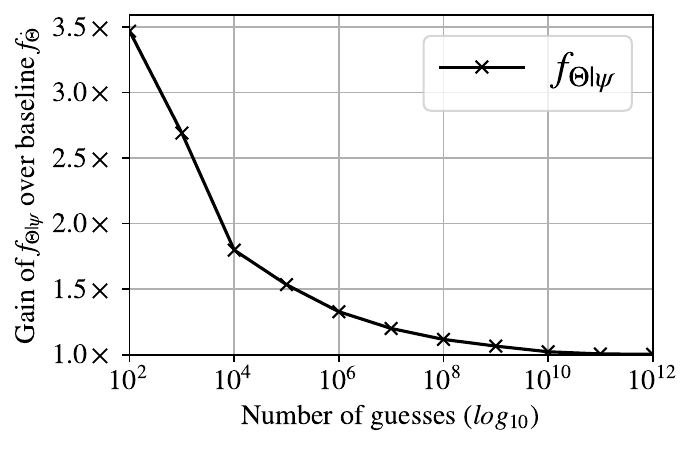}
		\caption{Average gain.}\label{fig:avggain}
	\end{subfigure}
}
	\caption{Panel~(a) reports the \textbf{average} performance of the seeded password models and the baseline on guessing attacks. Panel~(b) reports the number of guessed passwords of $\seeded$ over the baseline during the attacks. For instance, the coordinates $(10^2, 3.4)$ indicate that, on average, the seeded password models guessed $3.4$ times the number of passwords guessed by the baseline within $10^2$ guesses.}
	\label{fig:gn}

\end{figure}
Next, we empirically validate the capabilities of the password models produced by the~\uncm. We first compare them to the baseline. Then, we show that the seeded models produced by the pre-trained \uncm~can outperform password models that have been manually configured and trained for the target password distribution. 
\subsection{UNCMs vs. Universal approximation}
\label{sec:result_un}
We now compare the pre-trained \uncm~with the baseline model $\uni$~\ie the approximation of the universal password distribution (see Section~\ref{sec:untrain}).
\par

\paragraph{\textbf{Evaluation process}}
Given the pre-trained~\uncm, for each leak in $S\myeq (\ainf, X) \myin \lcv$ (where $\ainf$ are just the users' email addresses), we run Algorithm~\ref{alg:mspm} and generate a seeded password model~$\seeded$ for~$S$. Then, we use $\seeded$~and~$\uni$~to guess the unknown passwords in~$X$. In the process, we use the Monte Carlo method proposed by Dell'Amico~\etal~\cite {monte} to efficiently estimate the guess numbers for both models. Details on the hyper-parameters are given in Appendix~\ref{app:params}. 
\par

\paragraph{\textbf{Results}}
Figure~\ref{fig:ex_guess_global} reports the result of the guessing attack for $5$ example leaks in $\lcv$. The performance gain obtained by the seeded password model (green) over the baseline (black) is consistent but heterogeneous. As we discuss bellow, seeded models guess passwords faster than the baseline within the initial guesses. In the general case, this advantage tends to decrease as the attack progresses. In other cases, this gain remains consistent even after $10^{12}$ guesses \eg Figure~\ref{fig:gc}. A desirable property of the \uncm~is that it will never perform worse than the baseline. Indeed, when no suitable information can be inferred from $\ainf$, it seems that the \uncm~automatically falls back to model an approximation of the universal password distribution, performing on par with the baseline. Figure~\ref{fig:ge} shows an example of this behavior. 
\par

Regardless of these limit cases, the \uncm~performs better than the baseline with a substantial margin on average. This is shown in Figure~\ref{fig:avgg}, where we report the guessing performance of the seeded and baseline models averaged over $100$ different leaks uniformly sampled from~$\lcv$. Such a performance gain can be better understood in Figure~\ref{fig:avggain}, where we plot the average ratio of guessed passwords for the two models during the attacks (\ie $\frac{\# \text{passwords guessed by} \seeded }{\# \text{passwords guessed by} \uni}$). Within the initial $10^{2}$ guesses, the \uncm~guesses about $3.4$ times ($340\%$) more passwords than the universal approximation. Then, this gain gently decreases with the number of guesses. Additional results are provided in Appendices~\ref{app:pcfg}  and \ref{app:dynamic}, where we compare with other standard password models and dynamic dictionary attacks~\cite{pasquini2021usenix}, respectively.

\textbf{Remark:} It is important to emphasize that the passwords models~$\seeded$ and~$\uni$ have an equal number of parameters\ \iflong\footnote{For the \uncm, the configuration encoder is not used during the guessing attack; only the seeded password model is employed.}\fi and they have been trained on the same training data. The only difference is that the seeded models have access to the configuration seed computed over a subset of users' auxiliary information (only email addresses in the current setup). Additionally, recall that we removed all the users appearing in the training set from the test leaks in $\lcv$; thus, we can exclude that the \uncm~could memorize users' email addresses and simulate credential stuffing/tweaking attacks~\cite{stuffing}.

\subsubsection{\textbf{Evidence of adaptation to the target distribution}}
\begin{table}
	\centering
	\caption{Examples of weak passwords overestimated by the universal distribution approximation (guess numbers).}
	\label{table:peculiar}
	\resizebox{1\columnwidth}{!}{%
		\begin{tabular}{c|ccccc}
			\toprule
			
			\textbf{\textit{okmedicina.it}} & pippicalzelunghe & ascolipiceno & nonsaprei1 & baiaimperiale & piazzacairoli \\
			$f_{\dot{\Theta}}$ & $2 \cdot 10^{18}$ & $4 \cdot 10^{12}$ & $2 \cdot 10^{12}$ & $5 \cdot 10^{13}$ & $3 \cdot 10^{13}$ \\
			$f_{\Theta | \psi}$ & $4 \cdot 10^{9}$ & $2 \cdot 10^{6}$ & $3 \cdot 10^{6}$ & $2 \cdot 10^{9}$ & $9 \cdot 10^{9}$ \\
			\midrule
			
			\textbf{\textit{weryfikatorium.pl} }& rzeczpospolita & dywizjon303 & kurkawodna3 & samarytanin1 & kotwbutach28 \\
			$f_{\dot{\Theta}}$ & $6 \cdot 10^{15}$ & $1 \cdot 10^{15}$ & $8 \cdot 10^{12}$ & $2 \cdot 10^{13}$ & $1 \cdot 10^{15}$ \\
			$f_{\Theta | \psi}$ & $6 \cdot 10^{4}$ & $1 \cdot 10^{8}$ & $9 \cdot 10^{5}$ & $3 \cdot 10^{6}$ & $4 \cdot 10^{8}$ \\
			\midrule
			\textbf{\textit{atcenter.jp}} & kannkoku415 & 78hokuhoku & yutarikari7 & honbadakara & taikyokuken \\
			$f_{\dot{\Theta}}$ & $9 \cdot 10^{12}$ & $2 \cdot 10^{12}$ & $3 \cdot 10^{12}$ & $9 \cdot 10^{12}$ & $1 \cdot 10^{12}$ \\
			$f_{\Theta | \psi}$ & $2 \cdot 10^{9}$ & $7 \cdot 10^{8}$ & $2 \cdot 10^{9}$ & $6 \cdot 10^{9}$ & $1 \cdot 10^{9}$ \\
			\midrule
			
			
			 \textbf{\textit{le\_monde\_en\_enigmes.fr}} & mot de passe & ouarzazate & mvtmjsunp & 10ruedesroches & Mousquetaires \\
			$f_{\dot{\Theta}}$ & $4 \cdot 10^{11}$ & $1 \cdot 10^{11}$ & $2 \cdot 10^{13}$ & $2 \cdot 10^{14}$ & $1 \cdot 10^{11}$ \\
			$f_{\Theta | \psi}$ & $4 \cdot 10^{4}$ & $2 \cdot 10^{4}$ & $8 \cdot 10^{6}$ & $1 \cdot 10^{8}$ & $7 \cdot 10^{5}$ \\
			\midrule
			
			\textbf{\textit{seminarky.cz}} & ptakopysk & faktnevim123 & Vysokaskola1 & vyjebanec1 & milujulukase \\
			$f_{\dot{\Theta}}$ & $1 \cdot 10^{12}$ & $4 \cdot 10^{14}$ & $2 \cdot 10^{13}$ & $9 \cdot 10^{12}$ & $5 \cdot 10^{12}$ \\
			$f_{\Theta | \psi}$ & $3 \cdot 10^{4}$ & $1 \cdot 10^{8}$ & $1 \cdot 10^{7}$ & $5 \cdot 10^{7}$ & $3 \cdot 10^{7}$ \\
			\midrule
			
			\textbf{\textit{shop.santool.de}} & Maschinenbau & entwicklung & freyssinet99 & Modellbau+1 & baumwolle5 \\
			$f_{\dot{\Theta}}$ & $1 \cdot 10^{12}$ & $3 \cdot 10^{12}$ & $1 \cdot 10^{14}$ & $3 \cdot 10^{13}$ & $3 \cdot 10^{12}$ \\
			$f_{\Theta | \psi}$ & $4 \cdot 10^{4}$ & $4 \cdot 10^{6}$ & $1 \cdot 10^{9}$ & $4 \cdot 10^{8}$ & $1 \cdot 10^{8}$ \\
			\midrule
			
			\textbf{\textit{diarioelsur.cl}} & jehovaesmipastor123 & teamarexsiempre & serviciomovil & toyotires13 & estudiojuridico \\
			$f_{\dot{\Theta}}$ & $3 \cdot 10^{15}$ & $3 \cdot 10^{14}$ & $1 \cdot 10^{13}$ & $4 \cdot 10^{12}$ & $6 \cdot 10^{12}$ \\
			$f_{\Theta | \psi}$ & $1 \cdot 10^{9}$ & $3 \cdot 10^{9}$ & $1 \cdot 10^{8}$ & $2 \cdot 10^{8}$ & $3 \cdot 10^{8}$ \\
			\midrule
			
			\textbf{\textit{leilaoonlinerural.br}} & fellipeleao1 & harasoficial & MeuArquivo & damiaomatos & comunica?ao \\
			$f_{\dot{\Theta}}$ & $1 \cdot 10^{14}$ & $2 \cdot 10^{14}$ & $2 \cdot 10^{14}$ & $4 \cdot 10^{12}$ & $2 \cdot 10^{12}$ \\
			$f_{\Theta | \psi}$ & $2 \cdot 10^{9}$ & $6 \cdot 10^{9}$ & $8 \cdot 10^{9}$ & $2 \cdot 10^{8}$ & $3 \cdot 10^{8}$ \\
			\bottomrule
			
		\end{tabular}
	}
\end{table}
\begin{figure*}[t!]
	\centering
	\captionsetup{justification=centering}
	
	\begin{tabular}{c}

		\begin{subfigure}{.34\columnwidth}
			\centering
			\includegraphics[trim = 0mm 0mm 0mm 0mm, clip, width=.95\columnwidth]{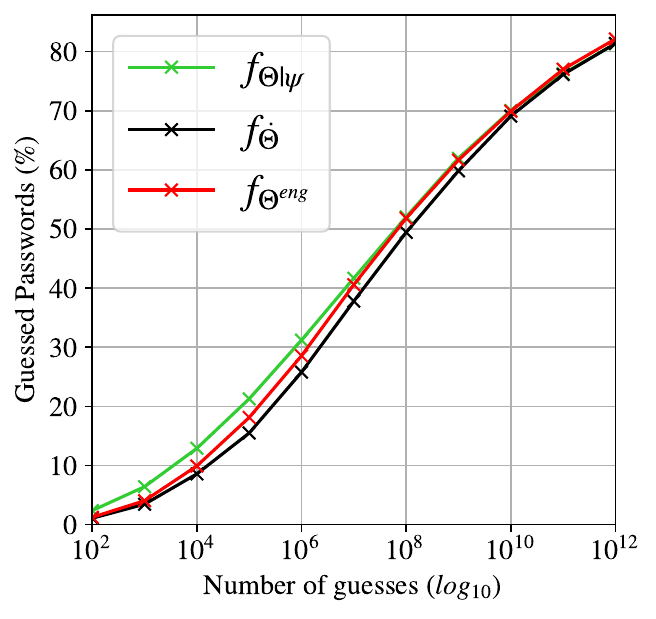}
		\end{subfigure}\begin{subfigure}{.33\columnwidth}
			\centering
			\includegraphics[trim = 0mm 0mm 0mm 0mm, clip, width=.95\columnwidth]{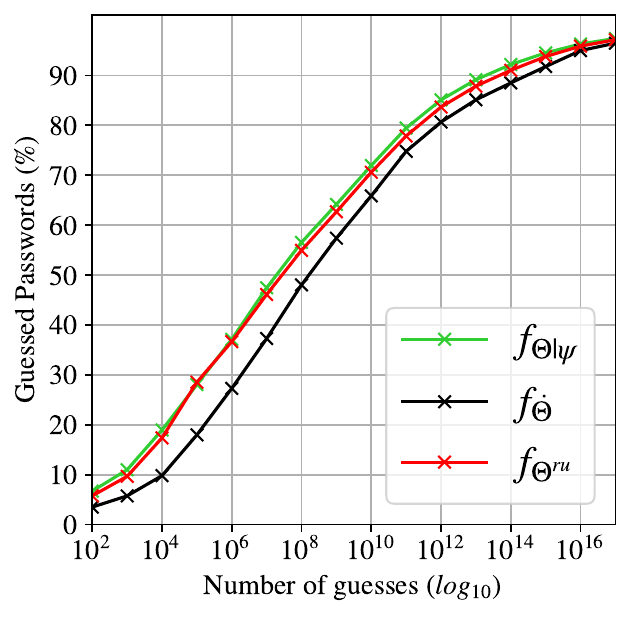}
		\end{subfigure}\begin{subfigure}{.34\columnwidth}
			\centering
			\includegraphics[trim = 0mm 0mm 0mm 0mm, clip, width=.95\columnwidth]{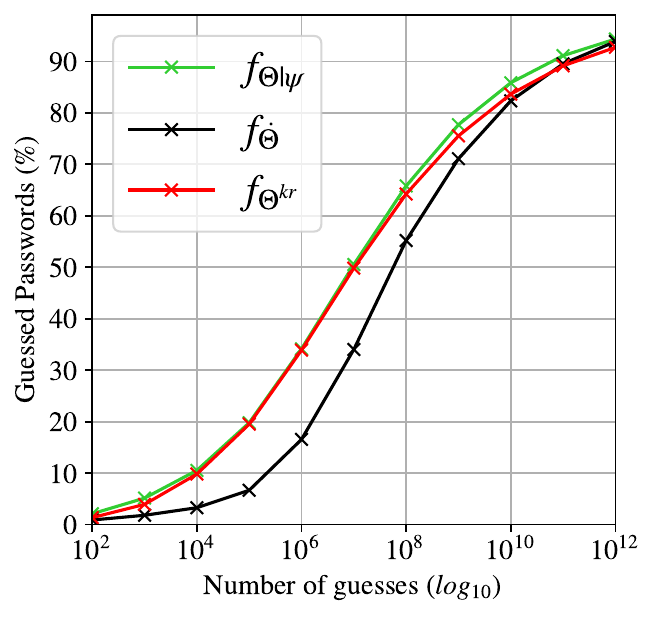}
	
		\end{subfigure}\begin{subfigure}{.33\columnwidth}
			\centering
			\includegraphics[trim = 0mm 0mm 0mm 0mm, clip, width=.95\columnwidth]{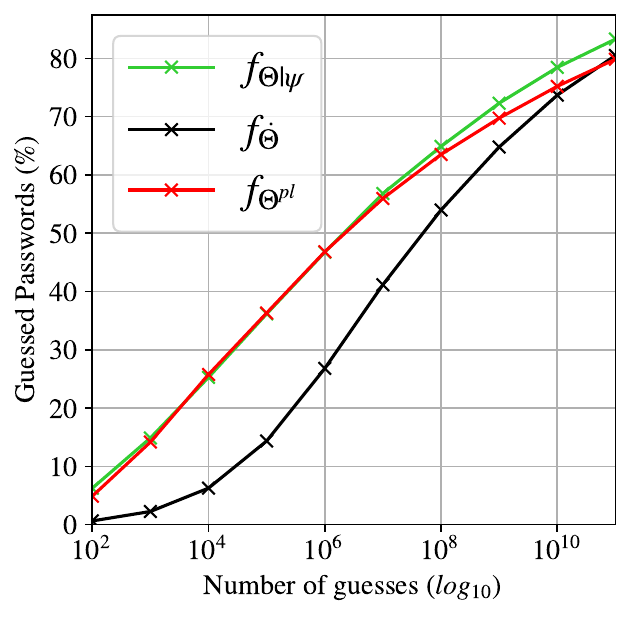}
		\end{subfigure}\begin{subfigure}{.34\columnwidth}
			\centering
			\includegraphics[trim = 0mm 0mm 0mm 0mm, clip, width=.95\columnwidth]{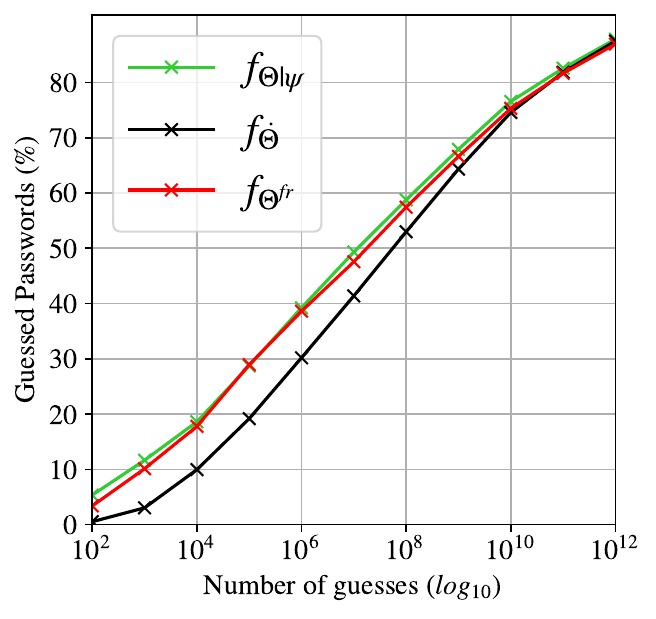}
		\end{subfigure}
		\begin{subfigure}{.34\columnwidth}
			\centering
			\includegraphics[trim = 0mm 0mm 0mm 0mm, clip, width=.95\columnwidth]{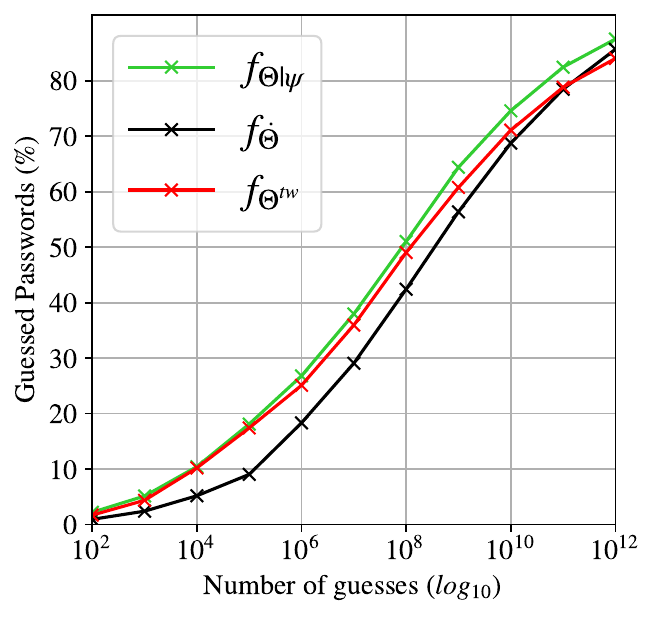}
		\end{subfigure}

\\
	\begin{subfigure}{.34\columnwidth}
		\centering
		\includegraphics[trim = 0mm 0mm 0mm 0mm, clip, width=.95\columnwidth]{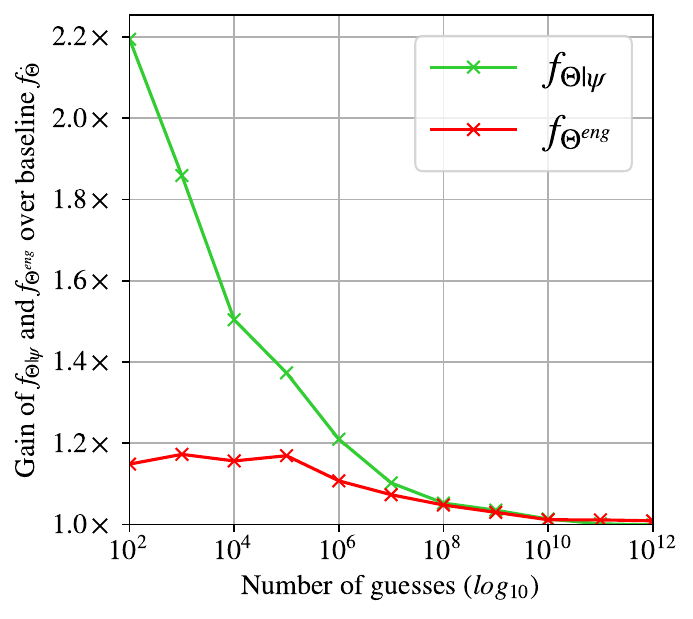}
		\caption{English \\ \scriptsize $|\mathbf{X}^{eng}|\myeq 24,616,140$} \label{fig:ru}
	\end{subfigure}\begin{subfigure}{.33\columnwidth}
		\centering
		\includegraphics[trim = 0mm 0mm 0mm 0mm, clip, width=.95\columnwidth]{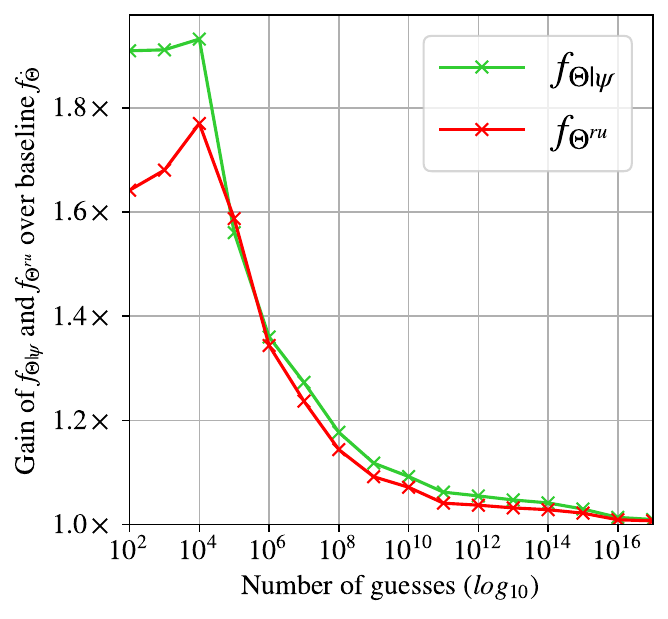}
		\caption{Russian \\ \scriptsize $|\mathbf{X}^{.ru}|\myeq 5,389,085$} \label{fig:ru}
	\end{subfigure}\begin{subfigure}{.34\columnwidth} \
		\centering
		\includegraphics[trim = 0mm 0mm 0mm 0mm, clip, width=.95\columnwidth]{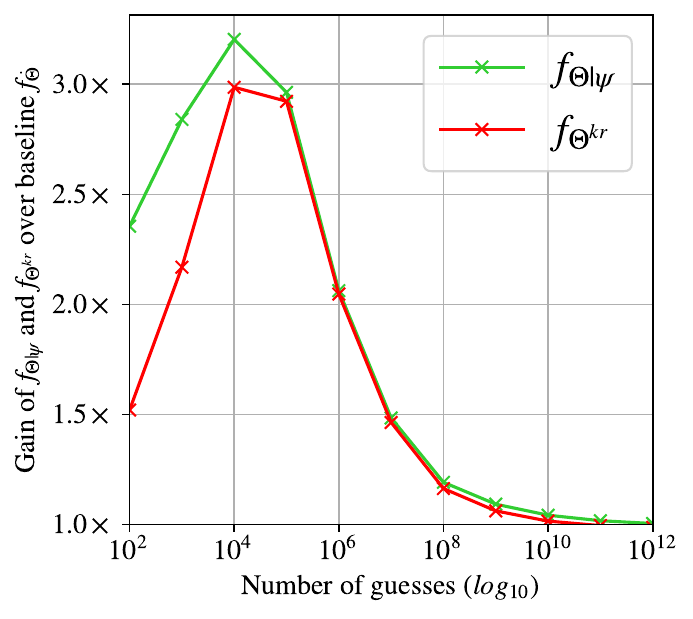}
		\caption{Korean \\ \scriptsize $|\mathbf{X}^{.kr}|\myeq 4,390,811$}\label{fig:defc}
	\end{subfigure}\begin{subfigure}{.34\columnwidth}
		\centering
		\includegraphics[trim = 0mm 0mm 0mm 0mm, clip, width=.95\columnwidth]{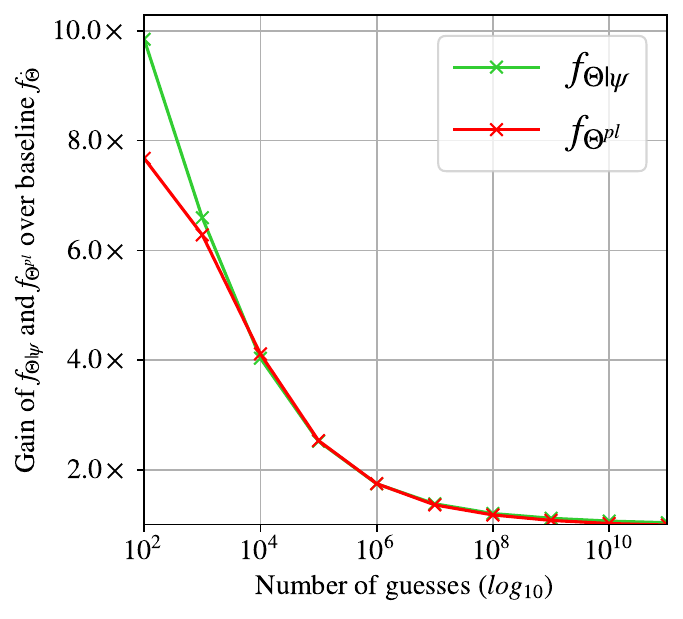}
		\caption{Polish \\ \scriptsize $|\mathbf{X}^{.pl}|\myeq 1,434,802$}\label{fig:defb}
	\end{subfigure}\begin{subfigure}{.34\columnwidth}
		\centering
		\includegraphics[trim = 0mm 0mm 0mm 0mm, clip, width=.95\columnwidth]{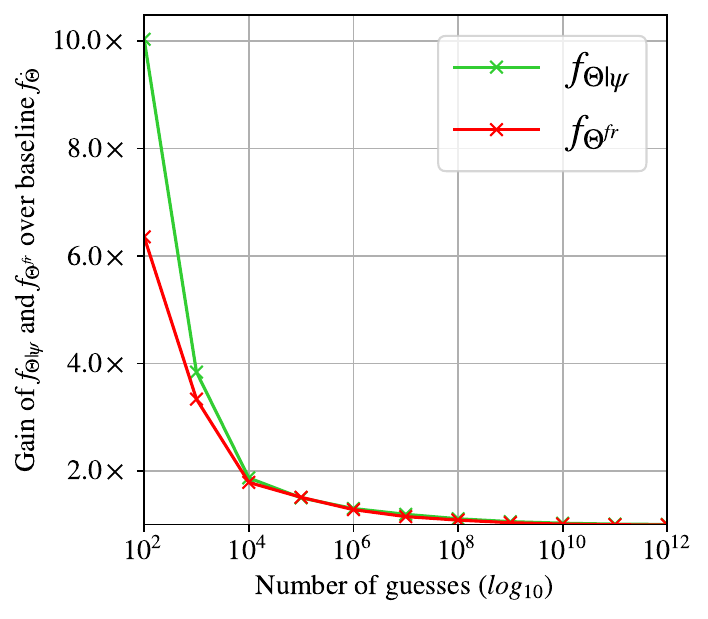}
		\caption{French \\ \scriptsize $|\mathbf{X}^{.fr}|\myeq 1,256,036$}\label{fig:defd}
	\end{subfigure}
	\begin{subfigure}{.33\columnwidth}
		\centering
		\includegraphics[trim = 0mm 0mm 0mm 0mm, clip, width=.95\columnwidth]{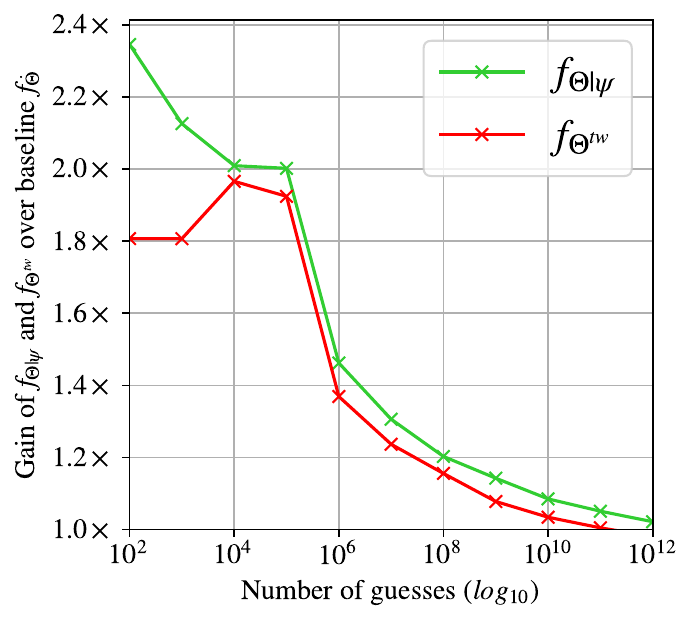}
		\caption{Taiwanese \scriptsize $|\mathbf{X}^{.tw}|\myeq 935,731$}\label{fig:deff}
	\end{subfigure}

	\end{tabular}
	\centering
	
	\captionsetup{justification=raggedright}
	\caption{
	\iflong For $6$ top domains $\mathtt{lan}$, each plot in the first row compares the average guessing performance of three password models (baseline, tailored model for $\mathtt{lan}$, and seeded password model generated by \uncm) computed over all the leaks with top domain~$\mathtt{lan}$ in~$\lcv$. The second row reports the relative ratio of the number of guessed passwords by $\dec_{\Theta^\mathtt{lan}}$ and $\seeded$ over the baseline. In the caption of each column, we report the cardinality of the leak collection used to train the tailored model $\dec_{\Theta^\mathtt{lan}}$.
	\else
	For $6$ language-specific password distributions, each plot in the first row compares the average guessing performance of three password models. The second row reports the relative ratio of the number of guessed passwords by $\dec_{\Theta^\mathtt{lan}}$ and $\seeded$ over the baseline ($\seeded$ is the seeded model, $\dec_{\dot{\Theta}}$ is the baseline, and $\dec_{\Theta^\mathtt{lan}}$ indicates the tailored model).
	\fi
	}
	\label{fig:uncm_vs_manual}
\end{figure*}

Recall that a password strength meter's main objective is to detect and prevent users from picking easily guessable passwords. Thus, correctly identifying passwords with low-guess numbers is the most valuable property of a meter~\cite{evpsm}. \ournew{The results above showcase the capability of the \uncm\ to detect weak passwords that are missed by the universal approximation model. In this section, we provide further intuition on the nature  of these mislabeled weak passwords.}



Table~\ref{table:peculiar} reports examples of weak passwords mislabeled by the baseline $\uni$~for $8$ leaks with different top domains in $\lcv$. For each leak, we report the first $5$ passwords such that the difference between the guess number attributed by $\uni$~and the one attributed by $\seeded$~is maximal{; formally, $\underset{x \in X}{top5}(\frac{\uni(x)}{\seeded(x)})$.} All the passwords reported in the table are considered relatively secure by the baseline (\ie the model assigns high guess numbers to them). However, as can be seen, most of those passwords are combinations of common nouns for the underlying community's language that, arguably, would be easily guessed by any realistic attack \eg a dictionary attack based on the source language's vocabulary.

As discussed in Section~\ref{sec:uncm}, the \uncm~is able to automatically adapt to the target password distribution by relying on users' auxiliary information. This enables the seeded model to correctly estimate the guessability of those peculiar weak passwords even if they lie in underrepresented modalities of the training set. To provide an intuitive example, the password \TT{pippicalzelunghe}\footnote{Italian name for \TT{Pippi Longstocking}.} (first row) requires $10^{18}$ guesses for the baseline (\ie computationally unguessable) while the \uncm~demonstrates that this password is weak and can be guessed within the first $10^{9}$ guesses even if \textit{Italian} passwords only account for $1.2\%$ of the whole \citday. The majority of the passwords listed in the table illustrate the same phenomenon but for different communities.

\subsection{\uncm\ vs. Manual configuration}
\label{sec:result_mc}
While we have shown that a pre-trained \uncm~performs better than a password model trained to approximate the universal password distribution, it is necessary to ask if a \uncm~can reach comparable performance to models that have been manually configured for the target system. As already discussed, given a target credential database~$S$, one can seek an (\textit{a priori}) optimal configuration for the password model by collecting and training on password leaks that share common characteristics with the target password distribution. In the following experiment, we use the target community's language as the main characteristic to define password similarity. This is universally recognized as the most impactful feature in shaping password distributions~\cite{psswd_lan}.
\par

\paragraph{\textbf{Evaluation process}}

 In the experiments, we consider $6$ different languages: \TT{English}, \TT{Russian}, \TT{Korean}, \TT{Polish}, \TT{French}, and \TT{Taiwanese} (listed in decreasing password count).
To automate the collection of tailored training sets from \citday, we use the top domain of the leaked website to infer the language of the underlying community; for instance, if a website has top-domain \TT{.pl}, we assume that the community has \textit{Polish} as the main language. For each top domain $\mathtt{lan}$, evaluation follows as:
\begin{enumerate}
	\item We collect all the leaks in $\lct$ (training collection) and $\lcv$ (test collection) with top domain $\mathtt{lan}$, creating two leak sub-collections~$\lct^\mathtt{lan}$ and~$\lcv^\mathtt{lan}$, respectively. For the \TT{English} passwords, we use a different filtering process which is described in Appendix~\ref{app:filter_eng_psswd}.
	\item We train the baseline model~\cite{melichera} as described in Section~\ref{sec:untrain} but with the training set~$\mathbf{X}^{\mathtt{lan}}\myeq \bigcup_{ \lct^\mathtt{lan}}X$, obtaining a model $\dec_{\Theta^\mathtt{lan}}$ \ie a tailored password model for the passwords with language~$\mathtt{lan}$.
	\item Then, we evaluate $\dec_{\Theta^\mathtt{lan}}$, $\uni$~and $\seeded$~on the collection of leaks $\lcv^\mathtt{lan}$ (password leaks from communities with language $\mathtt{lan}$). The evaluation follows the general procedure used in the previous setting.
\end{enumerate}

%

\paragraph{\textbf{Results}}
Results are summarized in Figure~\ref{fig:uncm_vs_manual}, where the number of guesses is averaged over all the leaks in $\lcv^\mathtt{lan}$ for any $\mathtt{lan}$, respectively. As expected, both the seeded password models $\seeded$ and the manually tailored ones $\dec_{\Theta^\mathtt{lan}}$ outperform the baseline. However, a valuable result is that not only do the seeded models match the performance of the manually configured ones, but they also manage to perform slightly better. In particular, the seeded models always guess more passwords correctly on the initial part of the attack and occasionally achieve higher performance also on high-guess-number passwords, as shown by the Korean, Polish, and Taiwanese cases. This suggests that the prior learned by the \uncm~from the users' email addresses can be more informative than the language prior exploited by the manually configured models~$\dec_{\Theta^\mathtt{lan}}$. Another decisive factor is that, while the \uncm\ adapts to the target distribution, it can still exploit the knowledge offered by the whole training set $\lct$. This additional knowledge can help the model to guess strong passwords that may be independent of the target distribution, while also improving its ability to generalize during the training process. We empirically confirmed this intuition by training a \uncm\ exclusively on $\mathbf{X}^{eng}$ and compare it against $\dec_\Theta^{eng}$. In this data constrained setting, the seeded model did not provide any significant advantage over the tailored model. Nonetheless, we stress that the additional knowledge offered by non-target distributions cannot be capitalized by standard password models as demonstrated by the low performance of the baseline model which has been trained on all available data.

The advantage of the seeded models seems to be inversely proportional to the size of the training set used to train~$\dec_{\Theta^\mathtt{lan}}$. Intuitively, the less data is available to train a tailored model, the larger is the potential performance gain achievable by the \uncm. In the results, the only exception is on the weakest passwords of the English setting (\ie up to $10^{6}$ guesses), where seeded models outperform the tailored one regardless of the size of the curated training set. We attribute this phenomenon to the inherent internal diversity of this distribution. Indeed, in contrast to the other settings in the figure, this is obtained as the aggregation of multiple sub-modalities (\eg English, American, Australian, etc.) which the \uncm\ can still potentially leverage for adaptation. However, it appears that, in this particular setting, the \uncm\ does not provide any advantage beyond the $10^{8}$ mark. 
\par

 In Appendix~\ref{app:categ}, we extend our analysis to other semantic factors such as the service provided by the web application (\eg Shopping, Education, Entertainment, etc.), achieving congruent results with what observed on the language-specific setting.

In conclusion, a pre-trained \uncm~is capable of achieving equivalent if no better performance than equivalent state-of-the-art password models, even if those are specifically configured for the target password distribution. To achieve this result, end-users do not need to manually retrieve tailored training sets or undertake any additional training phases; starting from the pre-trained model, a \uncm~only requires access to users' email address to automatically generate a password model configured for the target system.
%
%
\section{\underline{Private} \uncms}
\label{sec:puncm}
As discussed in Section~\ref{sec:deploy}, \textit{Bob} can download the pre-trained \uncm~and configure it for the target distribution by executing Algorithm~\ref{alg:mspm}. The obtained seeded password model~$\seeded$ can be then deployed as client-side PSM, for instance, by embedding it in the sign-up web page of $\mathcal{S}$.
 However, by releasing the model $\seeded$, \textit{Bob} may introduce privacy risk for the users whose information has been used to forge the seed~$\cs$. Indeed, having access to the seeded password model, and, so, to~$\cs$, a \textit{privacy-attacker} would be potentially able to infer information on $\sainf$. For instance, the adversary may be able to run membership inference attacks on the seed and infer whether a particular email address has been used to compute it, and, so, whether the user has an account in~$\mathcal{S}$. While, for the current setup, the introduced privacy risk is \iflong  marginal\footnote{Attackers can typically learn if an email address is in $\mathcal{S}$ in more efficient and reliable ways~\cite{userenum}.}\else limited\fi, the concern becomes relevant in the case other users' PII is available and used to forge the configuration seed. Ultimately, even if the configuration seed is a highly compressed representation of the auxiliary input data, it is not possible to exclude that it would leak arbitrary information on the web application's users. Next, we show that it is possible to \textbf{efficiently achieve formal user-level privacy guarantees} by designing a configuration encoder with differentially-private output. To this end, we introduce a differentially private attention mechanism that may have further applications in other contexts.
\iflong
\par

\textbf{Remark:}
It's important to note that the use of a private \uncm~is only necessary when the seeded model is made public, such as in the case of a PSM. In other standard settings, like reactive guessing attacks~\cite{6234434}, \textit{Bob} can use the original construction described in Section~\ref{sec:uncm} without further modifications, even when sensitive data is used for the configuration step.
 \fi
\subsection{Differentially-private configuration seeds}
\label{sec:dp}
%

\textbf{Threat model:}  Formally,  we model an adversary $\mathcal{A}$ who aims at inferring information on the set $U$ of users used to forge the configuration seed $\cs$ \ie the users associated to~$\sainf$. 
\iflong
Following the differential privacy model, $\mathcal{A}$  is a strong adversary who has arbitrary information on the distribution of users and unbounded computational capabilities.
\par
\fi
Given $\mathcal{A}$, we are challenged to produce a perturbed configuration seed $\widetilde{\cs}$ such that the adversary has a bounded advantage in inferring information about the users. We achieve this by making the seed differentially private (see Section~\ref{sec:dp}). We stress that the training set originally used to fit the \uncm~is public (\eg \textit{Cit0day} used in the paper) and it does not require to be protected from the adversary~$\mathcal{A}$.\footnote{Under the easily enforceable assumption that $\mathcal{S}$ and the training set are disjointed.} However, if the training set is assumed to be private, standard DP-SGD~\cite{dldp} can be used during the training process of the \uncm\ as for any other password model.
\subsubsection{A Differentially-Private Attention-Mechanism}
In order to achieve differentially-private configuration seeds, we need to act on the configuration encoder. As depicted in Figure~\ref{fig:model}, the information $\sainf$ we need to protect is individually projected through the sub-encoder, \textbf{becoming the set of value vectors for the mixing encoder}---an attention mechanism (AM). Thus, if we make the AM differentially private w.r.t. the value vectors, we obtain a differentially private configuration seed. We highlight that, unlike the typical ML setting,  \textbf{we want our model's outputs to be differentially private and not its parameters.}
\par

\iflong
\begin{figure}
	\centering
	\resizebox{.25\textwidth}{!}{
		\begin{tikzpicture}		
			
			\tikzstyle{chosen} = []
			\tikzstyle{other} = [rectangle, draw=red]
			\tikzstyle{arrow} = [->,>=stealth]
			
			\node (t0) [chosen] {$q_i$};
			\node (t1) [other, right of=t0] {$v_1$};
			
			\node (fk0) [above of=t0] {$g_Q$};
			\node (fq0) [above of=t1] {$g_K$};
			
			\draw [arrow] (t0) -- (fk0);
			\draw [arrow] (t1) -- (fq0);
			
			\node (d0) [above of=fk0, xshift=.5cm] {\Large{$\odot$}};
			
			\draw [arrow] (fk0) -- (d0);
			\draw [arrow] (fq0) -- (d0);
			
			\node (t2) [chosen, right of=t0, xshift=1.5cm] {$q_i$};
			\node (t3) [other, right of=t2] {$v_2$};
			
			\node (fk1) [above of=t2] {$g_Q$};
			\node (fq1) [above of=t3] {$g_K$};
			
			\draw [arrow] (t2) -- (fk1);
			\draw [arrow] (t3) -- (fq1);
			
			\node (d1) [above of=fk1, xshift=.5cm] {\Large{$\odot$}};
			
			\draw [arrow] (fk1) -- (d1);
			\draw [arrow] (fq1) -- (d1);
			
			\node (t4) [chosen, right of=t2, xshift=1.5cm] {$q_i$};
			\node (t5) [right of=t4, other] {$v_3$};
			
			\node (fk2) [above of=t4] {$g_Q$};
			\node (fq2) [above of=t5] {$g_K$};
			
			\draw [arrow] (t4) -- (fk2);
			\draw [arrow] (t5) -- (fq2);
			
			\node (d2) [above of=fk2, xshift=.5cm] {\Large{$\odot$}};
			
			\draw [arrow] (fk2) -- (d2);
			\draw [arrow] (fq2) -- (d2);
			
			\node (sf) [above of=d1, yshift=-.1cm] {\textit{sigmoid}};
			
			\draw [arrow] (d0) -- (sf);
			\draw [arrow] (d1) -- (sf);
			\draw [arrow] (d2) -- (sf);
			
			\node (w0) [above of=sf, yshift=-.4cm] {$\cdot w_1$};
			\node (w1) [above of=w0, yshift=-.5cm] {$\cdot  w_2$};
			\node (w2) [above of=w1, yshift=-.5cm] {$\cdot  w_3$};
			\draw [arrow] (sf) -- (w0);
			
			\node (v0) [other, left of=w0, xshift=-2cm] {$v_1$};
			\node (v1) [other, above of=v0, yshift=-.5cm] {$v_2$};
			\node (v2) [other, above of=v1, yshift=-.5cm] {$v_3$};
		
			\node (fv0) [right of=v0, xshift=.5cm] {$g_V$};
			\node (fv1) [right of=v1, xshift=.5cm] {$g_V$};
			\node (fv2) [right of=v2, xshift=.5cm] {$g_V$};
			
			\draw [arrow] (v0) -- (fv0);
			\draw [arrow] (v1) -- (fv1);
			\draw [arrow] (v2) -- (fv2);
			
			\draw [arrow] (fv0) -- (w0);
			\draw [arrow] (fv1) -- (w1);
			\draw [arrow] (fv2) -- (w2);
			
			\node (fv0) [right of=v0, xshift=.5cm] {$g_V$};
			\node (fv1) [right of=v1, xshift=.5cm] {$g_V$};
			\node (fv2) [right of=v2, xshift=.5cm] {$g_V$};
			
			\node (c0) [right of=w0, xshift=.1cm] {$\texttt{C}_s$};
			\node (c1) [right of=w1, xshift=.1cm] {$\texttt{C}_s$};
			\node (c2) [right of=w2, xshift=.1cm]  {$\texttt{C}_s$};
			
			\draw [arrow] (w0) -- (c0);
			\draw [arrow] (w1) -- (c1);
			\draw [arrow] (w2) -- (c2);

			\node (agg) [right of=c1] {\large{$\Sigma$}};
			
			\draw [arrow] (c0) -- (agg);
			\draw [arrow] (c1) -- (agg);
			\draw [arrow] (c2) -- (agg);
			
			\node (noise) [right of=agg, xshift=0.4cm] {$+\mathcal{N}(0, \mathbb{I}zs)$};
			\draw [arrow] (agg) -- (noise);
		
			\node (eq) [above of=noise, xshift=0cm, yshift=-.5cm, rotate=90] {$=$};
			\node (res) [above of=noise] {$\widetilde{\gamma}(q_i,\ \{v_1, v_2, v_3\})$};
		
		\end{tikzpicture}
	}
	\caption{Depiction of the introduced differentially-private attention-mechanism (DP-AM). In the computational graph, private inputs are highlighted in red.}
	\label{fig:attdp}
\end{figure}
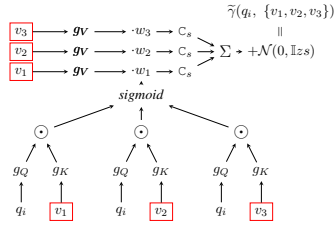
\fi

Luckily, the AM described in Section~\ref{sec:att} offers natural support to the application of DP. This becomes evident once we rename the weighted projection of the $i$'th value vector as $\pwv_i \myeq w_i \mycdot g_V(v_i)$. Then, the output of the attention mechanism can be rewritten as: $\gamma \myeq \sum_{i=1}^{n} \pwv_i$, which is a standard setting for the application of DP, \eg~\cite{dldp, brendan2018learning}. 
\par 

However, there is still a problem; the output of the \textit{softmax} used to compute the attention weights $w_i$
 is a function of all the value vectors as it normalizes every dot product ${d_i \myeq g_Q(q)\cdot g_K(v_i)^T}$ over their sum: $w_i \myeq \frac{e^{d_i}}{\sum_{j} e^{d_j}}$. This indirectly induces a correlation among the private vectors before the application of the noise, preventing us from achieving instance-level-DP (where \TT{instance} is the single value vector). To sidestep this issue, we replace the \textit{softmax} with an element-wise \textit{sigmoid} function. Similarly to the \textit{softmax}, the \textit{sigmoid} function maps the output of every dot product in the interval $[0,1]$ independently, making every attention weight $w_i$ a function of the sole corresponding value vector $v_i$.\footnote{However, those cannot be properly called \TT{attention weights} as their sum can be different from $1$.} 
 \par

Once we have independent attention weights, making the final sum ($\epsilon,\delta$)-DP reduces to bound the sensitivity of~$\gamma$ and then adds calibrated Gaussian noise to its output. In particular, we bound the sensitivity of the weighted values by clipping the $L_2$ norm of each vector $\pwv_i$ to a chosen value~$s$ via the clipping function:
	$\texttt{C}_s(\pwv_i) \myeq  \frac{\pwv_i}{max(1, \frac{\|\pwv_i\|}{s})}$~\cite{dldp}.

Then, Gaussian noise proportional to~$s$ is applied to the sum of the clipped values. Putting all together, the output of the differentially private attention mechanism is defined~as:
\iflong
\begin{equation}
	\widetilde{\gamma} \myeq \sum_{i=1}^{n} \left[ \texttt{C}_s (\pwv_i) \right] +\mathcal{N}(0, \mathbb{I}zs),
	\label{eq:dpatt}
\end{equation}
\else
$\widetilde{\gamma} \myeq \sum_{i=1}^{n} \left[ \texttt{C}_s (\pwv_i) \right] +\mathcal{N}(0, \mathbb{I}zs),$
\fi
where $z$ is the noise multiplier---a second hyper-parameter that controls the amount of applied noise, and $n$ is the number of value vectors. The resulting DP-AM is depicted in Figure~\ref{fig:attdp}.
\par

The attention-mechanism $\widetilde{\gamma}$ provides instance-level-DP for the value vectors. \textbf{As every value vector represents auxiliary information for a single user, instance-level-DP implies user-level privacy guarantees}. Here, the only requirement is that no user should be associated with more than a value vector; that is, no user should have multiple accounts on the same service.

Returning to \textit{Bob}, he can now download the private version of the \uncm~and obtain a private seeded model by just replacing $\gamma$ with $\widetilde{\gamma}$ in Algorithm~\ref{alg:mspm}. Hereafter, we refer to a differentially-private configuration seed as $\widetilde{\cs}$ and to a \mbox{DP-seeded} model as $\dec_{\Theta | \widetilde{\cs}}$. 

%

%
\begin{figure}[t!]
	\centering
	\resizebox{1\columnwidth}{!}{
	\begin{subfigure}{.45\columnwidth}
		\centering
		\includegraphics[trim = 0mm 3mm 0mm 0mm, clip, width=.95\columnwidth]{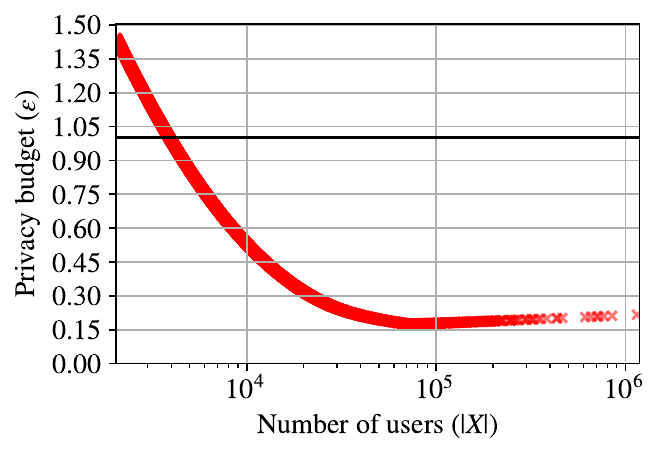}
		\caption{Privacy budget.}\label{fig:epsi}
	\end{subfigure}\begin{subfigure}{.45\columnwidth}
		\centering
		\includegraphics[trim = 0mm 3mm 0mm 0mm, clip, width=.95\columnwidth]{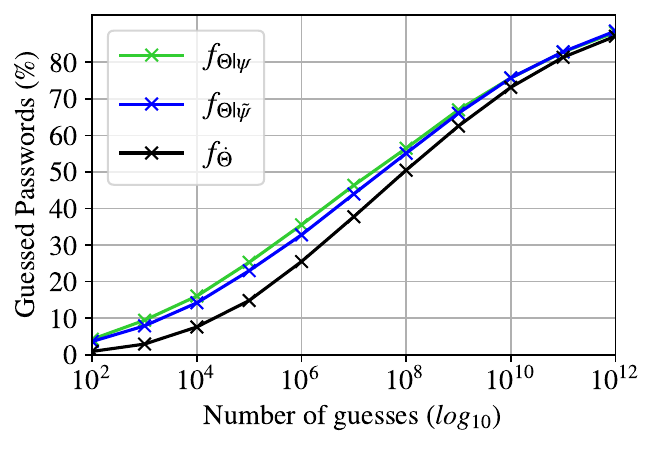}
		\caption{Average performance.}\label{fig:dpgp}
	\end{subfigure}
}	
	\caption{Panel (a) reports the value $\epsilon$ ($y$ axes) for every leak in $\lcv$ plotted against the leak size ($x$ axes). Panel~(b) is the average guessing performance of the seeded models, private seeded models and baseline on $100$ leaks from $\lcv$.}
	\label{fig:dpall}
\end{figure}
\subsection{Privacy budget and Utility loss}
\label{sec:dp_res}
Here, we quantify the privacy level as well as the utility loss induced by the differential private configuration encoder. In our setup, we seek to attain strong privacy guarantees for the users' auxiliary information \ie $\epsilon \leq 1$~\cite{Dwork_Smith_2010}. In this regard, we set the noise multiplier to $z \myeq 3$, while computing $\delta$ individually for each tested credential database~$X$ as $\delta \myeq \frac{10^{-2}}{|X|}$. We also evaluate stricter privacy guarantees by considering smaller~$\delta$~values: $\frac{10^{-3}}{|X|}$ and $\frac{10^{-4}}{|X|}$.

\par

\begin{figure}[t!]
	\centering
	\resizebox{.9\columnwidth}{!}{
	\begin{subfigure}{.45\columnwidth}
		\centering
		\includegraphics[trim = 0mm 3mm 0mm 0mm, clip, width=.95\columnwidth]{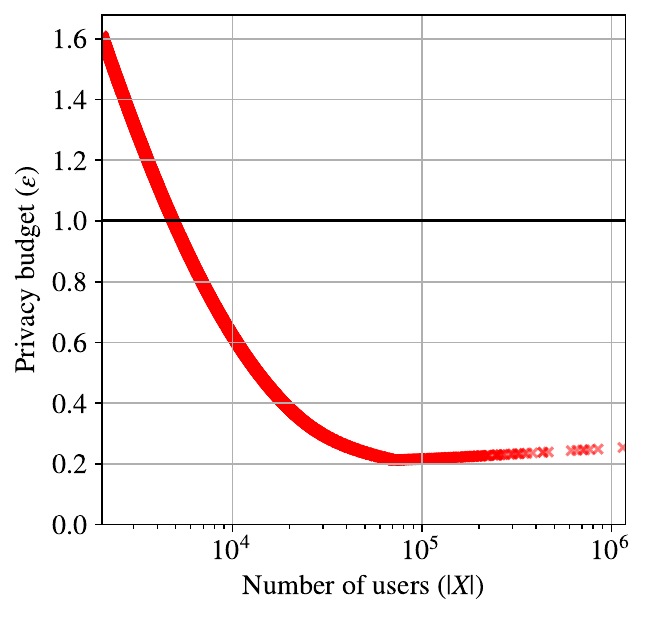}
		\caption{$\delta \myeq \frac{10^{-3}}{|X|}$}\label{fig:epsism0}
	\end{subfigure}\begin{subfigure}{.45\columnwidth}
		\centering
		\includegraphics[trim = 0mm 3mm 0mm 0mm, clip, width=.95\columnwidth]{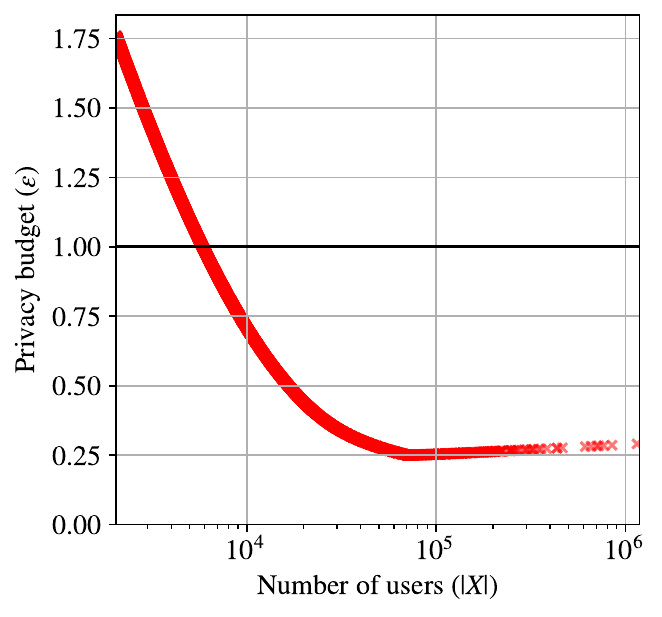}
		\caption{$\delta \myeq  \frac{10^{-4}}{|X|}$}\label{fig:epsism1}
	\end{subfigure}
}	
	\caption{Values of $\epsilon$ ($y$ axes) for every leak in~$\lcv$ plotted against the leak size ($x$ axes) given two different values of~$\delta$.}
	\label{fig:smallerdeltadp}
\end{figure}

\paragraph{\textbf{Privacy budget}}
Figure~\ref{fig:epsi} reports the value of~$\epsilon$ associated with the differential private configuration seed computed for every leak in~$\lcv$. Recall that the configuration seed $\widetilde{\cs}$ is computed only on a uniformly sampled subset of $\ainf$ of size at most~$k$. Thus, the $\epsilon$ associated with the configuration seed of a credential database $S$ is a function of the credential database size (\ie the number of users) due to \textit{privacy amplification by sub-sampling}~\cite{dpss0, dpss1}. Privacy amplification by subsampling tells us that we can scale our privacy budget~$\epsilon$ proportionally to the ratio between the size of the used input subset and the whole population $\ainf$:~$\frac{k}{|\ainf|}$. As shown in Figure~\ref{fig:epsi}, on average, {we achieve $\epsilon \myeq 0.76$}. In the worst-case scenario (\ie for the smallest leaks), we achieve  $\epsilon \myeq ~1.44$. \iflong While $1.44$ still provides meaningful privacy guarantees \iflong(see Appendix~\ref{app:mia})\fi, in these cases, the noise multiplier $z$ can be increased at inference time to push $\epsilon$ under $1$ (at the cost of additional utility loss). \fi Results for smaller values of $\delta$ are reported in Figure~\ref{fig:smallerdeltadp}.
\par

\paragraph{\textbf{Utility}}
Surprisingly, while achieving a remarkably small $\epsilon$ (on average), the utility loss induced by the DP-seeds and the sigmoid-based attention mechanism is limited. This is shown in Figure~\ref{fig:dpgp} where the average guessing performance of the non-private seeded password models (green) and the ones with differential private configuration seeds (blue) is reported. While demonstrating a slight performance degradation compared to the non-private model, the DP-seeded models still maintain a consistent advantage over the baseline observed for the non-private setup. 
\iflong 
We attribute the success of the private model (low $\epsilon$ value at low utility loss) to two main factors: (1) The privacy gain obtained through amplification via sub-sampling. (2) Informally, the configuration seed is only required to model a general description of the target password distribution; therefore, intuitively, the impact of the single user on the final seed is limited. That is, we can still model the core distributional properties of the target community even if information about individuals is destroyed.

In order to substantiate the validity of our differentially private attention mechanism and its implementation, we conducted an empirical evaluation utilizing Membership Inference Attacks (MIA), as recommended practice for the introduction of novel differential-privacy-based mechanisms~\cite{tramer2022debugging}. Results and details on the setup are given in Appendix~\ref{app:mia}.\fi

  In conclusion, we demonstrated that a seeded password model can be efficiently made private with a limited loss in utility. This property paves the way for future deployments of  \uncms~that exploit a broader spectrum of auxiliary input information without threatening users' privacy. 

\iflong
\section{Future work and \uncms\ extensions}
In this section, we provide a brief overview of some potential extensions of the \uncm\ framework, which could serve as a foundation for future investigations.
\subsection{Dynamic \uncms}
As described in Appendix~\ref{app:dynamic}, dynamic password guessing techniques such as those presented in~\cite{pasquini2021improving, pasquini2021usenix} may not be directly applicable to password strength metering. Nonetheless, these techniques offer a useful means of constructing more precise adversary models. Fortunately, integrating dynamic attacks into the \uncm\ framework can be achieved through a relatively straightforward process. The subsequent section provides a high-level overview of the implementation steps involved in creating a Dynamic \uncm.
\par

In essence, during an attack with a seeded password model, the intermediate set of guessed passwords $X_{\leq t}$ can serve as supplementary auxiliary information to generate an updated configuration seed at each time step $t$. For each password $x \in X_{\leq t}$, it is sufficient to concatenate a projected version of $x$ with the encoded email address of the associated user, similar to the approach illustrated for the user's \TT{name} in Figure~\ref{fig:model}. As the attack progresses, the configuration seed can be gradually refined by recomputing it at regular intervals using the accumulated guessed passwords up to that point. In a similar fashion, the seed can also be bootstrapped using prior knowledge of users' passwords, such as leveraging \textit{sister passwords} obtained from previously breached services. 
\par

Nonetheless, there are certain technical challenges that arise in deployment of dynamic autoregressive password models, such as adapting the enumeration (e.g., beam search) to operate on dynamic probability distributions in a efficient way. We defer this research avenue to future work.

\iflong
\subsection{Unbounded \uncms\ (\uncms\ with unlimited attention span)}
\begin{figure}[t]
	\centering
	\resizebox{.5\columnwidth}{!}{
		\begin{tikzpicture}		
		\tikzstyle{cell} = [circle, minimum width=.5cm, minimum height=1cm, text centered, draw=black, fill=green!10]
		\tikzstyle{vector} = [rectangle, minimum width=1cm, minimum height=.3cm,text centered, draw=black, fill=white, minimum width=4.5cm, minimum height=.4cm, rotate=90]
		\tikzstyle{arrow0} = [->,>=stealth]
		
		\tikzstyle{new} = [text=blue]
		
		\node (box) [rectangle, draw=black, dotted,  minimum width=2cm, minimum height=7cm, yshift=2.6cm, xshift=0cm, opacity=.5, pattern=north west lines, pattern color=gray, dotted]{};
		
		\node (cell00) [cell]{\small{$\text{LSTM}_{\text{c}_0}$}};%
		\node (cell20) [cell, above of=cell00, yshift=1cm]{\small{$\text{LSTM}_{\textit{c}_1}$}};%
			
		\node (cell10) [cell, above of=cell20, yshift=1cm]{\small{$\text{LSTM}_{\textit{c}_2}$}};%
		\node (t0) [below of=cell00, yshift=-.5cm]{};%

		\node (att) [above of=cell10, yshift=.5cm]{\Huge$\gamma$};%
		
		\node (d0) [above of=att, yshift=0cm]{};%

		\draw [arrow0] (cell00) -- (cell20) node[] {};
		\draw [arrow0] (cell20) -- (cell10) node[] {};
		\draw [arrow0] (t0) -- (cell00) node[] {};

		\draw [arrow0] (cell10) -- (att) node[] {};
		\draw [arrow0] (att) -- (d0) node[] {};

		\node (v0) [vector,  left of=cell00, yshift=2cm, xshift=2.5cm]{$\senc(\TT{johnsmith@mymail.us})$};
		\node (v1) [vector,  above of=v0]{$\senc(\TT{ferrari@posta.it})$};
		\node (v2) [vector,  above of=v1]{$\senc(\TT{jean.dupont@email.fr})$};
		\node (v3) [vector,  above of=v2, opacity=.5]{$\senc(\dots\dots\dots\dots\dots\dots\dots)$};
		\node (v4) [vector,  above of=v3, opacity=.25]{$\senc(\dots\dots\dots\dots\dots\dots\dots)$};
		\node (v5) [vector,  above of=v4, opacity=.1]{$\senc(\dots\dots\dots\dots\dots\dots\dots)$};

		\draw [arrow0] (v0.east) -- (att.west) node[] {};
		\draw [arrow0] (v1.east) -- (att.west) node[] {};
		\draw [arrow0] (v2.east) -- (att.west) node[] {};
		\draw [arrow0,  opacity=.5] (v3.east) -- (att.west) node[] {};
		\draw [arrow0, opacity=.25] (v4.east) -- (att.west) node[] {};
		\draw [arrow0, opacity=.1] (v5.east) -- (att.west) node[] {};

		\draw [->] (cell00) edge[loop right] (cell00);
		\draw [->] (cell10) edge[loop right] (cell00);
		\draw [->] (cell20) edge[loop right] (cell00);

		\end{tikzpicture}
	}
	\caption{Simplified depiction of a \textit{Unbounded} \uncm. Rectangles on the left represent the outputs of the sub encoder~$\senc$ for all the provided auxiliary information.}
	\label{fig:ununcm}
\end{figure}
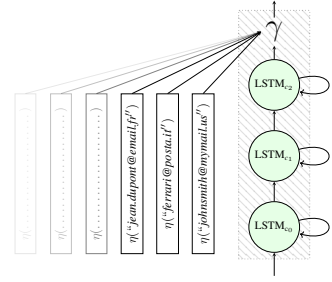
In the form presented in the paper, a \uncm\ condenses the available auxiliary information on the target set in a single, fixed-size, vector via the mixing encoder. This choice offers some fundamental advantages. Mainly, it allows the deployment of compact seeded password models and efficiently ensures differential privacy. However, this also limits the amount of information that can be extracted from the auxiliary data, and, ultimately, exploited, by the \uncm\ during guessing attack. Nonetheless, this is not a fundamental requirement of a \uncm\ and, at the cost of sacrificing performance and applicability, one can implement a \uncm\ with no auxiliary information bottleneck. Next, we briefly sketch this model which we dub \TT{Unbounded} \uncms.
\par

The idea is that, instead of training and using the mix. encoder to compress auxiliary information in a single vector, we allow the password model to access individual pieces of auxiliary information during the inference process. Figure~\ref{fig:ununcm} sketches this architecture. In particular, as in the standard setting (see Section~\ref{sec:imp}), we use the sub encoder $\senc$ to parse all the auxiliary information of users and project them in a set of equal-sized vectors. Then, a \TT{Bahdanau-like} attention mechanism~\cite{bahdanau2014neural} is used to dynamically bias the password model's behavior at inference time. More concretely, all the encoded auxiliary information are value vectors, while the query is dynamically generated as the output of the last layer of the LSTM. Upon every character generation step, the password model can attend to all the encoded pieces of auxiliary information and dynamically focus on / retrieve the required knowledge from them.}

We anticipate that enabling the conditional password model to attend to all auxiliary information and dynamically directing its attention based on the generated password will enhance the model's effectiveness in guessing attacks. However, this improvement would be paid in terms of efficiency, as inference would now have overhead linear in the number of auxiliary pieces of information. Moreover, the model would be too cumbersome to run in a browser, thereby limiting its applicability. Nonetheless, the model could prove useful for obtaining more accurate strength estimates in reactive studies.

\subsection{Different Password Models}
\new{In our study, we focused on using an LSTM network to implement the conditional password model of the \uncm. This choice was based on the model's proven effectiveness in estimating password strength and its familiarity within the password security community. However, it is worth noting that the password model architecture could potentially be replaced by any other neural network. For instance, one can rely on transformer-based or similar novel constructions~\cite{pasquini2021usenix}. All that is required is to attach the new password model to the configuration encoder at training time and define a suitable way to provide the configuration seed as additional input to~it.}
 \fi

\subsection{Additional auxiliary data modalities}
\label{app:seaddm}
Although the current implementation only utilizes email addresses, the sub-encoder can be extended to process various types of additional input data. In these cases, it is enough to introduce additional sub-models suited to handle the new auxiliary data modalities. In Figure~\ref{fig:model}, we offer the example of the name of the user (the second replica of the sub-encoder). Similarly to the username of the email address, this string can be parsed by introducing an additional RNN in the sub-encoder.

The only requirement here is that all the information associated with a user should be mapped to a single vector. That is, the output of any new sub-model should not be used as direct input for the mixing-encoder, but, as shown in the figure, it should be combined/merged (\eg via element-wise addition) with the output of the email-encoder for the associated email address. This requirement offers two main advantages: (1) it allows the mixing-encoder to model the (possible) semantic relation between the different sources of input data provided and (2) it drastically simplifies achieving differential privacy for the configuration seed as we discuss in Section~\ref{sec:dp}. 
\fi

\section{Ethical considerations}
\iflong

\new{In our study, we utilized a collection of previously leaked credentials to develop and validate the effectiveness of our password model. The use of this data raises ethical concerns. However, the information we used is already publicly available online, and the compromised accounts have already been added to breach alert systems~\cite{cit0dayart1}.}

\new{For the model training and evaluation, we only considered passwords that appeared in plaintext in the original leak and did not attempt to recover additional credentials stored as hashes. In the same direction, we did not augment the data in \citday\ by combining it with other public sources of information or preprocessing it in a way that would provide additional knowledge to an adversary. Therefore, our analysis does not exacerbate or extend the harm already caused to users by the leakage.}

\new{Although adversaries could use our techniques to improve attacks performance,  the introduced models enable security practitioners to automatically generate accurate password models for their systems, regardless of their lack of expertise and/or resources. By doing so, our approach will contribute to stronger password security practices and ultimately help protect users from cyber threats. Given that using leaks is the only way of carrying out this research, we believe that this benefit outweighs the possible risk introduced by our study.}
\else
We used leaked credentials to develop and validate our password model, which raises ethical concerns. However, the information we used is already publicly available online, and the compromised accounts used in the study already appear in breach alert systems~\cite{cit0dayart1}. 
For the model training and evaluation, we only considered passwords that appeared in plaintext in the original leak and did not attempt to recover additional credentials stored as hashes. In the same direction, we did not augment the data in \citday\ by combining it with other public sources of information. Therefore, our analysis does not exacerbate or extend the harm already caused to users by the leakage.

While adversaries could use our techniques to improve attacks, our approach also enables security practitioners to generate accurate password models for their systems, regardless of expertise or resources limitations. Thus, our work contributes to stronger password security practices and ultimately help protect users from cyber threats. We believe the benefits of this research outweigh the possible risks.
\fi

\section{Conclusion}
We present the first self-configurable password model that uses auxiliary data to adapt to the target password distribution at inference time. This addresses a major problem in the application of password security techniques in the real world. We make our pre-trained models public, making accurate password models more widely accessible.

Our framework is general and can be extended to benefit from future improvements in password models and deep learning techniques, as well as data availability, without compromising user privacy (see Section~\ref{sec:puncm}). Our work also demonstrates the potential of using auxiliary data to improve the performance of password guessing models in the classic trawling offline setting, paving the way for further research in this field.

\iflong
\section*{\textbf{Acknowledgements}}
We thank Marco Cianfriglia for the support during the initial part of the project and the anonymous reviewers for their valuable feedback and guidance. 
\fi

\bibliographystyle{plain}
\bibliography{bib}

\appendices

\iflong
\else
\begin{table*}[ht!]
	\centering
	\caption{Number of leaks for the top-$25$ top-level domains appearing in \textit{Cit0day.in} before and after pre-processing.}
	\label{table:cit}
	\resizebox{1\textwidth}{!}{%
		\begin{tabular}{c|ccccccccccccccccccccccccc}

 \toprule
    
    \textbf{Raw} & \textit{.com} & \textit{.ru} & \textit{.net} & \textit{.org} & \textit{.de} & \textit{.kr} & \textit{.br} & \textit{.pl} & \textit{.uk} & \textit{.it} & \textit{.fr} & \textit{.cz} & \textit{.tw} & \textit{.in} & \textit{.hu} & \textit{.jp} & \textit{.info} & \textit{.ca} & \textit{.th} & \textit{.eu} & \textit{.nl} & \textit{.ua} & \textit{.au} & \textit{.es} & \textit{.ar}\\
    \# leaks & $9539$ & $1208$ & $1167$ & $982$ & $874$ & $798$ & $609$ & $551$ & $533$ & $467$ & $438$ & $400$ & $330$ & $322$ & $305$ & $226$ & $223$ & $193$ & $191$ & $175$ & $172$ & $163$ & $142$ & $139$ & $134$ \\
    \midrule
    \textbf{Clean} & \textit{.com} & \textit{.ru} & \textit{.net} & \textit{.org} & \textit{.de} & \textit{.kr} & \textit{.pl} & \textit{.br} & \textit{.uk} & \textit{.it} & \textit{.cz} & \textit{.fr} & \textit{.hu} & \textit{.tw} & \textit{.in} & \textit{.jp} & \textit{.fj} & \textit{.info} & \textit{.ca} & \textit{.eu} & \textit{.ua} & \textit{.ar} & \textit{.au} & \textit{.nl} & \textit{.ro}\\
    \# leaks & $4741$ & $567$ & $560$ & $464$ & $426$ & $403$ & $389$ & $299$ & $286$ & $238$ & $210$ & $200$ & $192$ & $175$ & $172$ & $112$ & $102$ & $102$ & $99$ & $91$ & $79$ & $79$ & $72$ & $72$ & $71$ \\
    \bottomrule

\end{tabular}
	}
\end{table*}
\fi

\iflong
\else
\section{Handling additional data modalities}
\label{app:seaddm}
Although the current implementation only utilizes email addresses, the sub-encoder can be extended to process various types of additional input data. In these cases, it is enough to introduce additional sub-models suited to handle the new auxiliary data modalities. In Figure~\ref{fig:model}, we offer the example of the name of the user (the second replica of the sub-encoder). Similarly to the username of the email address, this string can be parsed by introducing an additional RNN in the sub-encoder.

The only requirement here is that all the information associated with a user should be mapped to a single vector. That is, the output of any new sub-model should not be used as direct input for the mixing-encoder, but, as shown in the figure, it should be combined/merged (\eg via element-wise addition) with the output of the email-encoder for the associated email address. This requirement offers two main advantages: (1) it allows the mixing-encoder to model the (possible) semantic relation between the different sources of input data provided and (2) it drastically simplifies achieving differential privacy for the configuration seed as we discuss in Section~\ref{sec:dp}. 
\fi

\iflong
\else
\begin{figure}
	\centering
	\resizebox{.25\textwidth}{!}{
		\begin{tikzpicture}		
			
			\tikzstyle{chosen} = []
			\tikzstyle{other} = [rectangle, draw=red]
			\tikzstyle{arrow} = [->,>=stealth]
			
			\node (t0) [chosen] {$q_i$};
			\node (t1) [other, right of=t0] {$v_1$};
			
			\node (fk0) [above of=t0] {$g_Q$};
			\node (fq0) [above of=t1] {$g_K$};
			
			\draw [arrow] (t0) -- (fk0);
			\draw [arrow] (t1) -- (fq0);
			
			\node (d0) [above of=fk0, xshift=.5cm] {\Large{$\odot$}};
			
			\draw [arrow] (fk0) -- (d0);
			\draw [arrow] (fq0) -- (d0);
			
			\node (t2) [chosen, right of=t0, xshift=1.5cm] {$q_i$};
			\node (t3) [other, right of=t2] {$v_2$};
			
			\node (fk1) [above of=t2] {$g_Q$};
			\node (fq1) [above of=t3] {$g_K$};
			
			\draw [arrow] (t2) -- (fk1);
			\draw [arrow] (t3) -- (fq1);
			
			\node (d1) [above of=fk1, xshift=.5cm] {\Large{$\odot$}};
			
			\draw [arrow] (fk1) -- (d1);
			\draw [arrow] (fq1) -- (d1);
			
			\node (t4) [chosen, right of=t2, xshift=1.5cm] {$q_i$};
			\node (t5) [right of=t4, other] {$v_3$};
			
			\node (fk2) [above of=t4] {$g_Q$};
			\node (fq2) [above of=t5] {$g_K$};
			
			\draw [arrow] (t4) -- (fk2);
			\draw [arrow] (t5) -- (fq2);
			
			\node (d2) [above of=fk2, xshift=.5cm] {\Large{$\odot$}};
			
			\draw [arrow] (fk2) -- (d2);
			\draw [arrow] (fq2) -- (d2);
			
			\node (sf) [above of=d1, yshift=-.1cm] {\textit{sigmoid}};
			
			\draw [arrow] (d0) -- (sf);
			\draw [arrow] (d1) -- (sf);
			\draw [arrow] (d2) -- (sf);
			
			\node (w0) [above of=sf, yshift=-.4cm] {$\cdot w_1$};
			\node (w1) [above of=w0, yshift=-.5cm] {$\cdot  w_2$};
			\node (w2) [above of=w1, yshift=-.5cm] {$\cdot  w_3$};
			\draw [arrow] (sf) -- (w0);
			
			\node (v0) [other, left of=w0, xshift=-2cm] {$v_1$};
			\node (v1) [other, above of=v0, yshift=-.5cm] {$v_2$};
			\node (v2) [other, above of=v1, yshift=-.5cm] {$v_3$};
		
			\node (fv0) [right of=v0, xshift=.5cm] {$g_V$};
			\node (fv1) [right of=v1, xshift=.5cm] {$g_V$};
			\node (fv2) [right of=v2, xshift=.5cm] {$g_V$};
			
			\draw [arrow] (v0) -- (fv0);
			\draw [arrow] (v1) -- (fv1);
			\draw [arrow] (v2) -- (fv2);
			
			\draw [arrow] (fv0) -- (w0);
			\draw [arrow] (fv1) -- (w1);
			\draw [arrow] (fv2) -- (w2);
			
			\node (fv0) [right of=v0, xshift=.5cm] {$g_V$};
			\node (fv1) [right of=v1, xshift=.5cm] {$g_V$};
			\node (fv2) [right of=v2, xshift=.5cm] {$g_V$};
			
			\node (c0) [right of=w0, xshift=.1cm] {$\texttt{C}_s$};
			\node (c1) [right of=w1, xshift=.1cm] {$\texttt{C}_s$};
			\node (c2) [right of=w2, xshift=.1cm]  {$\texttt{C}_s$};
			
			\draw [arrow] (w0) -- (c0);
			\draw [arrow] (w1) -- (c1);
			\draw [arrow] (w2) -- (c2);

			\node (agg) [right of=c1] {\large{$\Sigma$}};
			
			\draw [arrow] (c0) -- (agg);
			\draw [arrow] (c1) -- (agg);
			\draw [arrow] (c2) -- (agg);
			
			\node (noise) [right of=agg, xshift=0.4cm] {$+\mathcal{N}(0, \mathbb{I}zs)$};
			\draw [arrow] (agg) -- (noise);
		
			\node (eq) [above of=noise, xshift=0cm, yshift=-.5cm, rotate=90] {$=$};
			\node (res) [above of=noise] {$\widetilde{\gamma}(q_i,\ \{v_1, v_2, v_3\})$};
		
		\end{tikzpicture}
	}
	\caption{Depiction of the introduced differentially-private attention-mechanism (DP-AM). In the computational graph, private inputs are highlighted in red.}
	\label{fig:attdp}
\end{figure}
\fi

\iflong
\else 
\section{Additional Background on Password Models}
\label{app:addbkpsswwdml}
Existing password models construct over a heterogeneous set of assumptions and implementations. 

{Probabilistic password models} assume that real-world password distributions are sufficiently smooth to be accurately described by a suitable parametric probabilistic model~$\dec_\Theta$. Here, a password mass function is explicitly \cite{backoffmc, melichera, MM} or implicitly \cite{pasquini2021improving} derived from a set of observable data (\ie previously leaked passwords), typically, via maximum likelihood estimation~\cite{melichera, MM}. This is then used to assign a probability to each element of the key space.  During the guessing attack, guesses are produced by traversing the key space following the decreasing probability order imposed by~$\dec_\Theta$.

 Other hybrid approaches such as Probabilistic Context-Free Grammars (PCFG) and dictionary attacks~\cite{unix, 190978, reasoning, pasquini2021usenix}, instead, rely on simpler and more intuitive constructions, which tend to be closer to human logic. Generally, those assume passwords as realizations of templates and generate novel guesses by abstracting and applying such patterns on token collections. Those tokens are either directly given as part of the model configuration (\eg the dictionary and rules-set for dictionary attacks) or extracted from observed passwords in a training phase (\eg terminals/grammar for PCFG). In contrast with parametric models, these can produce only a limited number of guesses, which is a function of the chosen configuration.
\subsection{Autoregressive password models}

\fi
\section{Comparison with other password models}
\label{app:pcfg}

\begin{figure}
	\centering	
	\includegraphics[trim = 0mm 3mm 0mm 0mm, clip, width=1.05\columnwidth]{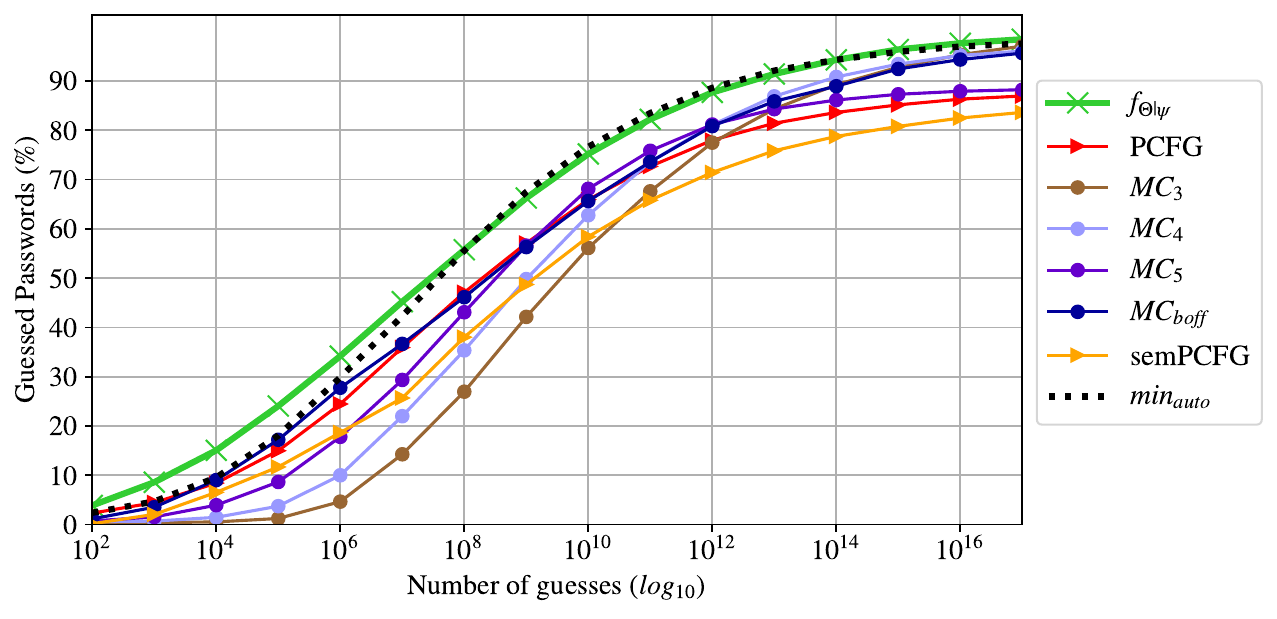}

	\caption{Guessing performance comparison  between the generated seeded password models (green) and $6$ other password models trained on~$\lct$ and their \textit{min-auto} setup~\cite{190978} (black dashed line). Results are averaged over $100$ leaks sampled from~$\lcv$ ($\seeded$ is the seeded model generated by the \uncm).}
	\label{fig:pcfg}
\end{figure}
For completeness, we report a direct comparison of guessing performance of the seeded password models produced by the \uncm~with other password models.  As for the baseline of Section~\ref{sec:baseline}, we train these password models using the union of the password leaks in $\lct$, while using the Monte Carlo method~\cite{monte} to estimate guess numbers. In particular, we compare with: 
\begin{itemize}
	\item \textbf{PCFG~\cite{pcfg}:} We use the model with the default setting available in the implementation.\footnote{\url{https://github.com/lakiw/pcfg_cracker}}
	\item \textbf{Markov Chains~\cite{MM, MM2}:} We evaluate a total of $4$ model variations. We train and test: a Markov chain of order $3$ ($MC_3$), a Markov chain of order $4$ ($MC_4$), a Markov chain of order $5$ ($MC_5$), and a Markov chain with backoff~\cite{MM2} ($MC_{boff}$). All the hyper-parameters are set to their default values as provided with the implementation~\cite{monte}.\footnote{\url{https://github.com/matteodellamico/montecarlopwd}}
	\item \textbf{semantic-PCFG~\cite{Veras2014}:} We use the model with the default setting available in the implementation.\footnote{\url{https://github.com/vialab/semantic-guesser}} Hereafter, we refer to this password model as~$\texttt{semPCFG}$.
	\item \textbf{min-auto} We additionally report results for the ideal \textit{min-auto} metric~\cite{190978}. As defined by Us~\etal~\cite{190978}, for every password~$x$, \textit{min-auto} is the minimum guess number across the pool of password models listed above: 
	\begin{multline}
		\footnotesize 
		\texttt{min}_{\texttt{auto}}(x)\myeq \min[\texttt{PCFG}(x), \texttt{MC}_3(x), \texttt{MC}_4(x),\\ \texttt{MC}_5(x),  \texttt{MC}_{bf}(x), \texttt{semPCFG}(x)],
	\end{multline}

	where the model evocation returns the estimated guess number for a password~$x$. Intuitively, this is the ideal/optimal combination of the tested models which provide a conservative estimation of password strength~\cite{190978}.
\end{itemize}

\paragraph{\textbf{Results}}
 Figure~\ref{fig:pcfg} reports the average guessing performance of the models over $100$ leaks sampled from $\lcv$. We additionally report a close-up of their performance on $5$ individual leaks in Figure~\ref{fig:ex_guess_global_adams}. Consistently with the findings outlined in Section~\ref{sec:result_un}, the seeded password models generated by the \uncm~perform better than the other approaches. Most interestingly,  the seeded models also achieves better performance than the ideal \text{min-auto} configuration (\ie the combination of all the other models) on the initial guesses, while performing overall \text{on par} after the $10^{8}$ mark. As discussed in Section~\ref{sec:result_mc}, the better performance on the initial guesses should be attributed to the capability of the~\uncm~of guessing weak passwords that are peculiar for the target distribution (\eg language-specific passwords).


\section{\uncm\ vs. Manual configuration: Evaluation on other semantic categories}
\label{app:categ}
Here, we extend the analysis of Section~\ref{sec:result_mc} by including results on categorization based on users' account types. Indeed, it is well-established that the type of service for which a password is created can impact users' password choices~\cite{flor}. In our analysis, we use the website category as an identifier, which refers to the type of service provided by the web application, such as shopping, education, entertainment, etc. Hereafter, we refer to it as \TT{website category}.

\paragraph{\textbf{Setup}}
The password leaks in the \citday\ collection have been already labeled according website category by the original data curator. Although the exact method used to derive these labels is unknown\iflong\footnote{For instance, it could have been obtained through an automatic tool or manually collected by the data curator.}\fi, we have verified their accuracy through manual inspection of a batch of randomly chosen examples.

In the experiments, we consider the top~$6$ categories with the most passwords in \citday\ (across all the languages); those are: \TT{Shopping} ($Sh$), \TT{IT} ($IT$), \TT{Entertainment} ($En$), \TT{Education} ($Ed$), \TT{Sports} ($Sp$), and \TT{Travel} ($Tr$).\footnote{We have excluded the category \TT{Business} from the plot as it is overly broad. Nonetheless, we evaluated this setting and the results are consistent with the other configurations.} Then, we follow the procedure described in Section~\ref{sec:result_mc}: we train a separate password model~\cite{melichera} for each training set that contains leaked passwords from a specific category (\eg only passwords from shopping websites), and test these models on a validation set with passwords from the same category.

\begin{figure*}[t!]
\resizebox{.97\textwidth}{!}{
	\centering
	\begin{subfigure}{.18\textwidth}
		\centering
		\includegraphics[scale=.3]{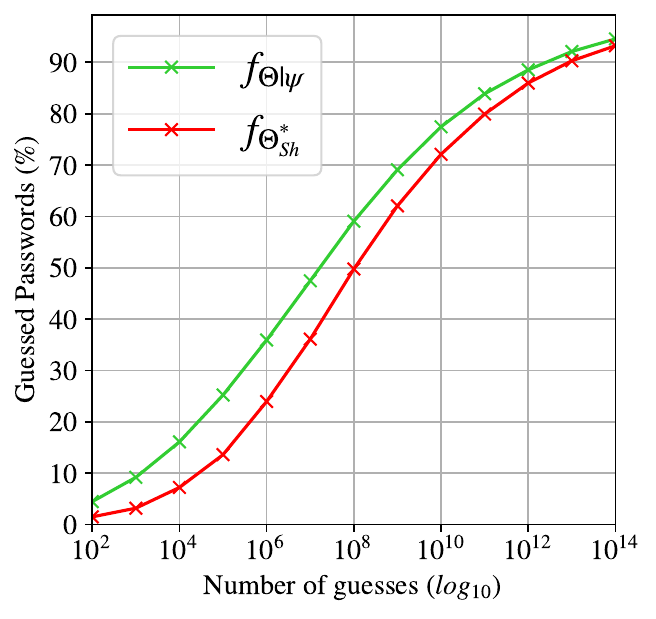}
		\caption{\scriptsize Shopping \\  \scalebox{.7}{$|\mathbf{X}^{*}_{Sh}|\myeq 8,305,366$}}
	\end{subfigure}~\begin{subfigure}{.18\textwidth}
		\centering
		\includegraphics[scale=.3]{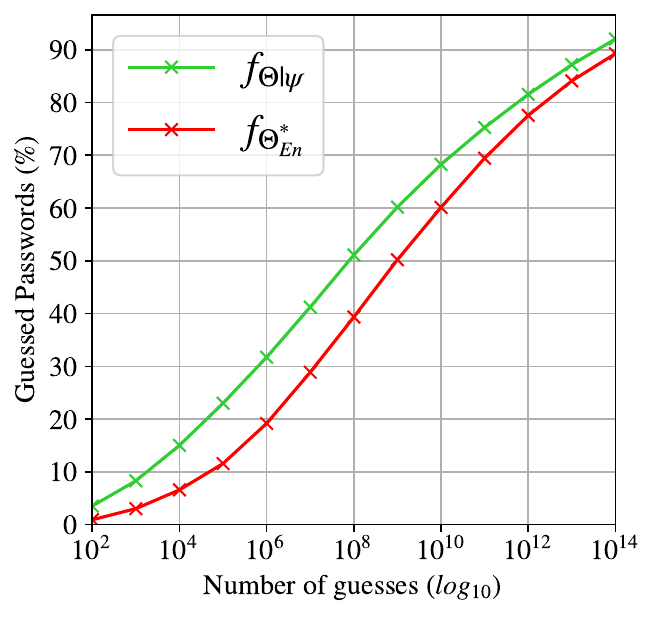}
		\caption{\scriptsize Entertain. \\ \scalebox{.7}{$|\mathbf{X}^{*}_{En}|\myeq 6,915,205$} }
	\end{subfigure}~\begin{subfigure}{.18\textwidth}
		\centering
		\includegraphics[scale=.3]{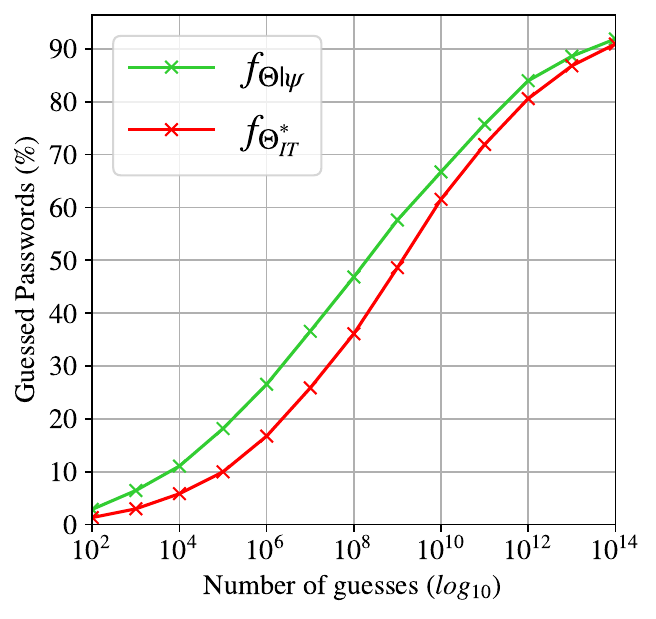}
		\caption{\scriptsize IT. \\ \scalebox{.7}{$|\mathbf{X}^{*}_{IT}|\myeq 6,564,410$} }
	\end{subfigure}~\begin{subfigure}{.18\textwidth}
		\centering
		\includegraphics[scale=.3]{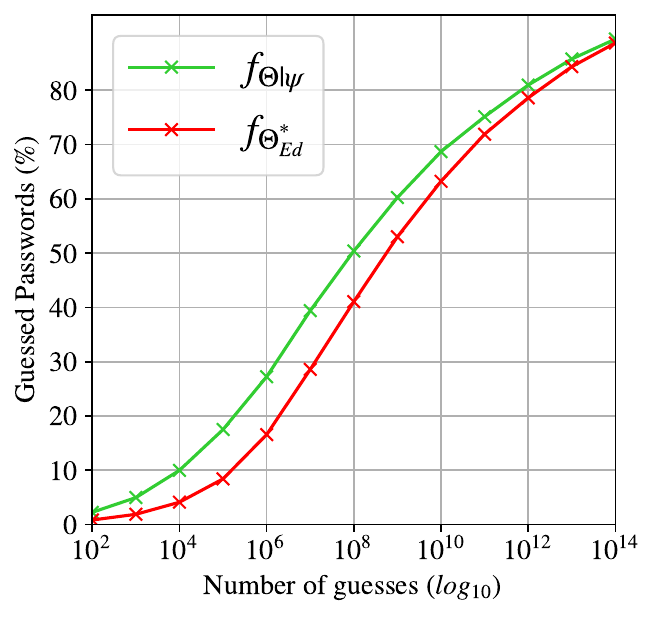}
		\caption{\scriptsize  Education \\ \scalebox{.7}{$|\mathbf{X}^{*}_{Ed}|\myeq 6,168,965$}}
	\end{subfigure}~\begin{subfigure}{.18\textwidth}
		\centering
		\includegraphics[scale=.3]{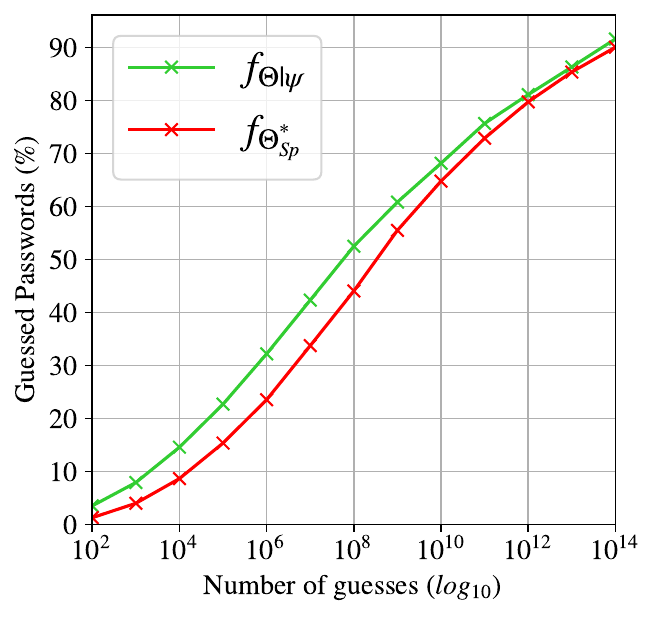}
		\caption{\scriptsize Sports \\ \scalebox{.7}{$|\mathbf{X}^{*}_{Sp}|\myeq 3,505,000$}}
	\end{subfigure}~\begin{subfigure}{.18\textwidth}
		\centering
		\includegraphics[scale=.3]{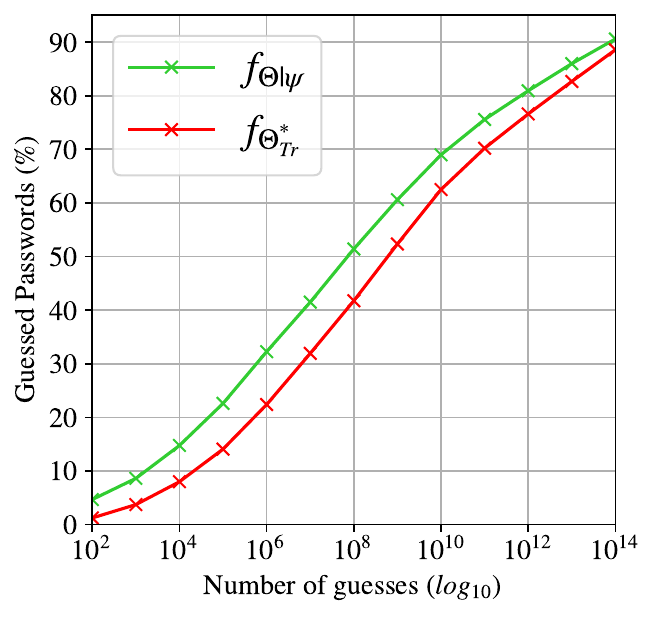}
		\caption{\scriptsize  Travel   \\ \scalebox{.7}{$|\mathbf{X}^{*}_{Tr}|\myeq 2,542,684$}}
	\end{subfigure}}\\ \resizebox{.97\textwidth}{!}{	\centering \begin{subfigure}{.18\textwidth}
		\centering
		\includegraphics[scale=.3]{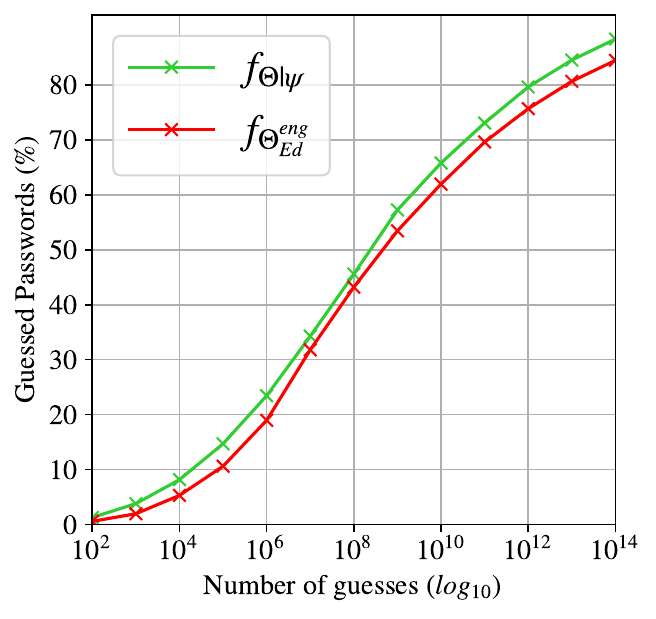}
		\caption{\scriptsize  Eng. Edu. \\ \scalebox{.7}{$|\mathbf{X}^{eng}_{Ed}|\myeq 2,068,472$}}
	\end{subfigure}~\begin{subfigure}{.18\textwidth}
		\centering
		\includegraphics[scale=.3]{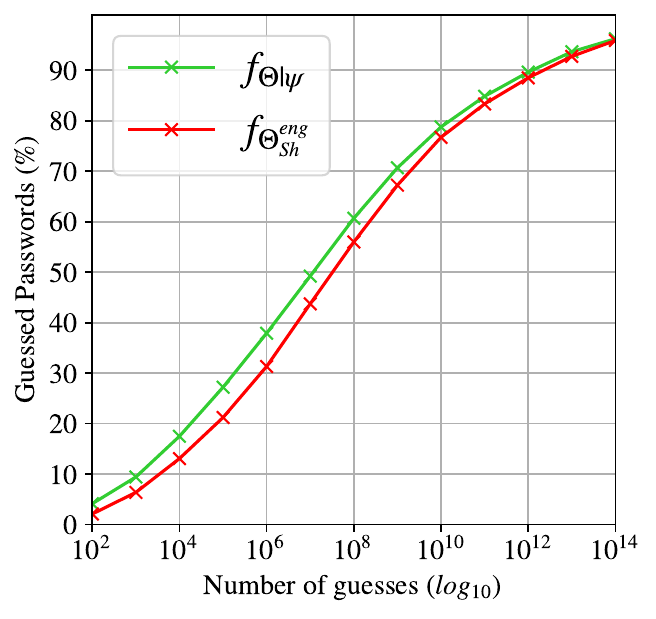}
		\caption{\scriptsize Eng. Shop. \\ \scalebox{.65}{$|\mathbf{X}^{eng}_{Sh}|\myeq 1,961,553$}}
	\end{subfigure}~\begin{subfigure}{.18\textwidth}
		\centering
		\includegraphics[scale=.3]{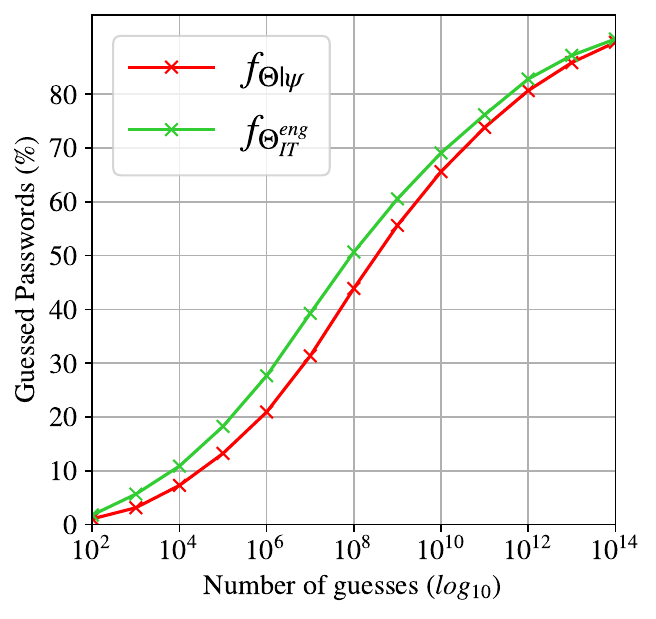}
		\caption{\scriptsize Eng. IT \\ \scalebox{.7}{$|\mathbf{X}^{eng}_{IT}|\myeq 1,855,932$}}
	\end{subfigure}~\begin{subfigure}{.18\textwidth}
		\centering
		\includegraphics[scale=.3]{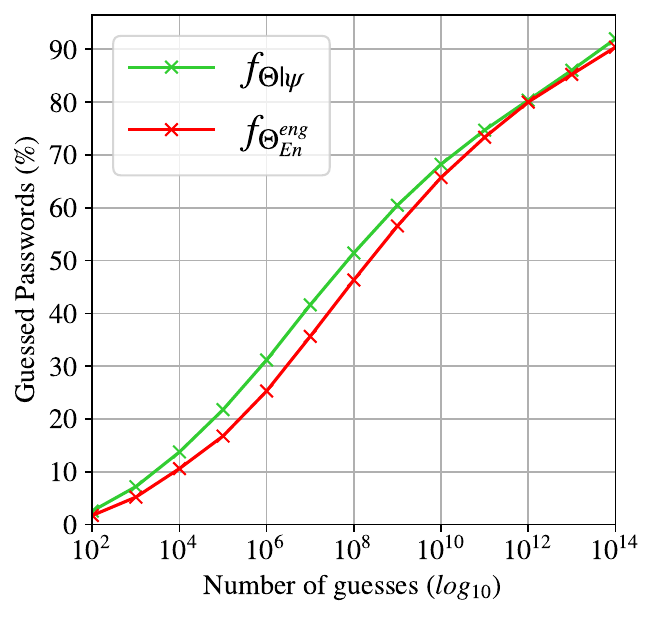}
		\caption{\scriptsize Eng. Ent. \\ \scalebox{.7}{$|\mathbf{X}^{eng}_{En}|\myeq 1,478,437$}}
	\end{subfigure}~\begin{subfigure}{.18\textwidth}
		\centering
		\includegraphics[scale=.3]{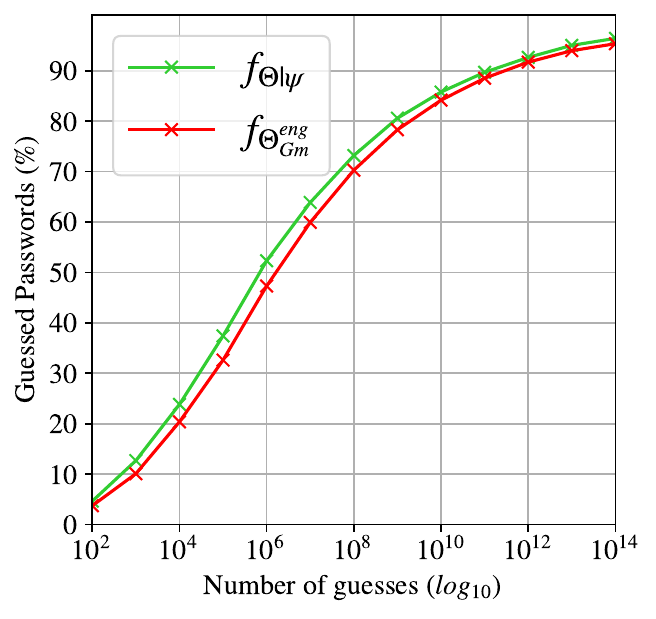}
		\caption{\scriptsize Eng. Games \\ \scalebox{.62}{$|\mathbf{X}^{eng}_{Gm}|\myeq 1,721,813$}}
	\end{subfigure}~\begin{subfigure}{.18\textwidth}
		\centering
		\includegraphics[scale=.3]{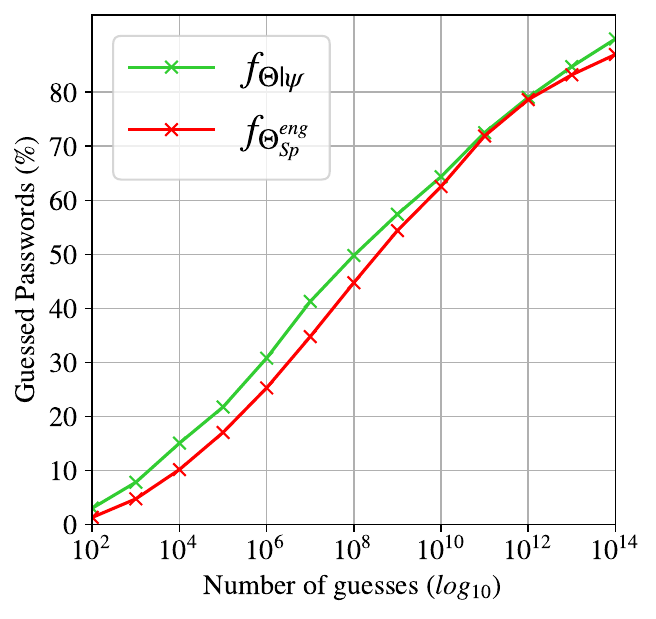}
		\caption{\scriptsize Eng. Sports \\ \scalebox{.62}{$|\mathbf{X}^{eng}_{Sp}|\myeq 1,155,522$}}
	\end{subfigure}}
\caption{Average guessing performance of seeded password models and manually tailored password models evaluated on $6$ different semantic subsets of leaks (grouped by website categories). The caption of each plot reports the size of the training set used to fit the manually tailored models ($\seeded$ is the seeded model and $\dec_{\Theta^\mathtt{lan}_{cat}}$ indicates the tailored model).}
\label{fig:categories}
\end{figure*}

\paragraph{\textbf{Results}}
Results are reported in the first row of Figure~\ref{fig:categories}. The seeded models exhibit superior performance compared to the manually tailored models, with a margin that surpasses that seen in language-specific distributions (Figure~\ref{fig:uncm_vs_manual}). This can be attributed to the fact that, as expected, the community language has a greater influence on shaping the password distribution than the website category.

To eliminate the influence of language on testing the \uncm's adaptability to service categories, we conduct the experiments again by fixing the language. Specifically, we filter the leaks by categories only in the English pool.\iflong\footnote{We used English leaks as those account for the highest number of passwords in total.}\fi\ Also in this case, we considered the six categories with the highest number of passwords. In the English subset, these categories are the same as in the previous case, except for \TT{Travel}, which was replaced with \TT{Games} ($Gm$). The results of these experiments are presented in the second row of Figure~\ref{fig:categories}.

As per our hypothesis, the community language played a major role in the previous results, and, in this language-fixed  setting, the difference in performance between the manually tailored models and the \uncm\ reduces. Nonetheless, the \uncm's generated models still outperform the tailored ones, supporting the argument made in Section~\ref{sec:result_mc}. However, it is important to note that the limited availability of training data for tailored models in such cases could be a significant factor that restricts their performance.

\iflong
\begin{figure*}[t]
	\centering
	\includegraphics[trim = 0mm 3mm 0mm 0mm, clip, width=1\textwidth]{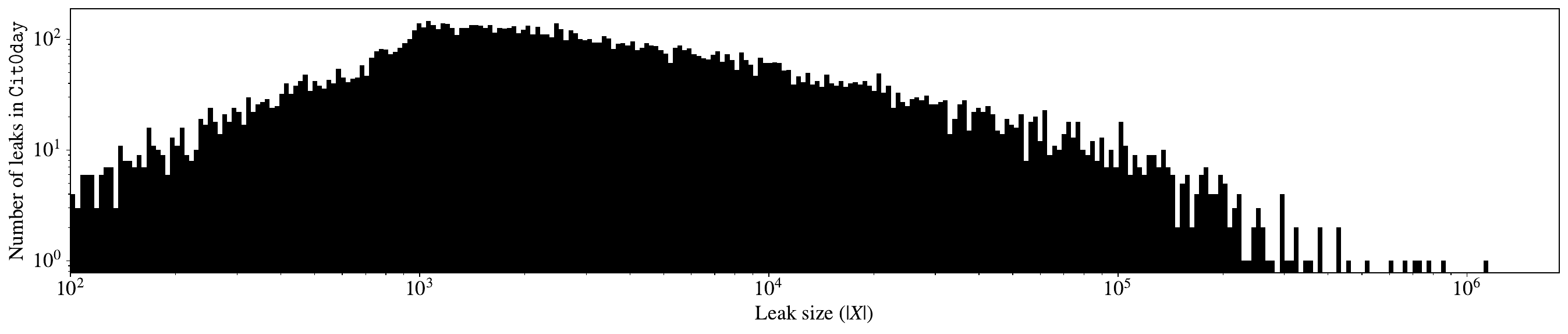}

	\caption{Distribution of the cardinality (number of users) of the leaks in \citday.}
	\label{fig:leakdis}
\end{figure*}
\fi

\section{Hyper-parameters and evaluation setting}
\label{app:params}
Follow the list of hyper-parameters used for training and evaluation of the tested models: 
\begin{itemize}
\footnotesize{
	\item \textbf{Number LSTM cells ($\dec$)}: 1000
	\item \textbf{Char. embedding size ($\dec$)}: 512
	\item \textbf{(Virtual) batch size (\uncm)}: 16
	\item \textbf{AM's projection width ($\gamma$)}: 756
	\item \textbf{Emb. size provider ($\senc$)}: 256
	\item \textbf{Emb. size domain ($\senc$)}: 256
	\item \textbf{Output size username GRU ($\senc$)}: 256
	\item \textbf{Char. embedding size username GRU ($\senc$)}: 16
	\item \textbf{Batch size $\uni$}: 64
	\item \textbf{Patience early stopping}: 5
	\item \textbf{Size $\Theta$ Monte Carlo guess number~\cite{monte}}: 100,000
}
\end{itemize}
Values of $\epsilon$ reported in Section~\ref{sec:dp_res} are computed using the implementation of~\cite{dldp} shipped with \textit{TensorFlow Privacy}.

\iflong
\subsection{Setting toy-example Figure~\ref{fig:toy_example}}
\label{app:toyparams}
To produce Figure~\ref{fig:toy_example}, we use the same general procedure described in Section~\ref{sec:result_mc}. That is, we train the model of Melicher~et al.\cite{melichera} on the leaks with top-domain \TT{.pl},  \TT{.hu}, and  \TT{.pl$||$.hu} from~$\lct$. The attack sets are respectively  \TT{985ad8d0.pl} and \TT{e1e222a8.hu} from~$\lcv$. 
\fi 

\subsection{Filtering english passwords}
\label{app:filter_eng_psswd}
{To identify password leaks of English speaking communities, we used a simple, yet, effective, heuristic. We started by considering all the leaks in \citday\ with top domains: \{\TT{us}, \TT{uk}, \TT{au}, \TT{net}, \TT{org}, \TT{com}\}. Then, we removed all the leaks that had more than $2\%$ of users’ email addresses with a top domain different from \{\TT{us}, \TT{uk}, \TT{au}, \TT{net}, \TT{org}, \TT{com}\}. The threshold \TT{$2\%$} is a parameter that we derived from a manually annotated subset of 50 examples. To validate the quality of the filtering process, we then, manually checked a random subsample of $50$ leaks. All the inspected web applications associated to the leaks but one were targeted to an English-speaking user community.}

We applied this process on both the training ($\lct$) and validation ($\lcv$) leak collection, obtaining sub-collections of $1776$ and $359$ leaks, respectively.

\section{On Dynamic \uncms}
\begin{figure*}[t!]
	\centering
	\captionsetup{justification=centering}

	\includegraphics[trim = 0mm 115mm 0mm 0mm, clip, width=.9\textwidth]{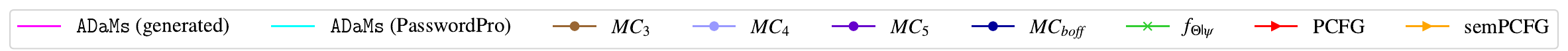}
	
\resizebox{1\textwidth}{!}{
	
	\begin{subfigure}{.4\columnwidth}
		\centering
		\includegraphics[trim = 0mm 0mm 0mm 0mm, clip, width=.95\columnwidth]{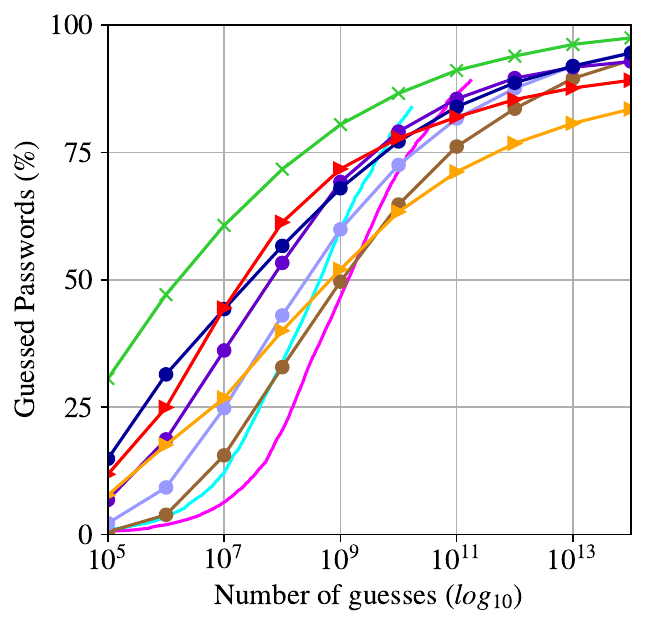}
		\caption{\footnotesize \textit{edj.pl}}
		\end{subfigure}
	\begin{subfigure}{.40\columnwidth}
		\centering
		\includegraphics[trim = 0mm 0mm 0mm 0mm, clip, width=.95\columnwidth]{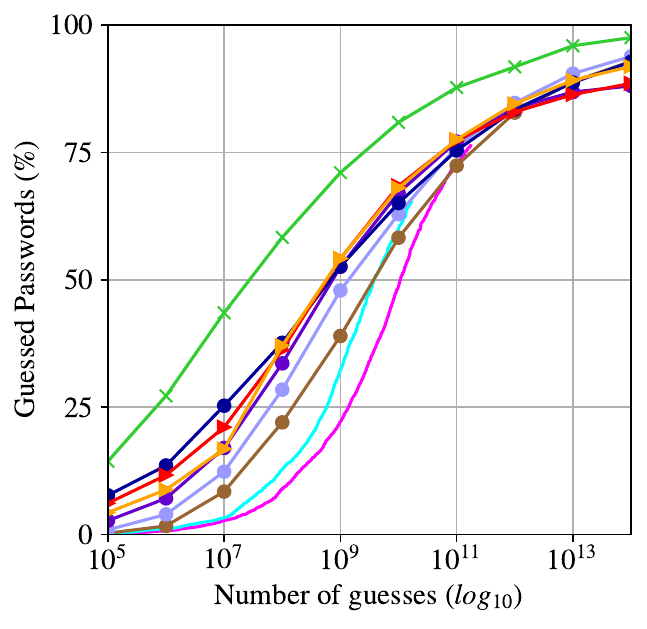}
		\caption{\footnotesize \textit{atcenter.or.jp} }
			\end{subfigure}\begin{subfigure}{.42\columnwidth}
		\centering
		\includegraphics[trim = 0mm 0mm 0mm 0mm, clip, width=.95\columnwidth]{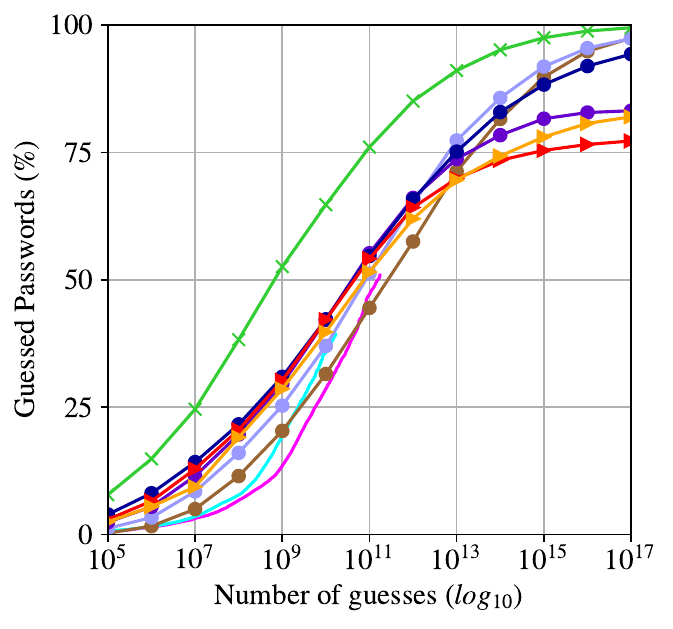}
		\caption{\footnotesize \textit{fvchinese.com}} 
	\end{subfigure}\begin{subfigure}{.4\columnwidth}
		\centering
		\includegraphics[trim = 0mm 0mm 0mm 0mm, clip, width=.95\columnwidth]{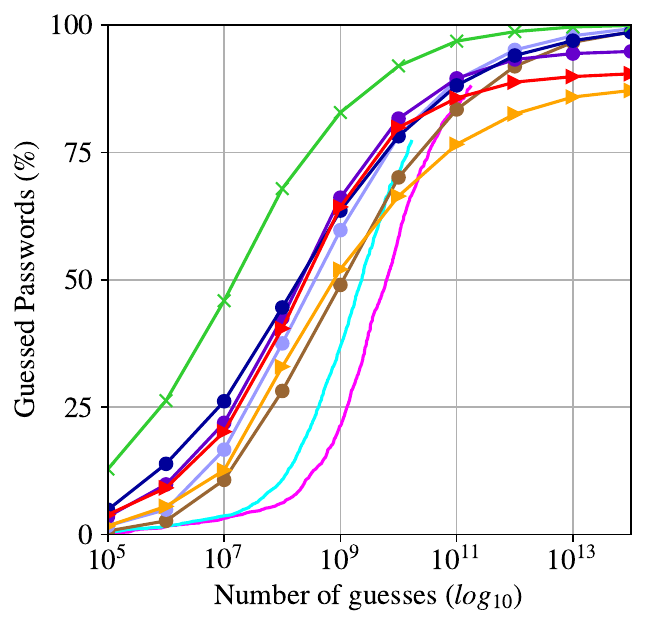}
		\caption{\footnotesize \textit{chinaguide.co.kr}}
	\end{subfigure}\begin{subfigure}{.42\columnwidth}
	\centering
	\includegraphics[trim = 0mm 0mm 0mm 0mm, clip, width=.95\columnwidth]{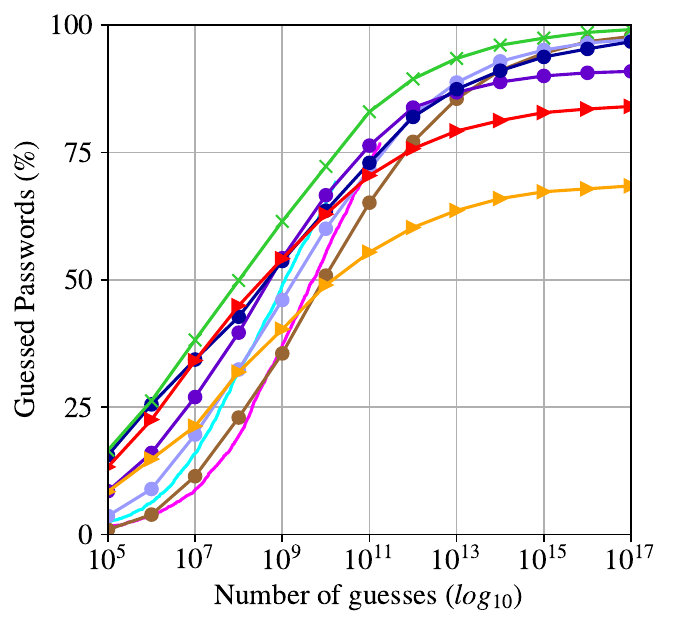}
	\caption{\footnotesize \textit{everdrybas$\dots$.mobi}}
	\end{subfigure}
}
\captionsetup{justification=raggedright}
	\caption{Comparison between the seeded password models and \textit{ADaMs} attack~\cite{pasquini2021usenix} (in addition to the models considered in Appendix~\ref{app:pcfg}). The notation $\seeded$ refers to the seeded model generated by the \uncm.}
	\label{fig:ex_guess_global_adams}
\end{figure*}
\subsection{Comparison with Dynamic password models}
\label{app:dynamic}
Dynamic password models~\cite{pasquini2021improving, pasquini2021usenix} can be seen as another attempt to solve the universal password model problem.  Those models, starting from a prior independent from the target password distribution, use the passwords that are guessed during the running attack as means to dynamic adapt the initial distribution to the target one. By design, this strategy requires to access plaintext passwords from the target set, making it applicable only in the guessing attacker setting. Indeed, in contrast with the \uncm, dynamic passwords models cannot be used as PSMs as no plaintext passwords are available besides the one tested by the meter. 
\par

For completeness, we report a comparison with the dynamic password model based on dictionary attack proposed in~\cite{pasquini2021usenix}. For this model, we use the two default models shipped with the official implementation: the adaptive models for the mangling rule-sets \textit{PasswordPro} and \textit{generated}. As input dictionary for the attacks, we use the union of all passwords in the leak collection $\lct$. From the latter, we remove all the duplicate entries and sort the passwords by decreasing frequency order. We run the guessing attack on the example leaks considered in Section~\ref{sec:result_un}. Results are reported in Figure~\ref{fig:ex_guess_global_adams}. We also report results for the other models considered in Appendix~\ref{app:pcfg}.
 The seeded password models outperform the dynamic dictionary attacks. Nonetheless, it is worth mentioning that the \textit{ADaMs} attacks results in a consistently more efficient implementation that support real-world guessing attacks, while the model of Melicher~\cite{melichera} is known to have limited application in this setting~\cite{pasquini2021usenix}. Thus, we recognize these two approaches to be complementary.    

\iflong

\else

\fi

\iflong
\section{Data cleaning}
\label{app:cleaning}
The cleaning procedure for the leak collection $\citday$ encompasses different steps.

\begin{enumerate}
	\item We built regular expressions for the most common hashes and removed all the accounts with passwords matching the regex(s).  
	\item We removed all the accounts that appear more than $150$ times in the different leaks (to filter out bots). 
	\item After the previous steps, we removed all the leaks with less than $100$ entries.
	\item We train a password model~\cite{melichera} on a manually curated subset of (clean) leaks and used it as anomaly detector. We applied the model on every leak in $\citday$ and removed the leaks with abnormal low-probability.
	\item Some of the leaks in the collection were duplicated. We computed the intersection between each set of passwords and removed the leaks with intersection larger than $90\%$
\end{enumerate}

Finally, a manual inspection on the test collection $\lcv$ has be carried out to remove abnormal leaks. The code for the cleaning pipeline as well as the list of leaks in $lct$ and $lcv$ is available in our repository. 
\fi

\iflong
\section{Membership inference attacks on $\cs$}
\label{app:mia}
\begin{figure}
\centering
	\resizebox{.8\columnwidth}{!}{
		\begin{tikzpicture}
		
			\tikzstyle{rnn} = [rectangle, minimum width=2.2cm, minimum height=1cm, text centered, draw=black, fill=green!10, rotate=90]
			\tikzstyle{embedding} = [rectangle, minimum width=2.2cm, minimum height=2.2cm,text centered, draw=black, fill=cyan!10]
			\tikzstyle{string} = []
			\tikzstyle{arrow0} = [->,>=stealth]
			\tikzstyle{arrow1} = [-, draw=black!70]
			\tikzstyle{vector} = [rectangle, minimum width=1cm, minimum height=.35cm,text centered, draw=black, fill=white]
			\tikzstyle{vectorv} = [vector, preaction={fill, white}, pattern=north west lines, pattern color=blue!15]

			\tikzstyle{encoder} = [rectangle, minimum width=1cm, minimum height=1cm,text centered, draw=black, fill=gray!10]

			

			\node (useramernn) [rnn,  xshift=0cm, yshift=7cm]{ \makecell{[RNN] \\ \textit{username}} };
			\node (d0embedding) [embedding,  right of=useramernn, xshift=1cm]{\makecell{[Emb. matrix] \\ \textit{provider}} };
			\node (d1embedding) [embedding,  right of=d0embedding, xshift=1.5cm, fill=cyan!05]{\makecell{[Emb. matrix] \\ \textit{domain}}};

			\node (semail) [string,  below of=d0embedding, yshift=-1cm,]{\Large{$\texttt{email}$}};
			
			\draw [arrow0] (semail) -- (useramernn.west) node[] {};
			\draw [arrow0] (semail) -- (d0embedding.south) node[] {};
			\draw [arrow0] (semail) -- (d1embedding.south) node[] {};
			
			\node (vuserame) [vector, above of=useramernn, xshift=.8cm, yshift=.8cm]{};
			\node (con0) [string,  right of=vuserame, xshift=-.25cm]{\large{$\|$}};
			\node (vd0) [vector, right of=vuserame, xshift=.5cm]{};
			\node (con0) [string,  right of=vd0, xshift=-.25cm]{\large{$\|$}};
			\node (vd1) [vector, right of=vd0, xshift=.5cm]{};
			
			\draw [arrow0] (useramernn.east) -- (vuserame) node[] {};
			\draw [arrow0] (d0embedding.north) -- (vd0) node[] {};
			\draw [arrow0] (d1embedding.north) -- (vd1) node[] {};

			\node (vemail) [vector, above of=vd0, yshift=.2cm, minimum width=2.1cm]{};
			\node (dvemail) [vector, above of=vd0, yshift=.82cm, minimum width=2.1cm, opacity=0]{};

			\draw [arrow0, shorten <= .1cm, shorten >= .1cm] (vd0) -- (vemail) node[] {};

			\node (seed) [vector, xshift=-1.3cm, left of=useramernn, yshift=3cm, minimum width=3cm, rotate=0, pattern=north west lines, pattern color=red!30]{$\cs\myeq \enc(\sainf)$};

			\node (dsptoken) [embedding,  left of=useramernn, xshift=-.5cm, yshift=4.6cm, opacity=0]{};	
			
			\node(dist) [rectangle, minimum width=4.2cm, minimum height=2.2cm, text centered, draw=black, fill=black!10, xshift=-6.5cm, yshift=5.5cm] {\large{\makecell{[Fully connected] \\ \textit{Distinguisher $D$}}}};

			\draw [arrow0] (seed.north) -- (dist.south) node[] {};
			\draw [arrow0] (vemail.north) -- (dist.south) node[] {};

			\node (outd) [above of=dist, yshift=1cm]{\Large{$P(\texttt{email} \in \sainf)$}};
			
			\draw [arrow0] (dist.north) -- (outd.south) node[] {};
						
			\end{tikzpicture}
	}

	\caption{Depiction of the attacker's model $\mathbf{A}$ trained to infer the membership of email addresses from a configuration seed~$\cs$.}
	\label{fig:miam}
\end{figure}
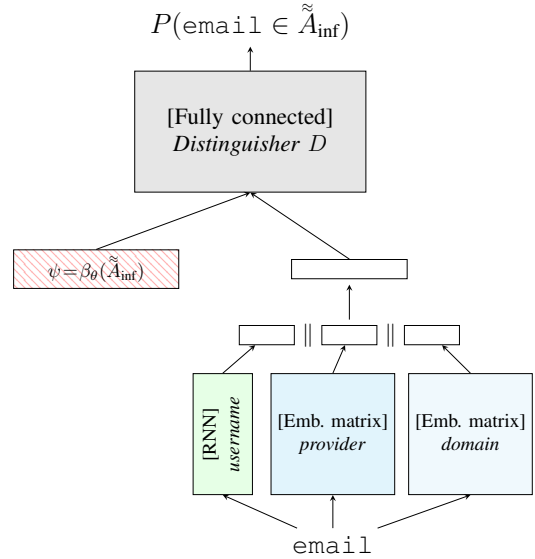
In this section, we introduce a membership inference attack (MIA) targeting the email addresses used to generate the configuration seed. Then, we use the attack to empirically validate the soundness of the proposed  \textit{private} \uncm.\\

\paragraph{\textbf{Setup}}
The setup for the attack involves a machine learning-based approach. Given a configuration seed $\cs \myeq \enc(\sainf)$ and an email address $\texttt{email}$, we train a neural network $\mathbf{A}$ to infer whether $\texttt{email} \myin \sainf$ \ie if the email address $\texttt{email}$ has been used as input to compute the seed. The network  $\mathbf{A}$ is depicted in Figure~\ref{fig:miam}. Here, we recycle the general architecture of the sub-encoder~$\senc$ (see Section~\ref{sec:subencoder}) to build a feature extractor for the input email address. The output of the sub-network, together with the configuration seed $\cs$, are then provided as input to a model~$D$. 
The distinguisher~$D$ is a fully-connected network with architecture:
\begin{enumerate}

	\item $\cs\ \| \ \senc(\texttt{email}) $ $\rightarrow$ \texttt{dense(512)}

	\item \texttt{dense(320)} $\rightarrow$ \texttt{batchnorm} $\rightarrow$ \texttt{ReLU}
	\item \texttt{dense(160)} $\rightarrow$ \texttt{batchnorm} $\rightarrow$ \texttt{ReLU}
	\item \texttt{dense(80)} $\rightarrow$ \texttt{batchnorm} $\rightarrow$ \texttt{ReLU}
	\item \texttt{dense(40)} $\rightarrow$ \texttt{batchnorm} $\rightarrow$ \texttt{ReLU}
	\item \texttt{dense(1)} $\rightarrow$ \texttt{sigmoid}
\end{enumerate}

The task of the distinguisher is to infer the probability that $\texttt{email} \in \sainf$ given the two inputs.

To train the model $\mathbf{A}$, we use $\lct$ (which is public). For every leak $S\myeq(\ainf,\ X) \in \lct$, we sample a subset of $k$ email address $\sainf$ from $\ainf$ and use it to compute a configuration seed $\cs$ via the pre-trained configuration encoder~$\enc$ (which is public as well). Then, we create a training set for the attacker's model $\mathbf{A}$ by generating triplets of the form:
$$
(\cs\myeq \enc(\sainf),\ \texttt{email},\ \texttt{email} \in \sainf),
$$ 
where the last element is a binary label $\{0,1\}$. During the training, the model is optimized to predict the correct label (\ie membership) given the input pair $(\cs,\  \texttt{email})$. When generating the training sets, we ensure that the two labels have the same proportion (\ie $50\%$ \TT{$0$} and  $50\%$ \TT{$1$}). We do the same for $\lcv$ and use the output as the validation set for the attack. We generate the databases for both the private \uncm~and the vanilla one (that we use as a baseline). In our setup, we study the attacks in the toy-setting with $k\myeq 10$ as MIAs fail on the non-private configuration seed when larger values for $k$ are used \eg $2048$. For the private setting, this results in $(\epsilon \myeq 1.44, \delta \myeq 5\mycdot 10^{-5})$-DP configuration seeds which is our worst-case setting.

For both the private and non-private settings, we train $\mathbf{A}$ using the \textit{Adam} optimizer~\cite{adam} in the default setting, a batch size of $256$, and early stopping. After the training, we tested the two models on the validation sets derived from~$\lcv$.

\paragraph{\textbf{Results}}
The accuracy of the attack on the non-private seed is $62.13\% \pm 0.60$, where random guessing is $50\%$. For the $(\epsilon \myeq 1.44, \delta \myeq 5\mycdot 10^{-5})$-DP seed, the accuracy of the MIAs is $50.26\% \pm 0.64$  which is marginally bellow the bound imposed by $(\epsilon \myeq 1.44, \delta \myeq 5\mycdot 10^{-5})$-DP.
\fi

\iflong
\else
\section{Meta-Review}

\subsection{Summary}
The paper introduces the notion of a ``Universal Neural-Cracking Machine" (UNCM) which can be trained on available password datasets and then quickly adjusted to generate a targeted model for a particular subpopulation e.g., Korean banking passwords. A UNCM uses deep learning to identify and exploit relationships between user passwords and available auxiliary data about the user account. Adjusting the UNCM only utilizes auxiliary data from the target subpopulation, and does not require additional training data.

\subsection{Scientific Contributions}
\begin{itemize}
\item Provides a Valuable Step Forward in an Established Field.
\item Creates a New Tool to Enable Future Science.
\end{itemize}

\subsection{Reasons for Acceptance}
\begin{enumerate}
\item Provides a Valuable Step Forward in an Established Field. There is a long line of research developing improved password cracking models. While many breached password datasets are available it is often unclear whether or not the model should be trained on all available datasets or only those that are most similar to the target population. The UNCM model is unique in that it can be trained on all available password datasets, and then efficiently tuned to obtain models targeting particular groups of users (e.g., Chinese, English, E-Commerce). The paper also shows how the tuning process can be performed in a differentially private manner to preserve privacy of auxiliary user data in defensive applications.
\item Creates a New Tool to Enable Future Science. The paper develops a new password cracking model by adapting recent ideas (Attention Mechanism) from the Natural Language Processing. The model outperforms or is equivalent to existing state-of-the-art and tailored password models.
\end{enumerate}

\subsection{Noteworthy Concerns}
\begin{enumerate} 

\item The term ``universal'' could easily be miss-interpreted. Thus, the reviewers want to clarify that a UNCM can only adapt to subpopulations that were well represented in the original training dataset. For example, if the training dataset does not include many passwords from Portuguese speaking users we would not expect a UNCM to effectively retarget the cracking model to Portuguese passwords without obtaining additional training data.
\end{enumerate}
\fi
	
\end{document}